%% file: K10_v3_arxiv.tex
\documentclass[iop]{emulateapj}
\usepackage{graphicx}
\usepackage{amssymb}
\usepackage{amsmath}
\usepackage{epstopdf}
\usepackage{enumerate}
\usepackage{url}
\usepackage{rotating}
\usepackage{appendix}
\usepackage{longtable}
\newcommand{\Kepler}{\textit{Kepler}}
\newcommand{\TTAgol}{\ensuremath{\rm TT_{\mathrm{Agol}}}}
\newcommand{\TTKip}{\ensuremath{\rm TT_{\mathrm{Kip}}}}
\newcommand{\PTTV}{\ensuremath{\rm P_{\mathrm{TTV}}}}
\newcommand{\ecosom}{\ensuremath{\sqrt{e_c}\mathrm{cos}\omega_c}}
\newcommand{\esinom}{\ensuremath{\sqrt{e_c}\mathrm{sin}\omega_c}}

\newcommand{\TT}{\ensuremath{T_{t,c}}}
\newcommand{\Kb}{\ensuremath{K_b}}
\newcommand{\Kc}{\ensuremath{K_c}}
\newcommand{\mbcirc}{\ensuremath{3.72\pm0.42~ \mearth}} 
\newcommand{\mccirc}{\ensuremath{13.98\pm1.79~ \mearth}}
\newcommand{\rhobcirc}{\ensuremath{6.46\pm0.73~ \gcc}} 
\newcommand{\rhoccirc}{\ensuremath{5.94\pm0.76~ \gcc}}
\newcommand{\mbecc}{\ensuremath{3.76 \pm0.43~ \mearth}}
\newcommand{\mcecc}{\ensuremath{14.59\pm1.90~ \mearth}}
\newcommand{\rhobecc}{\ensuremath{6.53\pm0.75~ \gcc}}
\newcommand{\rhocecc}{\ensuremath{6.21\pm0.81~ \gcc}}

\input{physical_quantities.tex}

\begin{document}
\title{Revised Masses and Densities of the Planets around Kepler-10$^*$}
\author{Lauren~M.~Weiss$^{1,\dagger,\star}$, Leslie~A.~Rogers$^{2,\ddagger}$, Howard~T.~Isaacson$^1$, Eric Agol$^{3,4}$, Geoffrey~W.~Marcy$^1$, Jason~F.~Rowe$^5$, David~Kipping$^6$, Benjamin J. Fulton$^7$, Jack~J.~Lissauer$^5$, Andrew~W.~Howard$^7$, Daniel Fabrycky$^8$.}
\affil{$^1$ Astronomy Department, University of California at Berkeley, 501 Campbell Hall, Berkeley, CA 94720, USA}
\affil{$^2$ California Institute of Technology Division of Geological and Planetary Sciences, 1200 East California Boulevard, Pasadena, CA 91125, USA}
\affil{$^3$ NASA Astrobiology Institute's Virtual Planetary Laboratory, Pasadena, CA 91125, USA}
\affil{$^4$ Department of Astronomy, Box 351580, University of Washington, Seattle, WA 98195, USA}
\affil{$^5$ NASA Ames Research Center, Moffett Field, CA 94035, USA}
\affil{$^6$ Harvard University Center for Astrophysics, 60 Garden Street, Cambridge, MA 02138, USA}
\affil{$^7$ Institute for Astronomy, University of Hawaii at Manoa, Honolulu, HI 96822, USA}
\affil{$^8$ Department of Astronomy and Astrophysics, The University of Chicago, 5640 South Ellis Avenue, Chicago, IL 60637, USA}
\altaffiltext{$*$}{\small W.M.O. Keck Observatory}
\altaffiltext{$\dagger$}{\small Ken \& Gloria Levy Fellow.}
\altaffiltext{$\ddagger$}{\small Hubble Fellow.}
\altaffiltext{$\star$}{\small lweiss@berkeley.edu}

\submitted{Accepted for publication in ApJ, Jan. 20, 2016.} 

\begin{abstract}
Determining which small exoplanets have stony-iron compositions is necessary for quantifying the occurrence of such planets and for understanding the physics of planet formation.  Kepler-10 hosts the stony-iron world Kepler-10b, and also contains what has been reported to be the largest solid silicate-ice planet, Kepler-10c.  Using 220 radial velocities (RVs), including 72 precise RVs from Keck-HIRES of which 20 are new from 2014-2015, and 17 quarters of \Kepler\ photometry, we obtain the most complete picture of the Kepler-10 system to date.  We find that Kepler-10b ($\rpl = 1.47 ~\rearth$) has mass \mbcirc\ and density \rhobcirc.  Modeling the interior of Kepler-10b as an iron core overlaid with a silicate mantle, we find that the iron core constitutes $0.17\pm0.11$ of the planet mass.  For Kepler-10c ($\rpl = 2.35 ~\rearth$) we measure mass \mccirc\ and density \rhoccirc, significantly lower than the mass computed in \citet[][$17.2\pm1.9~\mearth$]{Dumusque2014}.  Our mass measurement of Kepler-10c rules out a pure stony-iron composition.  Internal compositional modeling reveals that at least 10\% of the radius of Kepler-10c is a volatile envelope composed of hydrogen-helium ($0.2\%$ of the mass, $16\%$ of the radius) or super-ionic water ($28\%$ of the mass, $29\%$ of the radius). However, we note that analysis of only HIRES data yields a higher mass for planet b and a lower mass for planet c than does analysis of the HARPS-N data alone, with the mass estimates for Kepler-10 c being formally inconsistent at the $3 \sigma$ level.  Moreover, dividing the data for each instrument into two parts also leads to somewhat inconsistent measurements for the mass of planet c derived from each observatory. Together, this suggests that time-correlated noise is present and that the uncertainties in the masses of the planets (especially planet c) likely exceed our formal estimates. Transit timing variations (TTVs) of Kepler-10c indicate the likely presence of a third planet in the system, KOI-72.X.  The TTVs and RVs are consistent with KOI-72.X having an orbital period of 24, 71, or 101 days, and a mass from 1-7$~$\mearth.  
\end{abstract}

\section{Introduction}
The thousands of high-fidelity planet candidates between 1 and 4 Earth radii discovered by the \Kepler\ Mission \citep{Borucki2011,Batalha2013,Burke2014,Rowe2014,Mullally2015}, though absent from our solar system, are abundant in orbital periods $<100$ days around Sun-like stars \citep{Petigura2013, Fressin2013, Petigura2013a}.  To understand the formation of these common planets, we must constrain their compositions.  Are they terrestrial, or are they ``water worlds" that are primarily water by volume, or are they stony-iron cores overlaid with thick, hydrogen-rich envelopes of volatiles?  

In the last few years, the exoplanet community has measured the masses of dozens of small exoplanets, enabling the study of the compositions of individual planets and the identification of several stony-iron super-Earths.  Corot-7 b \citep[$\rpl=1.58\pm0.10 \rearth$, $\mpl=5.37\pm1.02~ \mearth$;][]{Bruntt2010, Haywood2014} and Kepler-10 b \citep[$\rpl=1.46\pm0.034~\rearth$, $\mpl=4.56\pm1.23~ \mearth$;][]{Batalha2011} were the first stony-iron planets discovered.  \citet{Carter2012} used transit timing variations to determine the mass of Kepler-36 b from orbital perturbations it induced on neighboring planet Kepler-36 c.  At the time of writing, Kepler-36 b has the best-determined mass and density of the known rocky exoplanets ($\mpl=4.56\pm1.23$, $\rhopl=8.8\pm2.5\gcc$).  \citet{Howard2013rocky}, \citet{Pepe2013}, and \citet{Grunblatt2015} measured the mass of the Earth-density planet Kepler-78 b ($\rpl = 1.20\pm0.09\rearth$, $\mpl = 1.87\pm0.26\mearth$, $\rhopl = 6.0\pm1.7\gcc$), the closest Earth-analog in terms of planet mass, radius, and density, although it is far too hot to support life as we know it.  

However, some small planets have definitively non-rocky surfaces and require hydrogen-helium envelopes to explain their low bulk densities.  For instance, three of the six planets orbiting Kepler-11 are smaller than 4 \rearth\ and have densities lower than 1.0 \gcc\  \citep{Lissauer2011,Lissauer2013}.  Likewise, two of four the planets orbiting Kepler-79 (a.k.a. KOI-152) are smaller than 4 \rearth\ and have densities lower than 1.0 \gcc\ \citep{Jontof-Hutter2014}.  In an intensive \Kepler\ follow-up campaign spanning 4 years, \citet{Marcy2014} measured or constrained the masses of 42 small exoplanets using Keck-HIRES, finding many planets that have volatiles and a few planets that might have stony-iron compositions.

The mass measurements listed above allowed the community to probe composition trends within the planet population.  Based on the density-radius distribution of 65 exoplanets smaller than Neptune, \citet{Weiss2014} found two empirical relations: among planets smaller than 1.5 \rearth, density increases nearly linearly with increasing planet radius in a manner consistent with a stony-iron composition like Earth's.  However, bulk density decreases with increasing radius for planets between 1.5-4.0 \rearth, implying an increasing admixture of volatiles above 1.5 \rearth. \citet{Rogers2015} used a hierarchical Bayesian framework to rigorously test the transition from stony-iron planets to planets with a gaseous envelope and found that at and above $1.6~\rearth$, the majority of planets have a volatile envelope, while the remaining minority are sufficiently dense to be comprised of iron and silicate only.  \citet{Dressing2015} measured the mass of Kepler-93 b ($\rpl=1.478\pm0.019$, $\mpl=4.02\pm0.68$) and determined that Kepler-93 b and the other known rocky planets (Kepler-78 b, Kepler-36 b, Kepler-10 b, and Corot-7 b) all have masses and radii that can be explained with an iron-silicate composition.  By contrast, KOI-273 b \citep[$\rpl=1.82 \pm 0.10\rearth$, $\mpl=5.46\pm2.50 \mearth$][]{Gettel2015} is too big to be rocky and requires a small volatile envelope.  \citet{Wolfgang2015} used a hierarchical Bayesian model to explore the diversity of planet mass, density, and composition as a function of planet radius.  They found that planets smaller than 1.5 \rearth\ are typically rocky, whereas planets larger than 1.5 \rearth\ typically require a small fraction of hydrogen gas or other volatiles to explain their densities.  Furthermore, small differences in the mass fraction of hydrogen in the planet's envelope explain the broad range of planet densities at a given radius for planets between 2-4\rearth.

Although planets smaller than 1.5 \rearth\ tend to be stony-iron and planets larger than 1.5 \rearth\ tend to have at least a small hydrogen envelope, there are exceptions to the pattern.  In the Kepler-138 system, which contains three planets smaller than 1.5 \rearth, at least one planet, Kepler-138 d ($\rpl = 1.212 \pm 0.075 \rearth$, $\mpl = 0.640^{+0.674}_{-0.387} \mearth$, $\rhopl = 2.1^{+2.2}_{-1.2} \gcc$), has a low enough density to require a volatile envelope \citep{Kipping2014, Jontof-Hutter2015_nature}.  Kepler-138 d is the smallest exoplanet that we know to contain a gaseous envelope.

The Kepler-10 system is a powerful testing ground for our understanding of the compositions of small planets.  Kepler-10 is a sun-like star with slow rotation and little stellar activity \citep{Dumusque2014}.  It has two planets discovered via transits in the \Kepler\ Mission: Kepler-10 b, which has an orbital period of 0.84 days and radius 1.47 \rearth, and Kepler-10 c, which has an orbital period of 45 days and radius 2.35 \rearth\ \citep{Batalha2011, Dumusque2014}.  \citet[][hereafter B11]{Batalha2011} measured the mass and bulk density of Kepler-10 b and determined that it was rocky, making this planet the first rocky planet discovered by the \Kepler\ Mission, and the second rocky exoplanet discovery.  More recently, \citet[][hereafter D14]{Dumusque2014} reported that Kepler-10 c has a radius of $2.35~\rearth$, a mass of $17.2\pm1.9~\mearth$, and a density of $7.1\pm~1.0 \gcc$.  Based on its position in the mass-radius diagram, D14 interpreted the composition of Kepler-10 c as mostly rock by mass, with the remaining mass in volatiles of high mean-molecular weight (likely water).  They noted, however, that compositional degeneracy prevented them from determining the precise water fraction.

Kepler-10 c is unusual in that the mass reported in D14 is large compared to other exoplanets its size.  Most exoplanets with radii 2.0-2.5$~$\rearth\ have much lower masses than 17$~$\mearth, with a weighted mean mass of 5.4$~$\mearth\ \citep{Weiss2014} in that size range.  For example, HD 97658 b, a planet discovered in RVs \citep{Howard2011} that was subsequently observed to transit its star, has a radius of $2.34\pm0.16 ~\rearth$ and a mass of $7.87\pm0.73 ~\mearth$ \citep{Dragomir2013}.  Kepler-68 b has a radius of $2.32\pm0.02~\rearth$ and a mass of $7.15\pm2.0~\mearth$ \citep{Marcy2014, Gilliland2013}.  Although there is a large scatter in the observed masses between 2 and 2.5$~$\rearth, this scatter results from a few low-mass planets of this size.  For example, Kepler-11 f, which has a radius of $2.49\pm0.06~\rearth$, has a mass of $1.94_{-0.88}^{+0.32} ~\mearth$ \citep[][Weiss et al. in prep.]{Lissauer2011,Lissauer2013}.  In contrast, the most massive planet in this size range other than Kepler-10 c is Kepler-131 b.  The initial mass measurement of Kepler-131 b \citep{Marcy2014} was $\rpl= 2.41\pm0.20~\rearth$, $\mpl=16.13\pm3.50~\mearth$, resulting in a bulk density of $6.0\pm1.98~\gcc$, but additional measurements obtained since publication show the mass to be much smaller; the confusion was from astrophysical rather than instrumental sources (personal communication, H. Isaacson in prep.).  Thus, Kepler-10 c seems to be unusual in its high mass for planets between 2-2.5$~\rearth$.  All of these planets except Kepler-10 c are included in the empirical mass-radius relation to exoplanets between 1.5 and 4 \rearth\ \citep[$\mpl/\mearth = 2.69 (\rpl/\rearth)^{0.93}$;][]{Weiss2014}, according to which a planet of size 2.3$~$\rearth\ should have a mass of 5.8$~$\mearth.

In this paper, we build on the data and analysis of D14, adhering to the techniques therein as completely as possible but with the addition of 72 RVs from Keck-HIRES, in an effort to calculate a new and improved two-planet orbital solution for the Kepler-10 system.  We also notice that Kepler-10 c exhibits transit timing variations (TTVs), i.e. perturbations to its orbit, as did \citet{Kipping2015}.  Because Kepler-10 b is dynamically distant from Kepler-10 c ($P_c/P_b=54$), Kepler-10 b cannot perturb Kepler-10 c sufficiently to reproduce the observed TTVs.  Therefore, we infer the existence of a third planet in the system, planet candidate KOI-72.X, which explains the observed TTVs.  We explore various dynamic configurations for KOI-72.X that reproduce the observed TTVs and are consistent with the RVs as well.  Finally, we comment on the compositions of Kepler-10 b and Kepler-10 c, and how their masses, radii, and densities compare to those of other small transiting planets.

\section{Radial Velocities of Kepler-10 from HIRES and HARPS-N}
\begin{figure}[htbp]
\begin{center}
\includegraphics[width=3in]{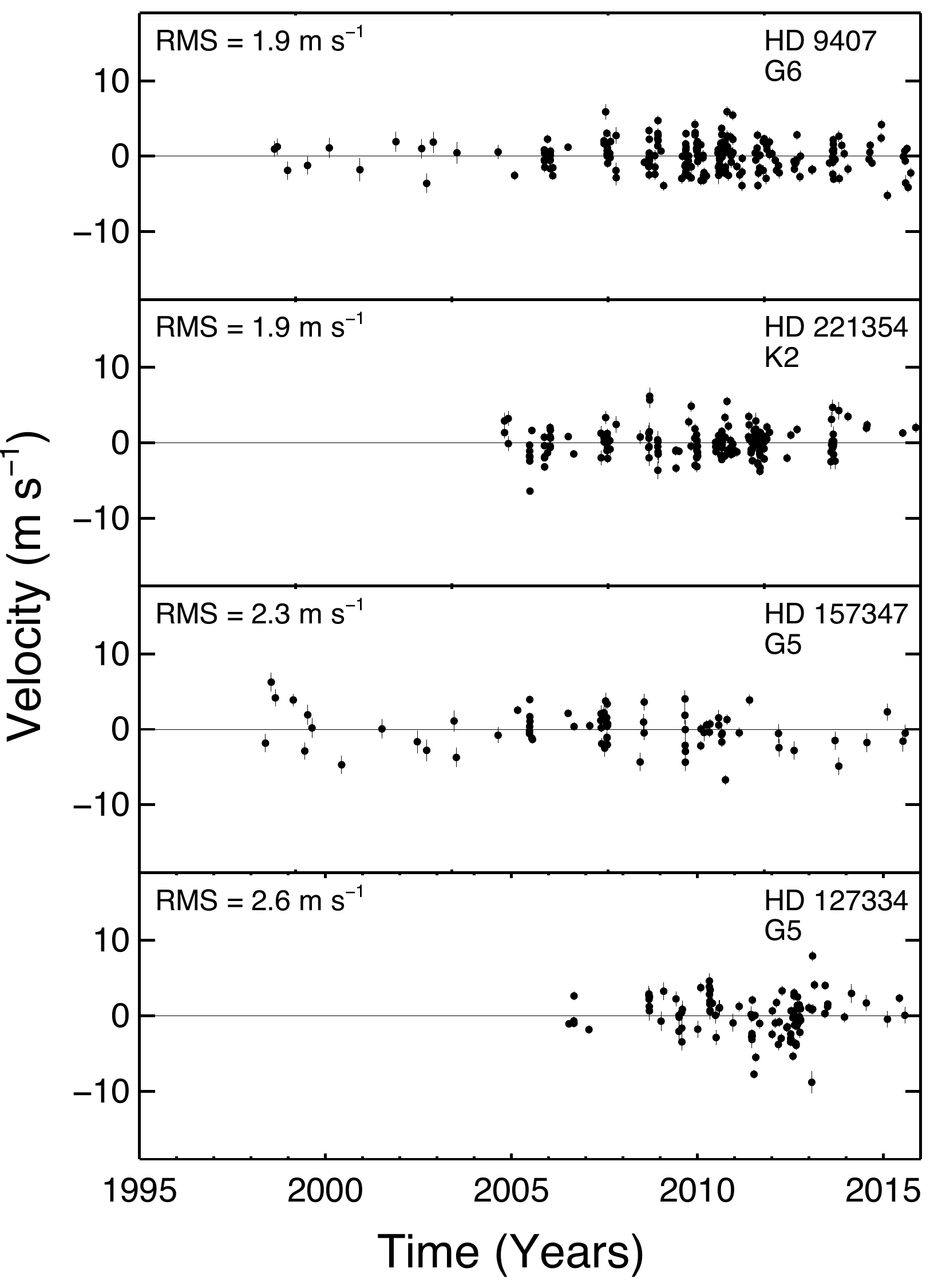}
\caption{RV vs. time for stable stars observed by Keck-HIRES.  The RMS of the RVs, stellar name, and stellar spectral type are shown.  The typical RMS of $\sim2~\ms$ achieved by Keck-HIRES over a decade for stars without planets demonstrates the ability of HIRES as a multi-season, precision-RV instrument.}
\label{fig:stable_stars}
\end{center}
\end{figure}

HIRES has a long history of achieving precision RVs with an RMS of $\sim2~\ms$ on quiet, sun-like stars over many years of observations \citep[][Figure \ref{fig:stable_stars}]{Howard2010_CPS, Howard2011}.  Our group has used HIRES to measure and place upper limits on the masses of many small planets, especially in the \Kepler\ era \citep[e.g.][]{Marcy2014}.  Because the planets transit and are vetted through a variety of astrophysical techniques, the burden of confirming the planet does not fall entirely to radial velocities.  In this case, \citet{Fressin2011} validated the planetary nature of Kepler-10 c, incorporating the transit shape and depth in multiple passbands, high-resolution imaging and spectroscopy of the host star, and stellar population synthesis to find a conservative false alarm probability of $1.6\times10^{-5}$.  \citet{Lissauer2012} demonstrated the very low probability of having a false alarm planet in a multi-planet system, further reducing the false alarm probability of Kepler-10 c by an order of magnitude.  Thus, it is not necessary to determine the mass of Kepler-10 c (or any other statistically validated small planet) with $3\sigma$ significance in order to confirm the planet's existence or to make statistically significant claims about the composition of the planet.  A mass upper limit might exclude a purely stony-iron composition with $3\sigma$ confidence while only being $1\sigma$ away from a mass of zero.  Such planets provide valuable information about the exoplanet population, and excluding them from population studies on the basis of their large fractional mass uncertainty ($\sigma_m/m$) will systematically exclude the low-mass exoplanets.  Modern exoplanet mass-radius relations \citep[e.g.][]{Weiss2014, Rogers2015, Wolfgang2015} incorporate the low-significance mass detections.  

Both the HIRES and HARPS-N spectrographs have successfully obtained high-precision RVs of \Kepler\ stars in the past \citep{Marcy2014, Dressing2015}.  Notably, two teams used each spectrograph to independently measure the RV signal from the low-mass planet Kepler-78 b.  \citet{Howard2013rocky} obtained a mass of $1.69\pm0.41 \mearth$ for Kepler-78 b, and \citet{Pepe2013} obtained a mass of $1.86\pm0.32$.  The independent detections of the RV signal in both spectrometers, and the agreement in the amplitude of that signal, demonstrate that both the HIRES and HARPS-N spectrometers are capable of accurately and precisely measuring low-amplitude RV signals.

Ground-based RV follow-up of Kepler-10 has been ongoing since Kepler-10 b and Kepler-10 c were discovered.  B11 presented 52 RVs obtained on Keck-HIRES in 2009--2010, the first seasons after Kepler-10 b and c were discovered, and D14 presented 148 RVs obtained on TNG-HARPS-N that span the summers of 2012--2013.  The early measurements presented in B11 targeted the quadrature times of planet b, whereas the later measurements from D14 targeted the quadrature times of planet c.

\begin{figure}[htbp] 
   \centering
\includegraphics[width=3in]{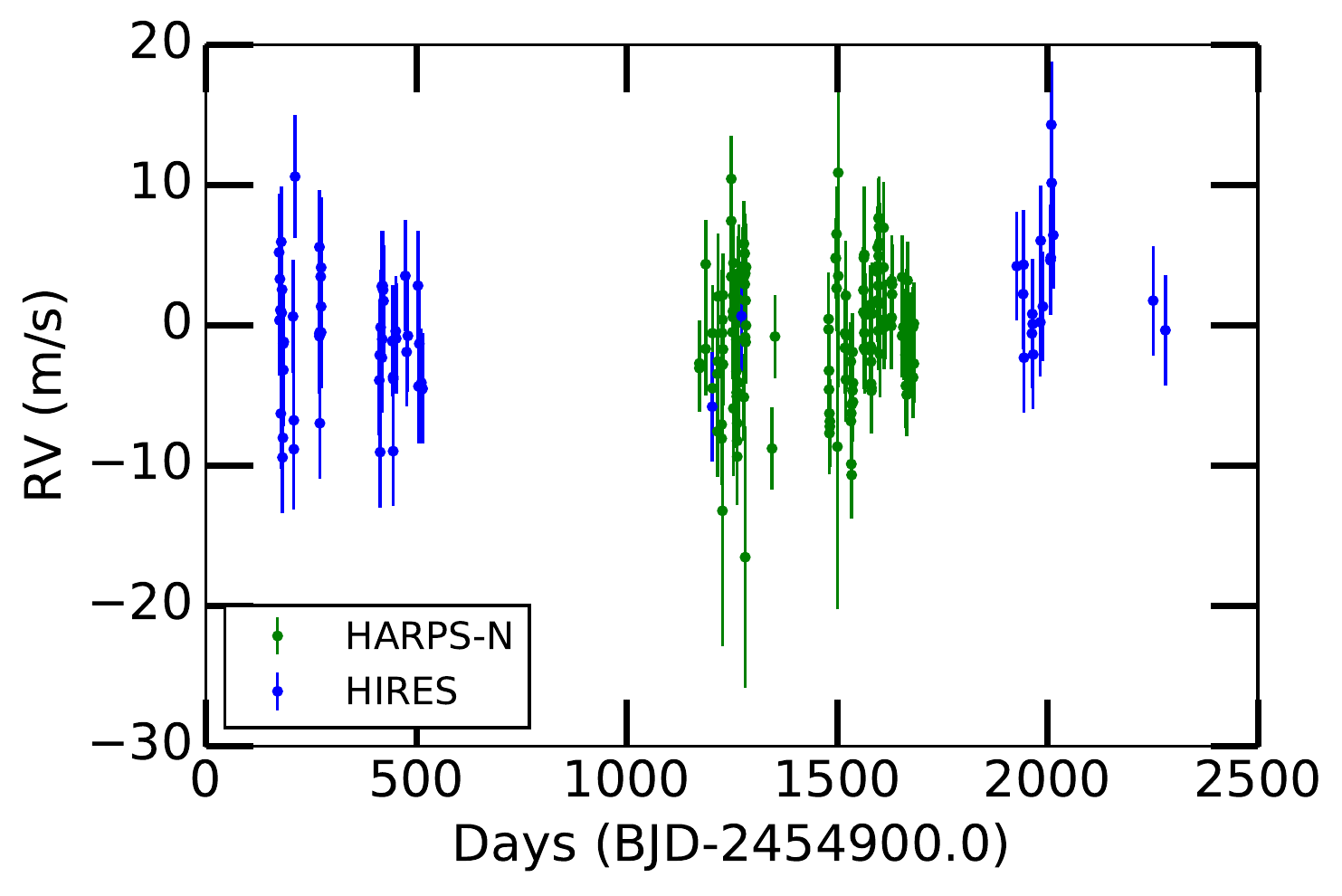}
   \caption{Radial velocity measurements of Kepler-10 from the HIRES (blue) and HARPS-N (green) high-resolution echelle spectrometers.}
   \label{fig:rvs}
\end{figure}

We present 20 additional RVs from 2014-2015 which, in combination with all the previous RVs, comprise the largest dataset of Kepler-10 RVs to date of 220 RVs total (Table 1 and Figure \ref{fig:rvs}).  HARPS-N, which is a fiber-fed, thermally stable spectrometer in a vacuum, achieves better velocity precision at given signal-to-noise than the HIRES spectrometer, but the larger aperture of the Keck telescope (10 m compared to 3.6 m) collects more photons.  Thus, both telescope-spectrometer setups achieve a velocity precision of a few \ms\ per half hour observing Kepler-10.  

\begin{figure}[hbtp] 
   \centering
\includegraphics[width=3in]{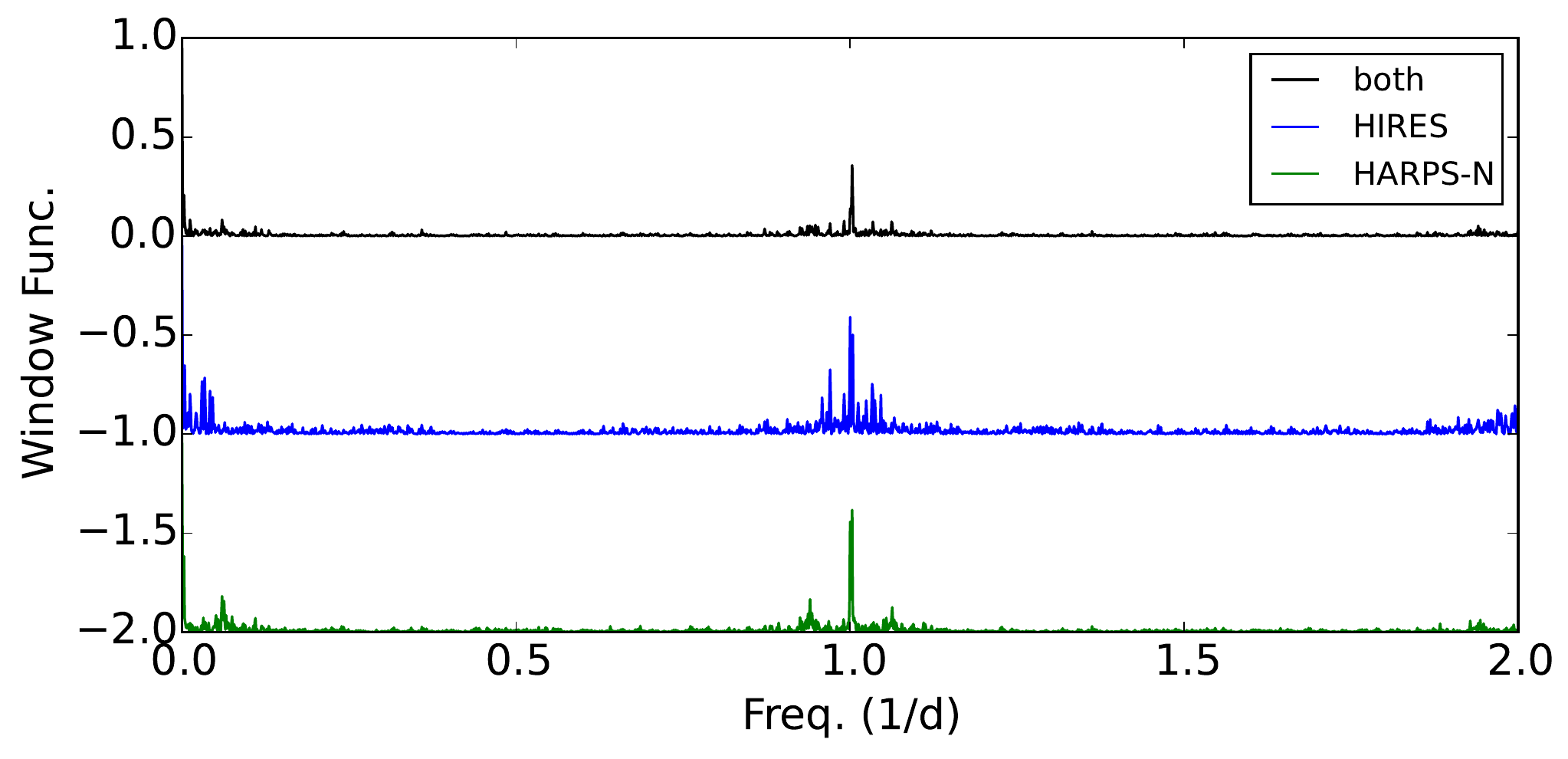}
\includegraphics[width=3in]{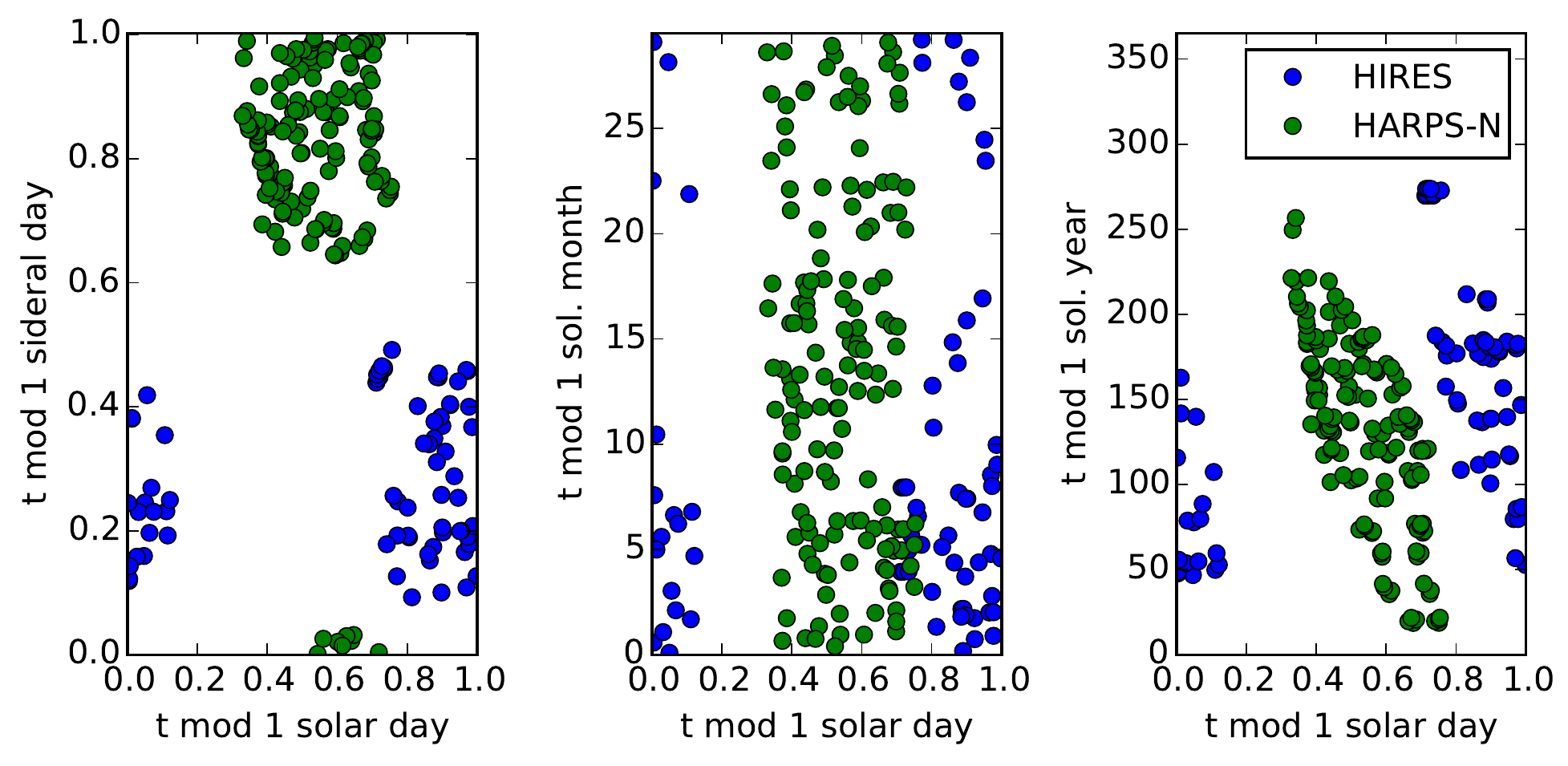}
   \caption{Top: window function of the combined RVs, HIRES RVs, and HARPS-N RVs, vertically offset for clarity.  The peaks near 1/day indicate the daily alias in each data set, and the peaks near 0.06/day are consistent with a monthly alias.  The combined RVs reduce the strength of the daily alias and remove the monthly alias.  Bottom left: time of observation modulo the sidereal day versus time of observation modulo the solar day shows the sidereal and solar daily phase coverage of the observations from HIRES (blue) and HARPS-N (green).  Bottom center: same as bottom left, but with time modulo the solar month on the dependent axis.  Bottom right: same as bottom left, but with time modulo the solar year on the dependent axis.}
   \label{fig:window}
\end{figure}

Although only 20 RVs were taken since the publication of D14, combining all the data provides several major advantages over either the B11 or D14 data alone.  Because the \Kepler\ field is best accessible during the summer, the data from both HIRES and HARPS-N are clumped in intervals of 2-3 months, a timescale barely longer than the orbital period of Kepler-10 c (45.3 days).  Observing just 1-2 orbits of planet c could be problematic if the stellar rotation period is comparable to the orbital period of planet c and temporarily phases with the orbit of planet c over a few rotation cyles.  Furthermore, incomplete observing phase coverage combined with noise can result in additional power in an alias of the planetary signal or a peak resulting from the window function \citep{Dawson2010, Rajpaul2015}.  The combined data cover observing phase as a function of sidereal day, solar day, and solar year better than either data set does alone (see Figure \ref{fig:window}), improving our resilience to noise manifesting in monthly and yearly aliases of planet c's orbit.  Furthermore, the combined baseline of 6 years (2009-2015) is much longer than the 2-year baseline achieved in either of the previous papers.  The long observing baseline helps to average out possible spurious signals that can arise from stellar activity on the timescales of stellar rotation and convection ($\sim1$ month).  The long baseline also improves our sensitivity to possible long-period signals.  These advantages motivate combining all of the reliable data.

\begin{center}
\begin{longtable}{lrrrl}
\tablecaption{RVs of Kepler-10 from HIRES and HARPS-N.}
\label{tab:rvs}
\tablenum{1}
\endfirsthead
\tablehead{\colhead{BJD} & \colhead{RV} & \colhead{unc. RV} & \colhead{SNR} & \colhead{Instrument} \cr 
\colhead{( - 2454900.0)} & \colhead{(m/s)} & \colhead{(m/s, inc. jitter)} & \colhead{} & \colhead{} } 
\endhead
\multicolumn{5}{r}{{Continued on next page}} \cr \hline
\endfoot
\multicolumn{5}{l}{{\bf Notes.}}\cr
\multicolumn{5}{l}{{Jitter has already been applied to the RV uncertainties.}}\\
\multicolumn{5}{l}{{Offsets between the RV data sets have been applied.}}\\
\multicolumn{5}{l}{{HARPS-N 1 refers to the pre-upgrade CCD (before 21 Sep. 2012).}}\\
\multicolumn{5}{l}{{HARPS-N 2 refers to the post-upgrade CCD (after 12 Nov. 2012).}}\\
\endlastfoot
173.900499 & 5.21 & 4.18 & 152.8 & HIRES \\
174.877797 & 0.37 & 3.94 & 217.3 & HIRES \\
175.773348 & 3.31 & 3.95 & 214.6 & HIRES \\
176.862854 & 1.10 & 3.96 & 217.1 & HIRES \\
177.923401 & -6.28 & 3.97 & 222.3 & HIRES \\
178.922398 & 5.96 & 3.96 & 217.8 & HIRES \\
179.972876 & 0.90 & 4.00 & 217.8 & HIRES \\
180.896063 & 2.56 & 3.94 & 216.3 & HIRES \\
181.969271 & -9.41 & 3.96 & 217.6 & HIRES \\
182.847887 & -8.01 & 3.97 & 215.8 & HIRES \\
183.760854 & -1.30 & 3.92 & 217.0 & HIRES \\
183.945387 & -3.17 & 3.99 & 218.2 & HIRES \\
184.877994 & -1.17 & 3.93 & 217.0 & HIRES \\
206.889914 & 0.64 & 4.04 & 220.6 & HIRES \\
208.885123 & -8.82 & 4.28 & 154.1 & HIRES \\
208.890922 & -6.76 & 5.28 & 80.0 & HIRES \\
211.830107 & 10.61 & 4.38 & 166.7 & HIRES \\
269.71177 & -0.59 & 4.22 & 152.7 & HIRES \\
269.720925 & -0.77 & 4.08 & 202.7 & HIRES \\
269.733899 & 5.59 & 4.04 & 215.5 & HIRES \\
270.715114 & -0.47 & 3.99 & 214.5 & HIRES \\
270.733655 & -6.97 & 3.98 & 213.4 & HIRES \\
272.756489 & 3.48 & 4.03 & 157.2 & HIRES \\
273.714025 & 1.35 & 3.97 & 198.7 & HIRES \\
273.720425 & 4.13 & 5.03 & 75.4 & HIRES \\
273.727555 & -0.49 & 4.00 & 193.4 & HIRES \\
412.04715 & -3.91 & 3.94 & 217.3 & HIRES \\
413.004124 & -2.11 & 3.94 & 215.9 & HIRES \\
414.004814 & -9.03 & 3.95 & 215.3 & HIRES \\
415.111272 & -0.13 & 4.12 & 205.1 & HIRES \\
417.998478 & 2.79 & 3.96 & 215.8 & HIRES \\
418.121283 & -2.30 & 3.93 & 215.4 & HIRES \\
419.027179 & -0.98 & 4.06 & 214.7 & HIRES \\
420.062974 & 2.83 & 3.95 & 218.5 & HIRES \\
421.006521 & 2.53 & 3.93 & 218.1 & HIRES \\
421.969467 & 1.75 & 3.97 & 217.7 & HIRES \\
443.050045 & -1.13 & 3.90 & 241.1 & HIRES \\
444.031958 & -1.08 & 3.93 & 242.3 & HIRES \\
444.964655 & -3.66 & 4.34 & 132.4 & HIRES \\
444.977237 & -8.96 & 3.93 & 241.4 & HIRES \\
445.068315 & -3.82 & 4.00 & 218.3 & HIRES \\
450.972744 & -0.41 & 3.92 & 243.6 & HIRES \\
451.987745 & -0.93 & 3.95 & 242.7 & HIRES \\
473.81361 & 3.54 & 3.97 & 241.1 & HIRES \\
476.86488 & -1.89 & 3.88 & 313.6 & HIRES \\
479.902118 & -0.74 & 3.98 & 242.6 & HIRES \\
503.897962 & 2.84 & 3.89 & 313.6 & HIRES \\
505.056086 & -4.34 & 4.06 & 211.1 & HIRES \\
507.012774 & -1.30 & 3.92 & 311.8 & HIRES \\
511.985723 & -4.08 & 3.86 & 314.8 & HIRES \\
512.805268 & -4.57 & 3.86 & 314.4 & HIRES \\
514.80255 & -4.50 & 3.93 & 281.8 & HIRES \\
1172.682384 & -2.70 & 3.05 & 56.2 & HARPS-N 1 \\
1172.704768 & -3.04 & 3.11 & 53.4 & HARPS-N 1 \\
1187.57572 & -1.67 & 3.32 & 45.0 & HARPS-N 1 \\
1187.596901 & 4.37 & 3.18 & 48.5 & HARPS-N 1 \\
1203.106739 & -5.79 & 3.90 & 312.2 & HIRES \\
1203.661644 & -0.55 & 3.45 & 38.7 & HARPS-N 1 \\
1203.689793 & -4.47 & 3.32 & 41.9 & HARPS-N 1 \\
1215.691945 & -3.45 & 3.04 & 60.6 & HARPS-N 1 \\
1215.713149 & -7.55 & 3.22 & 52.5 & HARPS-N 1 \\
1216.704755 & 2.06 & 4.47 & 28.4 & HARPS-N 1 \\
1216.719003 & -2.56 & 4.77 & 26.5 & HARPS-N 1 \\
1225.568254 & -7.06 & 3.76 & 35.2 & HARPS-N 1 \\
1225.589446 & -8.06 & 3.31 & 44.2 & HARPS-N 1 \\
1226.447899 & -2.83 & 2.94 & 59.4 & HARPS-N 1 \\
1226.664948 & -0.54 & 4.53 & 28.7 & HARPS-N 1 \\
1227.422428 & -13.21 & 9.66 & 13.9 & HARPS-N 1 \\
1227.441641 & 0.41 & 2.91 & 62.7 & HARPS-N 1 \\
1228.43499 & 2.16 & 2.95 & 60.1 & HARPS-N 1 \\
1228.560476 & -2.77 & 2.94 & 59.7 & HARPS-N 1 \\
1228.662918 & -1.70 & 3.14 & 49.7 & HARPS-N 1 \\
1248.408122 & 7.45 & 3.22 & 43.5 & HARPS-N 1 \\
1248.511779 & 10.45 & 3.06 & 48.5 & HARPS-N 1 \\
1248.617819 & 3.48 & 3.05 & 53.9 & HARPS-N 1 \\
1251.396386 & -0.48 & 3.33 & 42.5 & HARPS-N 1 \\
1252.407893 & 1.01 & 2.84 & 67.9 & HARPS-N 1 \\
1252.640275 & 0.59 & 3.36 & 45.4 & HARPS-N 1 \\
1253.395106 & 2.11 & 2.98 & 54.3 & HARPS-N 1 \\
1253.493404 & -5.91 & 4.83 & 25.1 & HARPS-N 1 \\
1253.647013 & 4.45 & 2.97 & 58.8 & HARPS-N 1 \\
1260.473487 & -1.01 & 4.25 & 31.4 & HARPS-N 1 \\
1260.626355 & -3.28 & 3.98 & 34.8 & HARPS-N 1 \\
1261.39719 & -4.78 & 3.42 & 40.7 & HARPS-N 1 \\
1261.573183 & -5.06 & 3.65 & 38.9 & HARPS-N 1 \\
1262.394398 & -8.23 & 3.33 & 43.4 & HARPS-N 1 \\
1262.487962 & -6.97 & 3.28 & 45.2 & HARPS-N 1 \\
1262.567972 & -9.35 & 3.47 & 42.1 & HARPS-N 1 \\
1264.385261 & 1.87 & 3.27 & 44.4 & HARPS-N 1 \\
1265.380836 & 3.11 & 3.27 & 45.0 & HARPS-N 1 \\
1266.384814 & 3.74 & 3.45 & 38.9 & HARPS-N 1 \\
1266.534418 & -4.07 & 3.36 & 42.0 & HARPS-N 1 \\
1266.601002 & -4.84 & 3.48 & 40.9 & HARPS-N 1 \\
1272.8013 & 0.67 & 3.91 & 283.2 & HIRES \\
1275.410615 & 4.18 & 2.84 & 67.5 & HARPS-N 1 \\
1275.521886 & 3.13 & 2.97 & 58.3 & HARPS-N 1 \\
1278.373888 & -5.11 & 3.09 & 49.5 & HARPS-N 1 \\
1278.494638 & 5.83 & 3.03 & 53.7 & HARPS-N 1 \\
1279.373462 & 3.53 & 2.91 & 58.8 & HARPS-N 1 \\
1279.521052 & 4.21 & 3.02 & 57.6 & HARPS-N 1 \\
1280.40035 & 5.13 & 2.84 & 70.7 & HARPS-N 1 \\
1280.543831 & 2.94 & 2.97 & 60.5 & HARPS-N 1 \\
1281.435709 & -0.84 & 2.81 & 73.4 & HARPS-N 1 \\
1281.529573 & -16.52 & 9.34 & 16.1 & HARPS-N 1 \\
1281.534515 & 3.71 & 2.98 & 61.3 & HARPS-N 1 \\
1282.398292 & 1.78 & 2.86 & 64.9 & HARPS-N 1 \\
1282.535095 & -1.19 & 2.98 & 58.3 & HARPS-N 1 \\
1283.373536 & 0.00 & 3.23 & 44.7 & HARPS-N 1 \\
1283.560153 & 4.16 & 3.08 & 53.2 & HARPS-N 1 \\
1345.332519 & -8.77 & 2.93 & 60.6 & HARPS-N 2 \\
1352.34146 & -0.79 & 2.98 & 66.6 & HARPS-N 2 \\
1479.677427 & 0.46 & 3.32 & 53.2 & HARPS-N 2 \\
1479.750484 & -0.28 & 3.25 & 54.8 & HARPS-N 2 \\
1480.663449 & -3.22 & 3.18 & 55.6 & HARPS-N 2 \\
1480.739863 & -4.57 & 3.06 & 61.1 & HARPS-N 2 \\
1481.685619 & -6.27 & 2.94 & 72.1 & HARPS-N 2 \\
1481.75076 & -7.68 & 2.94 & 72.0 & HARPS-N 2 \\
1482.671897 & -7.20 & 2.88 & 75.3 & HARPS-N 2 \\
1482.753519 & -6.84 & 3.00 & 66.5 & HARPS-N 2 \\
1496.608742 & 4.82 & 2.86 & 80.9 & HARPS-N 2 \\
1496.724487 & 4.77 & 2.84 & 83.1 & HARPS-N 2 \\
1498.614845 & 6.52 & 3.41 & 51.1 & HARPS-N 2 \\
1498.727696 & 2.65 & 3.09 & 60.7 & HARPS-N 2 \\
1500.594327 & -8.64 & 11.61 & 14.9 & HARPS-N 2 \\
1502.590476 & 10.89 & 5.82 & 26.6 & HARPS-N 2 \\
1502.707713 & 3.53 & 7.95 & 20.0 & HARPS-N 2 \\
1518.588709 & -1.60 & 3.27 & 49.9 & HARPS-N 2 \\
1518.689186 & -0.56 & 3.19 & 50.1 & HARPS-N 2 \\
1520.585042 & 2.13 & 3.89 & 37.1 & HARPS-N 2 \\
1520.698459 & -3.86 & 3.03 & 59.1 & HARPS-N 2 \\
1521.589725 & -1.62 & 2.92 & 68.8 & HARPS-N 2 \\
1521.686301 & -0.71 & 2.75 & 90.7 & HARPS-N 2 \\
1532.559497 & -2.57 & 2.81 & 78.4 & HARPS-N 2 \\
1532.704814 & -6.82 & 2.78 & 84.1 & HARPS-N 2 \\
1533.562499 & -6.26 & 2.79 & 83.9 & HARPS-N 2 \\
1533.708811 & -9.88 & 2.76 & 88.1 & HARPS-N 2 \\
1534.523174 & -10.66 & 3.12 & 57.1 & HARPS-N 2 \\
1534.689787 & -5.63 & 2.85 & 73.0 & HARPS-N 2 \\
1536.539212 & -4.64 & 2.83 & 78.5 & HARPS-N 2 \\
1536.69836 & -1.89 & 2.80 & 81.7 & HARPS-N 2 \\
1537.537063 & -5.45 & 2.85 & 75.3 & HARPS-N 2 \\
1537.698873 & -4.10 & 2.78 & 84.5 & HARPS-N 2 \\
1562.440914 & 0.95 & 3.05 & 60.9 & HARPS-N 2 \\
1562.595327 & 2.52 & 3.04 & 60.7 & HARPS-N 2 \\
1563.499671 & 4.82 & 3.07 & 58.3 & HARPS-N 2 \\
1563.673401 & -1.63 & 2.94 & 63.5 & HARPS-N 2 \\
1564.515083 & 5.04 & 4.85 & 29.9 & HARPS-N 2 \\
1564.675676 & -0.53 & 3.33 & 49.0 & HARPS-N 2 \\
1565.523526 & -1.72 & 3.13 & 56.4 & HARPS-N 2 \\
1566.47765 & -1.59 & 2.94 & 67.5 & HARPS-N 2 \\
1566.698684 & 0.74 & 2.96 & 67.4 & HARPS-N 2 \\
1578.422373 & -1.78 & 3.08 & 59.9 & HARPS-N 2 \\
1578.607224 & 1.29 & 2.91 & 71.5 & HARPS-N 2 \\
1579.468298 & 1.48 & 2.81 & 83.0 & HARPS-N 2 \\
1579.606263 & -1.48 & 2.81 & 84.3 & HARPS-N 2 \\
1580.550925 & -2.58 & 2.78 & 86.3 & HARPS-N 2 \\
1580.702582 & -4.15 & 2.85 & 79.9 & HARPS-N 2 \\
1581.443399 & -4.66 & 2.94 & 67.5 & HARPS-N 2 \\
1581.578226 & -4.45 & 3.23 & 53.2 & HARPS-N 2 \\
1582.444149 & -1.68 & 3.15 & 56.2 & HARPS-N 2 \\
1582.62878 & 0.83 & 3.67 & 42.6 & HARPS-N 2 \\
1595.439337 & 3.83 & 2.82 & 82.7 & HARPS-N 2 \\
1595.606467 & 1.79 & 2.95 & 68.4 & HARPS-N 2 \\
1596.385682 & 5.56 & 2.92 & 71.5 & HARPS-N 2 \\
1596.639015 & 4.24 & 2.87 & 77.4 & HARPS-N 2 \\
1597.498195 & -0.38 & 2.80 & 85.5 & HARPS-N 2 \\
1597.67948 & 2.83 & 2.89 & 75.6 & HARPS-N 2 \\
1598.494932 & 7.64 & 2.84 & 81.0 & HARPS-N 2 \\
1598.671622 & 4.95 & 3.06 & 62.1 & HARPS-N 2 \\
1599.444782 & 6.99 & 2.88 & 75.5 & HARPS-N 2 \\
1599.669099 & 7.67 & 2.97 & 69.9 & HARPS-N 2 \\
1600.450037 & 5.89 & 2.85 & 79.3 & HARPS-N 2 \\
1600.634886 & 1.53 & 3.08 & 62.8 & HARPS-N 2 \\
1601.425511 & -2.03 & 3.08 & 61.1 & HARPS-N 2 \\
1601.65852 & 1.59 & 3.08 & 62.6 & HARPS-N 2 \\
1610.395668 & 4.14 & 3.36 & 48.7 & HARPS-N 2 \\
1610.406339 & 6.97 & 3.29 & 50.2 & HARPS-N 2 \\
1611.547768 & -0.14 & 2.97 & 66.5 & HARPS-N 2 \\
1612.49135 & 0.09 & 2.84 & 79.8 & HARPS-N 2 \\
1613.483021 & 0.44 & 2.83 & 80.6 & HARPS-N 2 \\
1628.459887 & -0.04 & 3.08 & 60.1 & HARPS-N 2 \\
1628.564955 & 0.51 & 3.16 & 58.7 & HARPS-N 2 \\
1629.480763 & 0.57 & 2.83 & 83.6 & HARPS-N 2 \\
1629.614778 & 3.18 & 3.24 & 57.0 & HARPS-N 2 \\
1630.444186 & 2.22 & 2.82 & 84.2 & HARPS-N 2 \\
1630.53025 & 2.92 & 2.87 & 77.3 & HARPS-N 2 \\
1654.373345 & 3.44 & 3.01 & 56.8 & HARPS-N 2 \\
1654.46788 & -0.74 & 2.98 & 58.8 & HARPS-N 2 \\
1657.371082 & -0.12 & 2.77 & 84.4 & HARPS-N 2 \\
1657.503069 & -0.78 & 3.10 & 58.0 & HARPS-N 2 \\
1662.435891 & -2.10 & 3.61 & 45.0 & HARPS-N 2 \\
1663.373238 & -2.36 & 2.89 & 71.8 & HARPS-N 2 \\
1663.472796 & -4.30 & 3.04 & 62.5 & HARPS-N 2 \\
1665.352941 & -4.94 & 2.97 & 66.8 & HARPS-N 2 \\
1665.482578 & 1.24 & 2.82 & 81.1 & HARPS-N 2 \\
1667.347572 & 3.21 & 2.77 & 88.3 & HARPS-N 2 \\
1671.344738 & -1.73 & 3.37 & 48.5 & HARPS-N 2 \\
1671.455024 & -0.97 & 3.35 & 49.9 & HARPS-N 2 \\
1680.342329 & -0.12 & 2.89 & 70.8 & HARPS-N 2 \\
1680.43589 & -3.70 & 2.93 & 69.6 & HARPS-N 2 \\
1682.32927 & -2.72 & 2.80 & 81.4 & HARPS-N 2 \\
1682.376757 & 0.14 & 2.96 & 66.8 & HARPS-N 2 \\
1926.898038 & 4.23 & 3.89 & 283.4 & HIRES \\
1942.001946 & 2.24 & 3.95 & 283.1 & HIRES \\
1942.954468 & 4.33 & 3.92 & 283.4 & HIRES \\
1943.951042 & -2.30 & 3.89 & 284.7 & HIRES \\
1962.874534 & -0.57 & 3.89 & 286.0 & HIRES \\
1963.860172 & 0.82 & 3.90 & 283.8 & HIRES \\
1964.900473 & 0.11 & 3.93 & 283.4 & HIRES \\
1965.945739 & -2.07 & 3.90 & 282.6 & HIRES \\
1982.934778 & 0.23 & 3.86 & 281.7 & HIRES \\
1983.770615 & 6.05 & 3.93 & 280.2 & HIRES \\
1989.012469 & 1.35 & 3.90 & 282.4 & HIRES \\
2006.909888 & 4.64 & 3.88 & 280.8 & HIRES \\
2007.77185 & 4.84 & 3.81 & 283.8 & HIRES \\
2008.976767 & 14.31 & 4.49 & 131.4 & HIRES \\
2009.88478 & 10.16 & 3.85 & 280.9 & HIRES \\
2013.741571 & 6.44 & 3.84 & 285.1 & HIRES \\
2251.114854 & 1.77 & 3.90 & 280.9 & HIRES \\
2280.075021 & -0.34 & 3.93 & 277.9 & HIRES \\
\hline
\end{longtable}
\end{center}

\subsection{HIRES Doppler Pipeline}
We calculate precise radial velocities of Kepler-10 using the standard Doppler code of the CPS group \citep{Howard2011_730} with the inclusion of a new de-trending routine. Previoulsy published RVs (B11) were collected during the 2009-2010 observing seasons. Subsequent observations were taken in the 2014 observing season. Long term RV precision spanning 10 years is consistently achieved with HIRES as described in \citet{Howard2011_730}. Exposure times of $\sim$30 minutes are required to achieve SNR $\sim$200 in the iodine region, resulting in internal RV errors of 1.5 to 2.0 \ms. Each observation uses a slit with dimensions of 0.87\arcsec x 14.\arcsec0  yielding a resolving power of 60,000, and allowing for subtraction of night sky emission lines, and scattered moonlight.

During the initial RV extraction we model the instrumental PSF as a sum of 13 Gaussians with positions and widths fixed but their heights free to vary \citep{Butler1996}. Any correlations in the final RVs with the heights of these Gaussians (PSF parameters) likely indicate small inadequacies of our PSF model to completely describe the shape of the instrumental PSF. RV shifts caused by the gravitational influence of orbiting bodies should not be correlated with the shape of the instrumental PSF.

In order to clean the RVs of any possible systematic trends we detrend the final RVs by removing correlations with the instrumental PSF parameters, the magnitude of the RV uncertainty, and the S/N ratio of the spectrum.  After masking any $5\sigma$ outliers, we search for significant correlations by calculating the Spearman rank correlation coefficient for each of these variables with RV (Spearman1904). We take note of any parameters that show a correlation coefficient greater than 0.1 and include these variables in a multivariate ordinary least squares linear regression using the STATSMODELS\footnote{https://pypi.python.org/pypi/statsmodels} package in Python. This final multidimensional surface is then subtracted from the final RVs. This technique improves the RMS of the RV time series of Kepler-10 from 4.9 \ms\ to 4.6 \ms\ by detrending against nine PSF parameters and the S/N ratio of the spectra.  To check that the detrending algorithm does not accidentally remove the signal of the planets, we calculated a two-planet circular fit to both the detrended and non-detrended RVs.  The RMS of the change in RV introduced by the detrending algorithm was $1.2~\ms$ (i.e., less than the uncertainty in each RV), and the solutions for all parameters were consistent within $0.1~\ms$ (much less than our $1\sigma$ uncertainties in the parameters).  We do not add the random noise introduced by the detrending algorithm to our error budget because our technique for solving for the jitter (see Equation \ref{eqn:rv}) naturally incorporates the uncertainties that arise through this method.

\subsection{Analysis of the HIRES and HARPS-N RVs}
\begin{figure*}[htbp] 
   \centering
\includegraphics[width=3in]{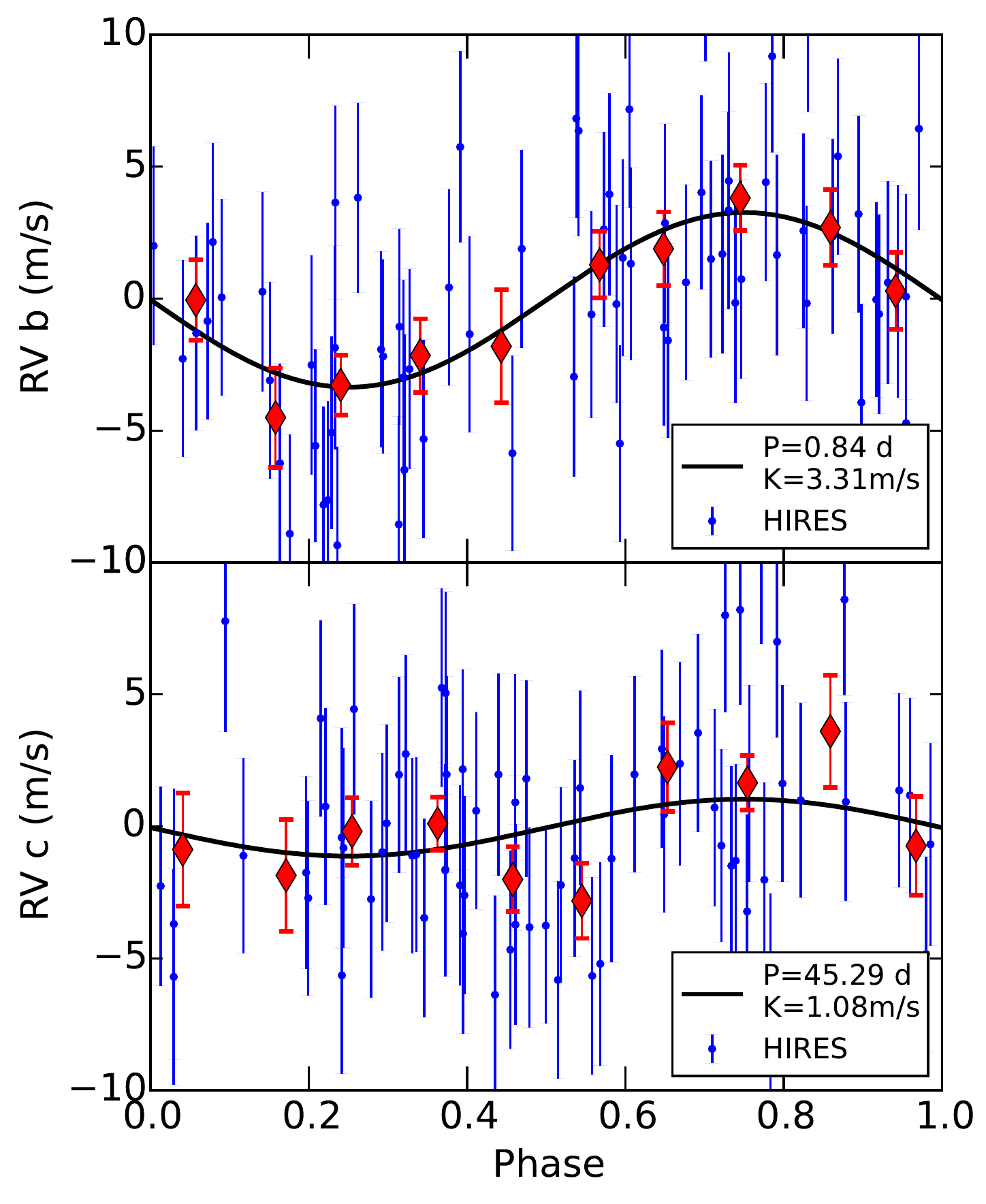}
\includegraphics[width=3in]{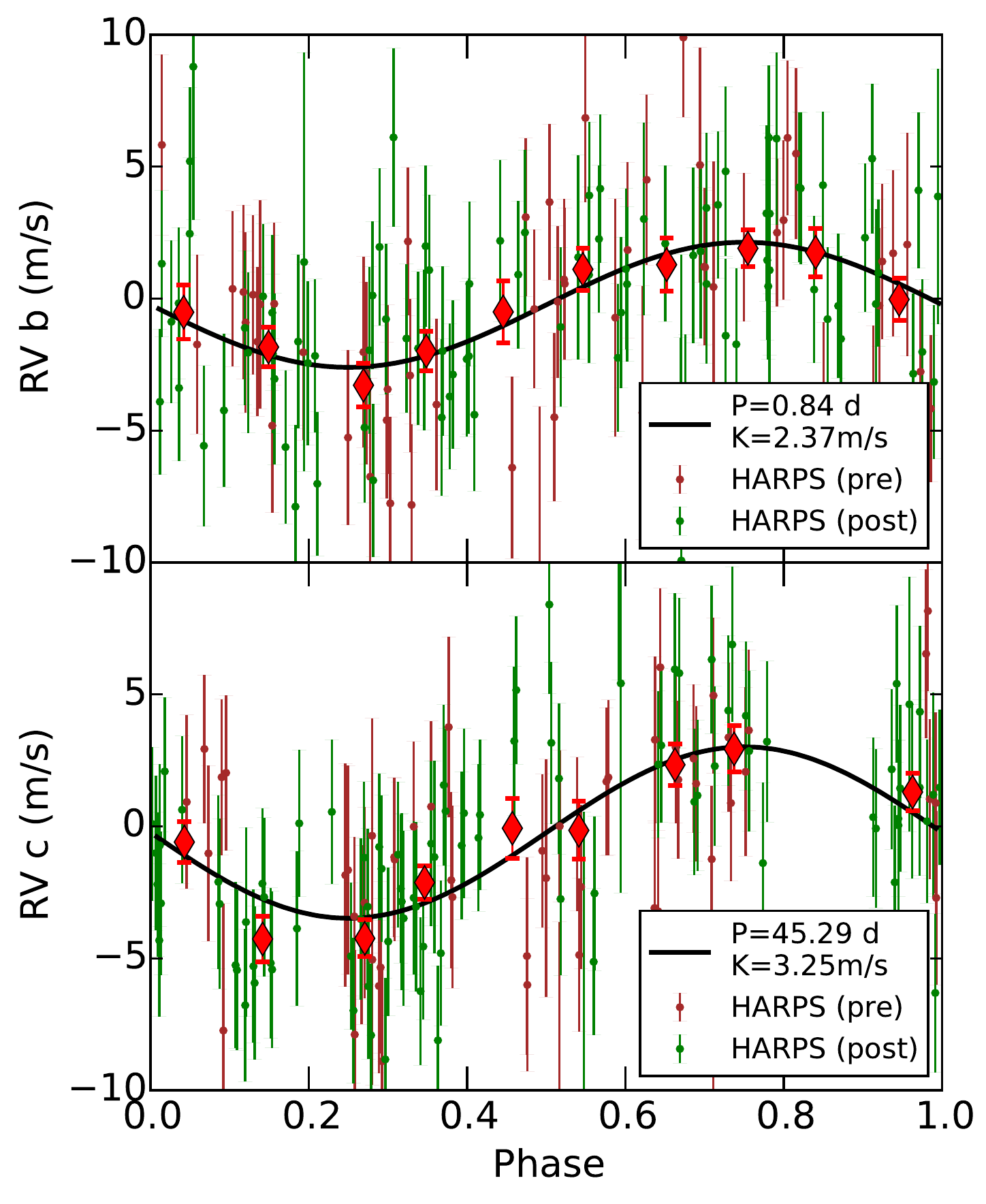}
\caption{Left: the RVs from HIRES phase-folded to the orbital periods of Kepler-10 b (top) and c (bottom).  The red diamonds show the weighted mean RV of the HIRES data in bins of 0.1 orbital phase.  The black curve shows the best two-planet circular fit: $K_b = 3.31~ \ms$, $K_c = 1.08~ \ms$.  Right: same as left, but using the pre-CCD upgrade (brown) and post-CCD upgrade (green) HARPS-N RVs.  Our best two-planet circular fit yields $K_b = 2.37~ \ms$, $K_c = 3.25~ \ms$, in agreement with \citet{Dumusque2014}.}
   \label{fig:hires_v_harps}
\end{figure*}

\begin{figure*}[htbp] 
   \centering
\includegraphics[width=6in]{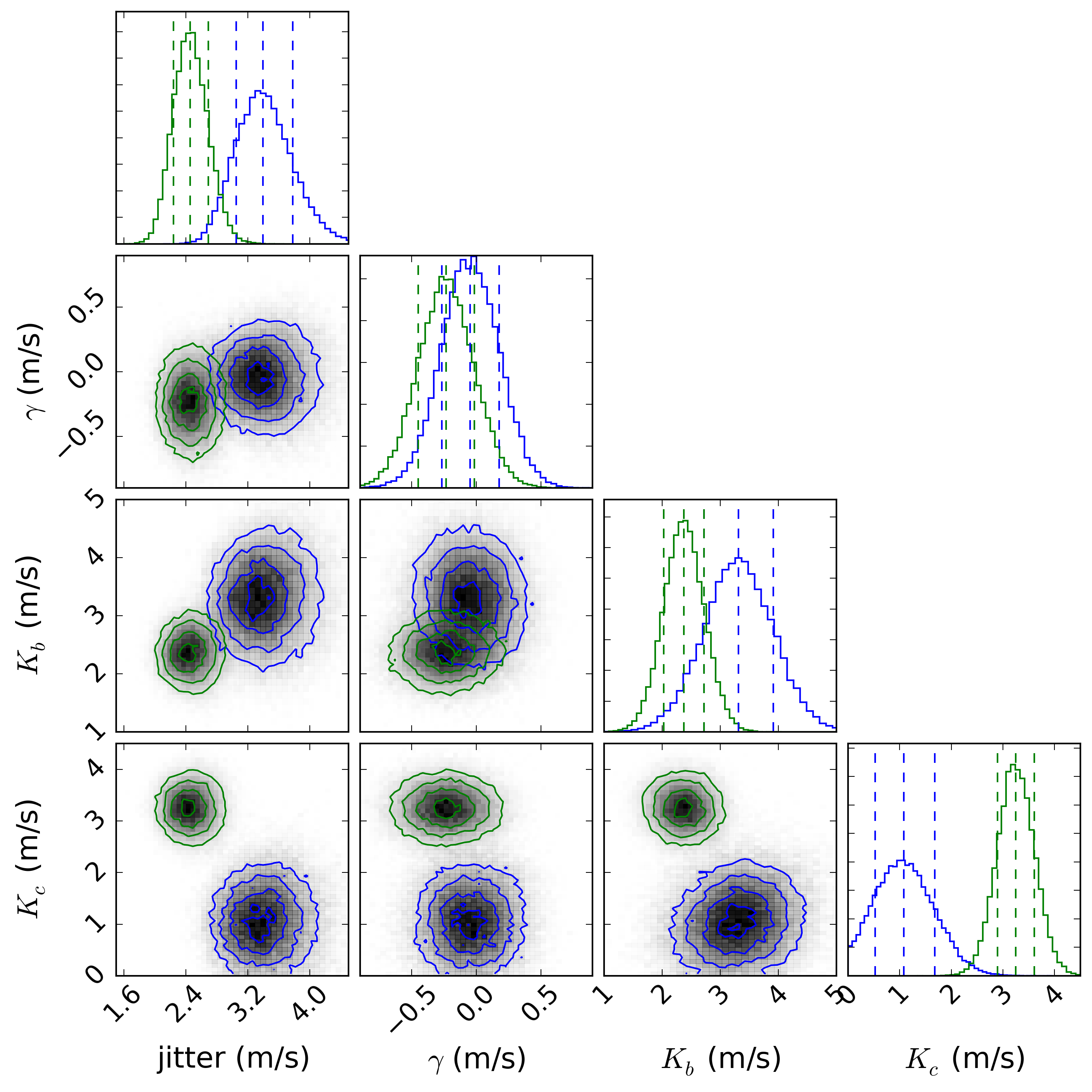}
\caption{The MCMC posterior distributions of two-planet circular fits to the HIRES RVs alone (blue) and the HARPS-N RVs alone (green).  The parameters are jitter, the RV zero-point offset $\gamma$, and the RV semi-amplitudes from planets b (\Kb) and c (\Kc).  Dashed lines denote the 16$^\mathrm{th}$, 50$^\mathrm{th}$, and 84$^\mathrm{th}$ percentiles.}
   \label{fig:hires_v_harps_mcmc}
\end{figure*}

Figure \ref{fig:hires_v_harps} shows the Kepler-10 RVs from HIRES and HARPS-N phase-folded to the orbital periods of the two transiting planets, with red diamonds indicating the weighted mean RV in bins of 0.1 orbital phases to guide the eye.  Figure \ref{fig:hires_v_harps_mcmc} shows the MCMC posterior distributions of two-planet circular fits to the HIRES RVs alone (blue) and the HARPS-N RVs alone (green).  A summary of the best fit parameters to the HIRES RVs alone are given in Table \ref{tab:HIRES_circ}.  The HIRES RVs yield $m_c = 5.69^{+3.19}_{-2.90}~\mearth$, a result that disagrees with the best fit to the HARPS-N RVs ($m_c = 17.2\pm1.9$) by $3.1~\sigma$.

\begin{table}[htbp]
\caption{Two-Planet Circular Fit to Only HIRES RVs}
\label{tab:HIRES_circ}
\tablenum{2}
\begin{tabular}{llllll}
\hline
\hline
\colhead{Paramter} & \colhead{Units} & \colhead{Median} & \colhead{$+1\sigma$} & \colhead{$-1\sigma$}& \colhead{Ref.}\cr
\hline
jitter & \ms & 3.41 & 0.39 & 0.34& A\cr
$\gamma$ & \ms & -0.05 & 0.22 & 0.22 & A \cr
$K_b$ & \ms & 3.31 & 0.59 & 0.59 & A\cr
$K_c$ & \ms & 1.09 & 0.61 & 0.55 & A \cr
$m_b$ & \mearth & 4.61 & 0.83 & 0.83 & A,B\cr
$m_c$ & \mearth & 5.69 & 3.19 & 2.9 & A,B \cr
$r_b$ & \rearth & 1.47 & 0.03 & 0.02 & B\cr
$r_c$ & \rearth & 2.35 & 0.09 & 0.04 &B\cr
$\rho_b$ & \gcc & 8.0 & 1.43 & 1.44 &A,B\cr
$\rho_c$ & \gcc & 2.42 & 1.36 & 1.24&A,B\cr
\hline
\end{tabular}
\tablecomments{All parameters were explored with uniform priors.}
\tablerefs{A. This work.  B. \citet{Dumusque2014}}
\end{table}

\subsection{Analysis of the Discrepancy between HIRES and HARPS-N RVs}
\label{sec:discrepancy}
What is the source of the discrepancies between $K_b$ and $K_c$ in the HIRES and HARPS-N data?  HIRES RVs are stable with an RMS of $2~\ms$ for various stars of spectral types without any known planets over decades (Figure 1).  Removing $2.5\sigma$ outliers outliers changes \Kb\ and \Kc\ by 1\%, an insignificant amount compared to our uncertainties.  We do not find any significant correlations between the RVs and barycentric correction, or between the RVs and stellar activity indices, in either the HIRES or HARPS-N data.

\begin{figure*}[htbp] 
   \centering
     \includegraphics[width=3in]{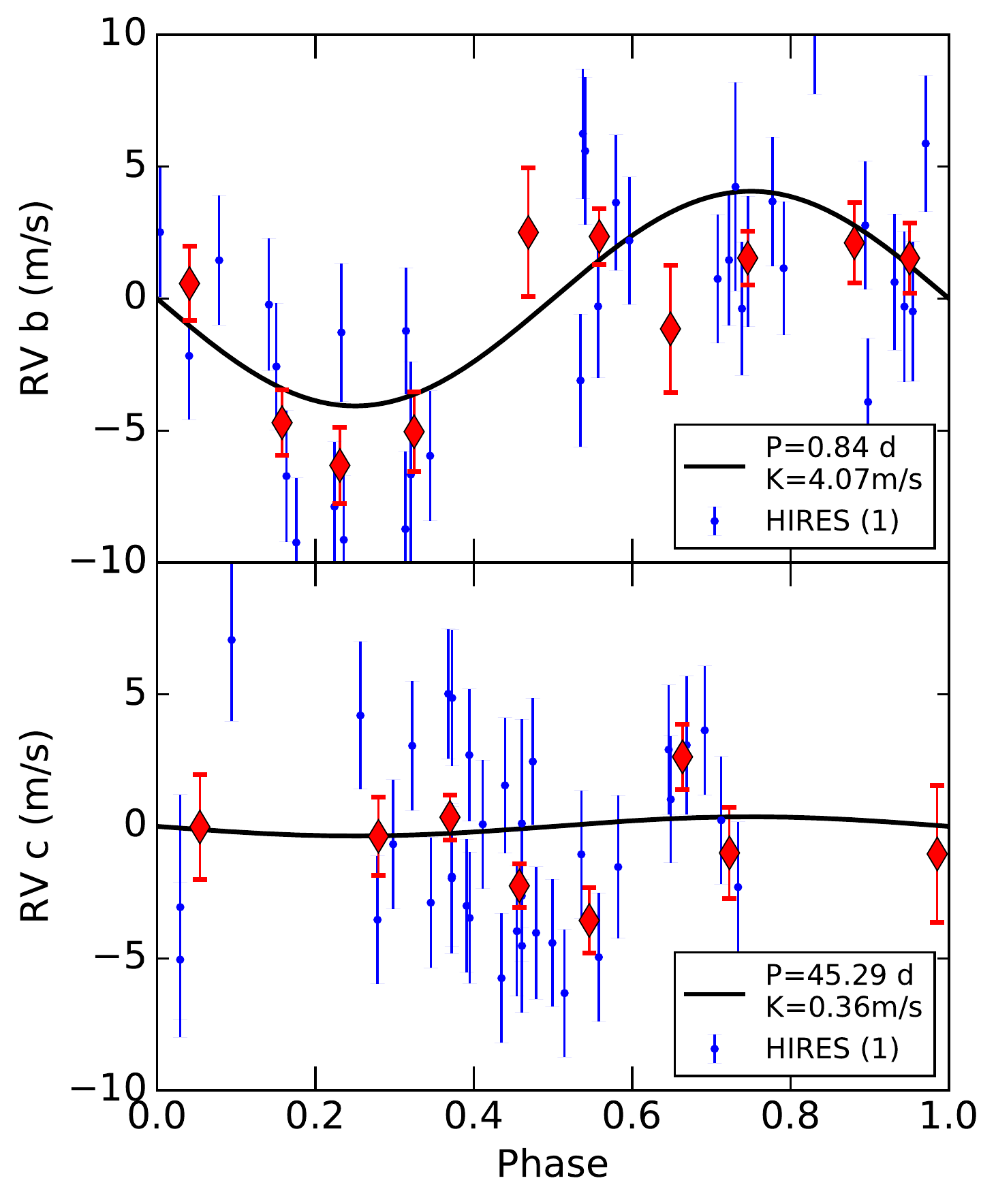}
       \includegraphics[width=3in]{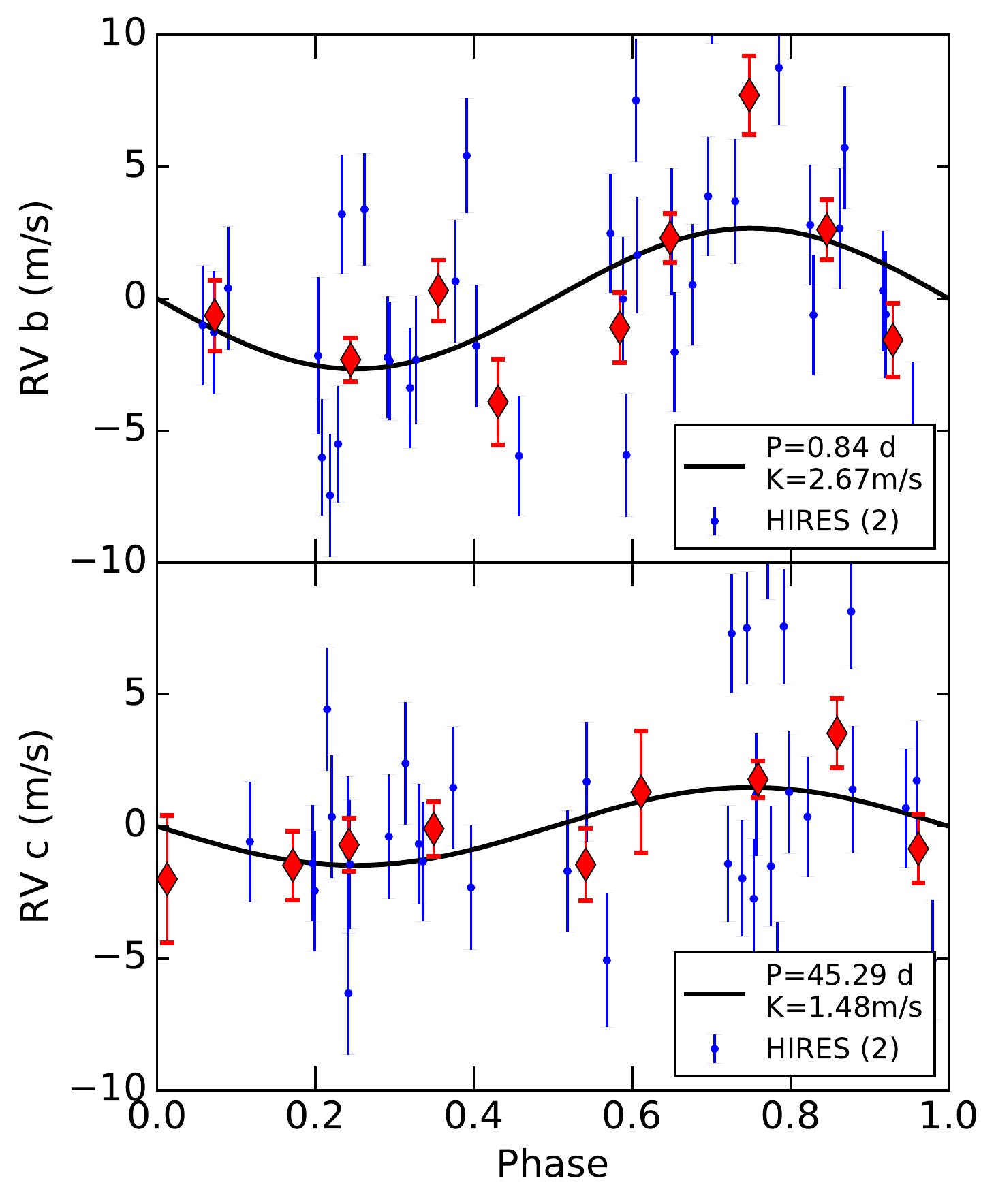}
  \includegraphics[width=3in]{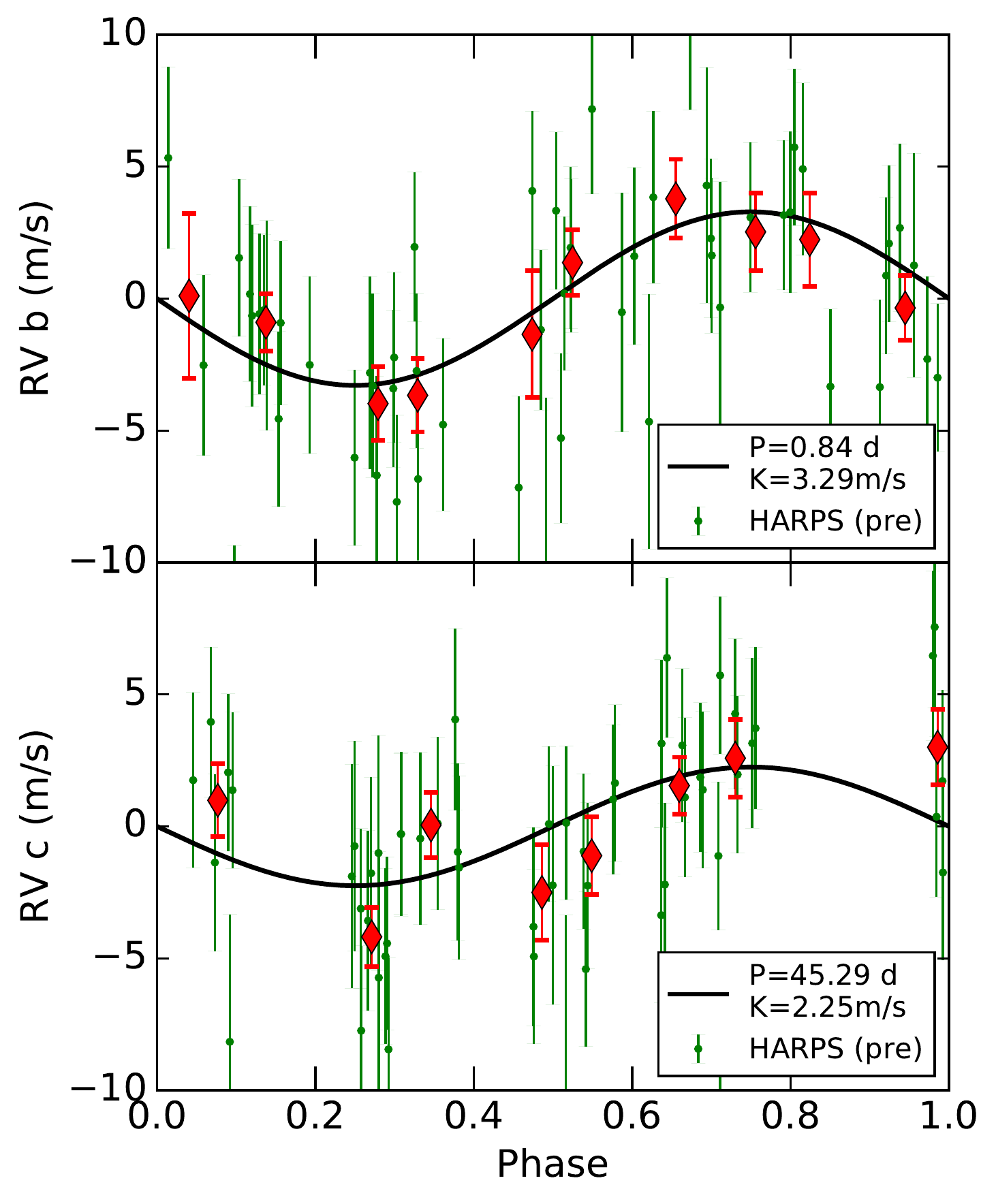}
  \includegraphics[width=3in]{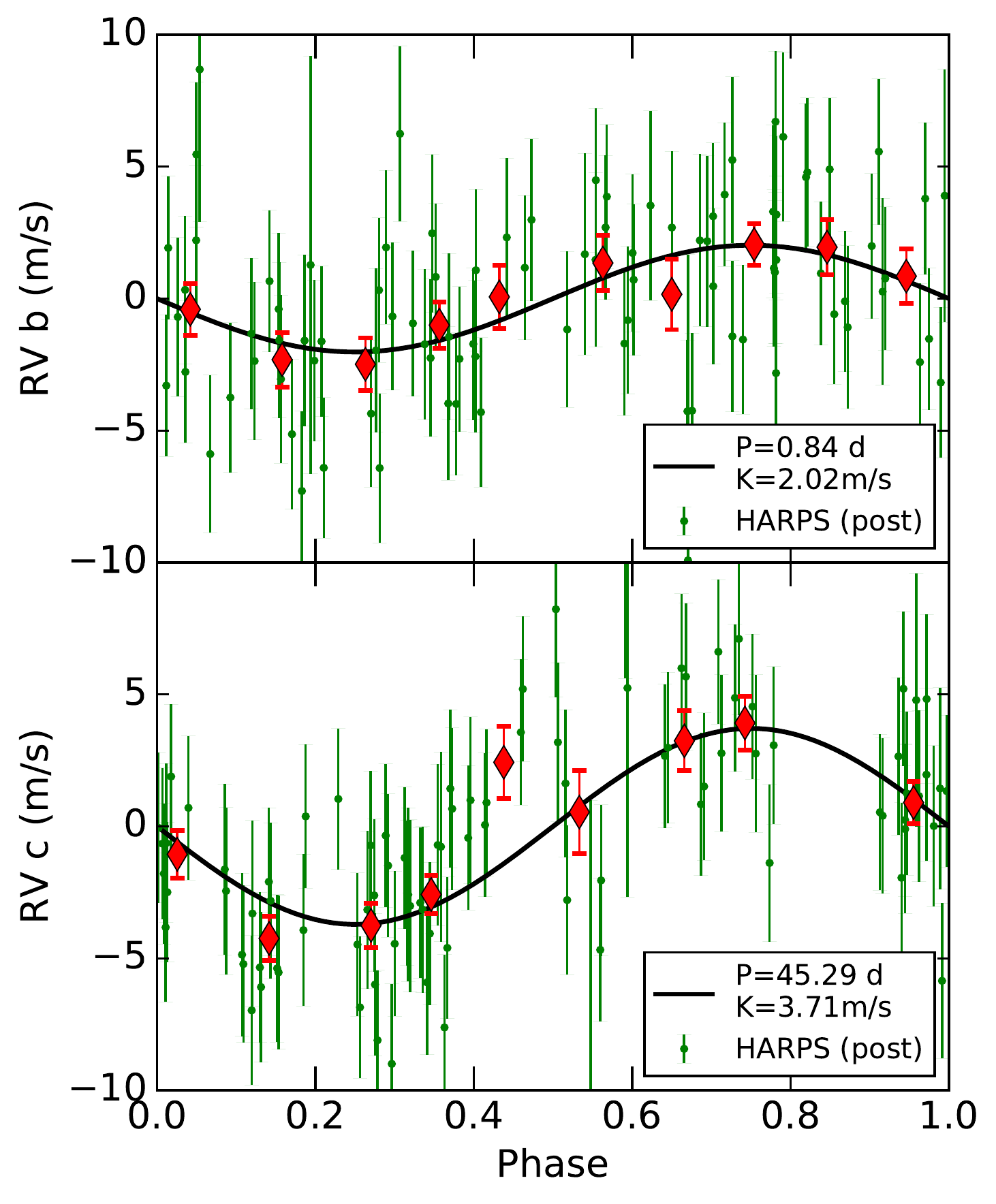}
   \caption{The RV curve of Kepler-10 phase-folded to the orbital periods of planets b and c, for four different subsets of the data: the first half of the HIRES data (top left, $\Kb=4.07\pm0.95\ms$, $\Kc=0.36\pm0.7\ms$), the second half of the HIRES data (top right, $\Kb=2.67\pm0.88\ms$, $\Kc=1.48\pm0.80\ms$), the HARPS-N data from before their CCD upgrade (bottom left, $\Kb=3.29\pm0.62\ms, \Kc=2.25\pm0.59\ms$), and the HARPS-N data from after their CCD upgrade (bottom right, $\Kb=2.02\pm0.37\ms, \Kc=3.71\pm0.41\ms$).  The derived values of the RV semi-amplitude, $K$, for both planets b and c are different by more than 1\ms  ($\sim30$\%) from the two halves of the RV data sets from each spectrometer.  These inconsistencies within each spectrometer indicate some time-correlated contribution to the RVs, perhaps from stellar activity, additional planets, or systematic RV errors at the level of $\sim$1\ms in the spectrometers.}
   \label{fig:changing_K}
\end{figure*}

When we break either the HIRES or HARPS-N data into two epochs (first half vs. second half of the acquired RVs), both spectrometers find significantly different values of $K_b$ and/or $K_c$ in the first versus the second half of their RV data.  Using just the first half of the HIRES data, we find $K_{b,1} = 4.07\pm0.95 \ms$, $K_{c,1} < 1.10 \ms$ (68\% confidence).  Using just the second half of the HIRES RVs, we get $K_{b,2} = 2.67\pm0.88 \ms$, $K_{c,2}=1.48\pm0.80 \ms$.  For HARPS-N, we divided the RVs into those taken before and after Nov. 12, 2012 (the date of their CCD upgrade, which is a convenient division time).  Using just the pre-upgrade HARPS-N data, we find $K_{b,1} = 3.29\pm0.62 \ms$, $K_{c,1} = 2.25\pm0.59 \ms$.  Using just the post-upgrade HARPS-N RVs, we get $K_{b,2} = 2.02\pm0.37 \ms$, $K_{c,2}=3.71\pm0.41 \ms$ (see Figure \ref{fig:changing_K}.  The $1.7\sigma$ difference in $K_b$ and $2.0\sigma$ difference $K_c$ between the pre- and post-upgrade RVs from HARPS-N is larger than we would expect from statistical fluctuations alone.  The apparent change in \Kb\ and \Kc\ suggests that an additional, time-correlated source, possibly from stellar activity or the presence of additional planets, confounds both the HIRES and HARPS-N spectrometers on short timescales.

\section{Planetary Properties of Kepler-10: Two-Planet Solutions}
The significant discrepancy between the best two-planet fits to the HIRES and HARPS-N data sets motivates a reanalysis of the data.  Since we cannot find evidence that either data set is compromised, we choose to combine all the available data from HIRES and HARPS-N to calculate the most up-to-date planet masses.  For consistency and relevant comparison to previous findings, we adopt the Kepler-10 stellar properties from D14, which are listed in Table \ref{tab:comparison}.

\subsection{Two-Planet Circular Fit}
The RV signals produced by small planets are often too low-amplitude, compared to the typical RV noise, to precisely measure their orbital eccentricities.  Thus, the RVs of many of the systems of small planets discovered by \Kepler\ and RV surveys (e.g. Eta-Earth) are consistent with planets in circular or nearly circular orbits.  We are motivated to explore a two-planet circular orbit because the RVs do not demand eccentric orbits for either Kepler-10 b or Kepler-10 c.  Furthermore, B11 and D14 model the Kepler-10 system with circular orbits, so we explore a two-planet circular fit to enable a direct comparison to their results.

The photometrically determined transit times from \Kepler\ constrain the orbital ephemerides of Kepler-10 b and c.  Specifically, the photometry precisely constrains the time of transit, orbital period, and inclination of each planet.  The remaining free dynamical parameter for each planet is the mass.  We solve for the mass via the observable semi-amplitude of the RV sinusoid, $K$.

In addition to the dynamical parameters, fitting RVs from two different spectrographs incurs several nuisance parameters.  There is a zero point offset for each set of RVs.  Each measured RV has some internal uncertainty, plus error of astrophysical origin (from stellar oscillations, plage, starspots, magnetic activity, etc.), plus additional errors from the spectrometer.  For each spectrometer, we report the combined astrophysically-induced error and spectrometer-induced error as a jitter term, $\sigma_\mathrm{jitter}$, which we add in quadrature with the internal uncertainty in the RV to obtain the total uncertainty of each RV.  The internal uncertainty of the RV varies from measurement to measurement, whereas the jitter term is the same for all RVs taken by a single spectrometer.

Therefore, the two-planet circular fit has seven free parameters: the semi-amplitude of the RVs resulting from planet b ($K_b$), the semi-amplitude of the RVs resulting from planet c ($K_c$), the velocity zero-point of the RVs ($\gamma$), an offset between the RVs taken by the HIRES spectrograph and the RVs taken on the pre-upgrade HARPS-N CCD (offset 1), an offset between the RVs taken by the HIRES spectrograph and the RVs taken on the post-upgrade HARPS-N CCD (offset 2), the jitter of the HIRES spectrograph ($j_1$) and the the jitter of the HARPS-N spectrograph ($j_2$).  The orbital period, time of transit, and orbital inclination were derived from photometry in D14, and we fix them at the values published therein.

To determine the best circular fit to the data, we adopt the same likelihood function as D14:
\begin{equation}
\mathcal{L} = \prod_{i} \frac{1}{\sqrt{2\pi(\sigma_i^2+\sigma_j^2)}}
\mathrm{exp}\big[-\frac{(\mathrm{RV}_{i} - 
\mathrm{RV}_{\mathrm{mod},i})^2}{2(\sigma_i^2+\sigma_j^2)}\big]
\label{eqn:rv}
\end{equation}
where $\mathrm{RV}_{i}$ is the $i^\mathrm{th}$ observed RV,
$\mathrm{RV}_{\mathrm{mod},i}$ is the  $i^\mathrm{th}$ modeled RV,
$\sigma_i$ is the uncertainty in the $i^\mathrm{th}$ observed RV, and
$\sigma_j$ is the jitter term from the instrument on which the observation was made (either HIRES or HARPS).  We minimized the negative log-likelihood via the Levenberg-Marquardt method with the Python package lmfit.

We performed a Markov Chain Monte Carlo (MCMC) analysis to understand the full posterior distribution of the dynamical parameters and their covariances.  We used the Python package emcee \citep{Foreman-Mackey2013}, an affine-invariant MCMC sampler.  We adopted uniform priors in $j_1$, $j_2$, $\gamma$, offset 1, offset 2, $K_b$, and $K_c$, while restricting $j_1$, $j_2, \Kb, \Kc > 0$.  Our dynamical equations, choice of parameters, and the priors on those parameters were chosen to replicate D14 as closely as possible, while allowing the inclusion of the HIRES RVs.

\begin{figure*}[hbtp]
\begin{center}
\includegraphics[width=6in]{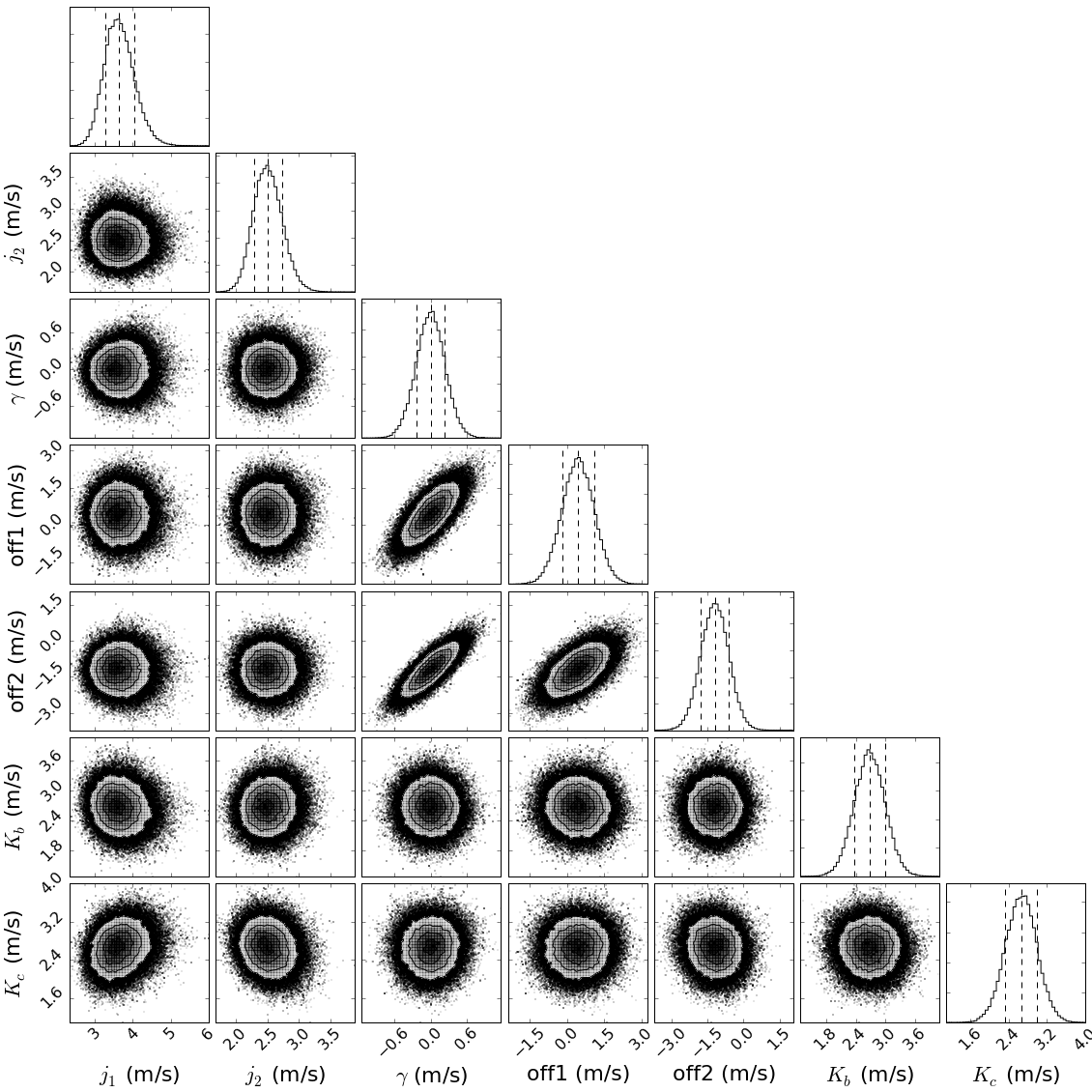}
\caption{Posterior distribution for the two-planet circular fit to Kepler-10 RVs.  Variables are the jitter of the HIRES instrument ($j_1$), jitter of the HARPS-N instrument ($j_2$), velocity zero point ($\gamma$), velocity offset between HIRES and the HARPS-N RVs from before the CCD upgrade (off1),  velocity offset between HIRES and the HARPS-N RVs from after the CCD upgrade (off2), and the semi-amplitudes of the RV curve for planets b ($K_b$) and c ($K_c$).  The dahsed lines indicate the 16$^{th}$, 50$^{th}$, and 84$^{th}$ percentiles.}
\label{fig:emcee2c}
\end{center}
\end{figure*}

\begin{figure}[hbtp] 
   \centering
\includegraphics[width=3in]{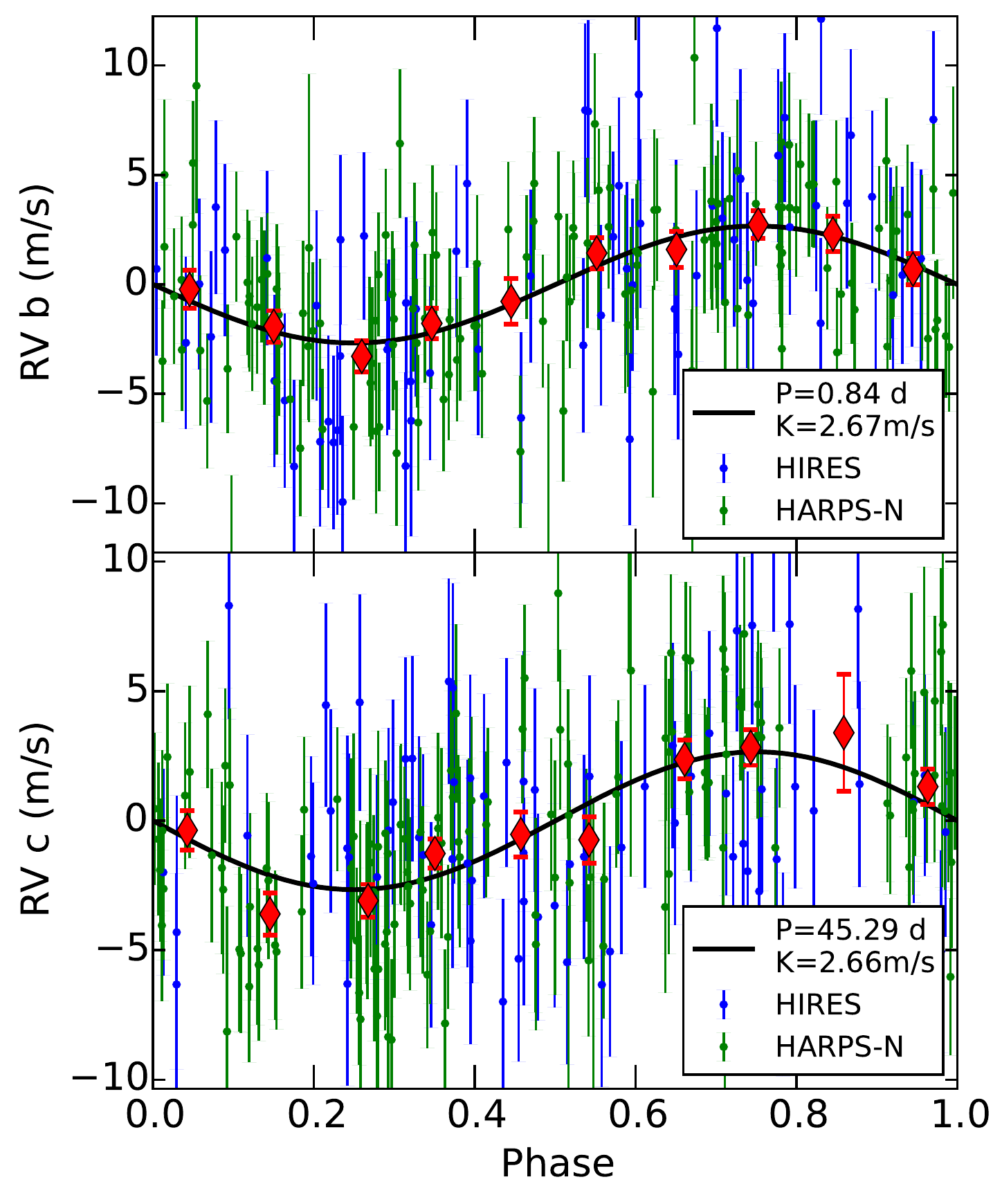}   
\caption{The RVs from HIRES (blue) and HARPS-N (green) phase-folded to the periods of Kepler-10 b (top) and c (bottom).  The red diamonds show the weighted mean RV of the HIRES and HARPS-N data combined in bins of 0.1 orbital phase.  The best two-planet circular fit is shown in black.  The orbits of both planets are constrained by the \Kepler-determined transit times.  The best two-planet circular fit yields $m_b = \mbcirc$, and $m_c = \mccirc$.}
   \label{fig:2plcirc}
\end{figure}

The posterior of the MCMC sampler is shown in Figure \ref{fig:emcee2c}.  The two-planet circular fit using the parameters from the median of the posterior distribution is shown in Figure \ref{fig:2plcirc}.  The best two-planet circular fit yields $m_b = \mbcirc$, $m_c = \mccirc$, $\rho_b=\rhobcirc$, and $\rho_c = \rhoccirc$.  Table \ref{tab:circ} lists the median and $1\sigma$ uncertainties of the marginalized parameters and derived planet masses and densities.

\begin{table}[htbp]
\caption{Two-Planet Circular Fit MCMC Parameters}
\label{tab:circ}
\tablenum{3}
\begin{tabular}{llllll}
\hline
\hline
\colhead{Paramter} & \colhead{Units} & \colhead{Median} & \colhead{$+1\sigma$} & \colhead{$-1\sigma$}& \colhead{Ref.}\cr
\hline
HIRES jitter &\ms& 3.62 & 0.41 & 0.37 &A\\ 
HARPS jitter &\ms& 2.49 & 0.24 & 0.21  &A\\ 
$\gamma$&\ms&-0.01 & 0.23 & 0.23 &A \\ 
offset 1&\ms&0.44 & 0.63 & 0.64  &A\\ 
offset 2&\ms&-1.21 & 0.56 & 0.57  &A\\ 
\Kb &\ms& 2.67 & 0.30 & 0.30 &A \\ 
\Kc &\ms& 2.67 & 0.34 & 0.34  &A\\ 
$m_b$ &\mearth &3.72 & 0.42 & 0.42  &A, B\\ 
$m_c$ &\mearth&13.98 & 1.77 & 1.80  &A, B\\ 
$r_b$ & \rearth & 1.47 & 0.03 & 0.02 & B\\
$r_c$ & \rearth & 2.35 & 0.09 & 0.04 &B\\
$\rho_b$ & \gcc & 6.46 & 0.72 & 0.74 &A,B\\ 
$\rho_c$ & \gcc &5.94 & 0.75 & 0.77 &A,B\\ 

\hline
\end{tabular}
\tablecomments{All parameters were explored with uniform priors.}
\tablerefs{A. This work.  B. \citet{Dumusque2014}}
\end{table}

\begin{figure}[htbp]
\begin{center}
\includegraphics[width=3in]{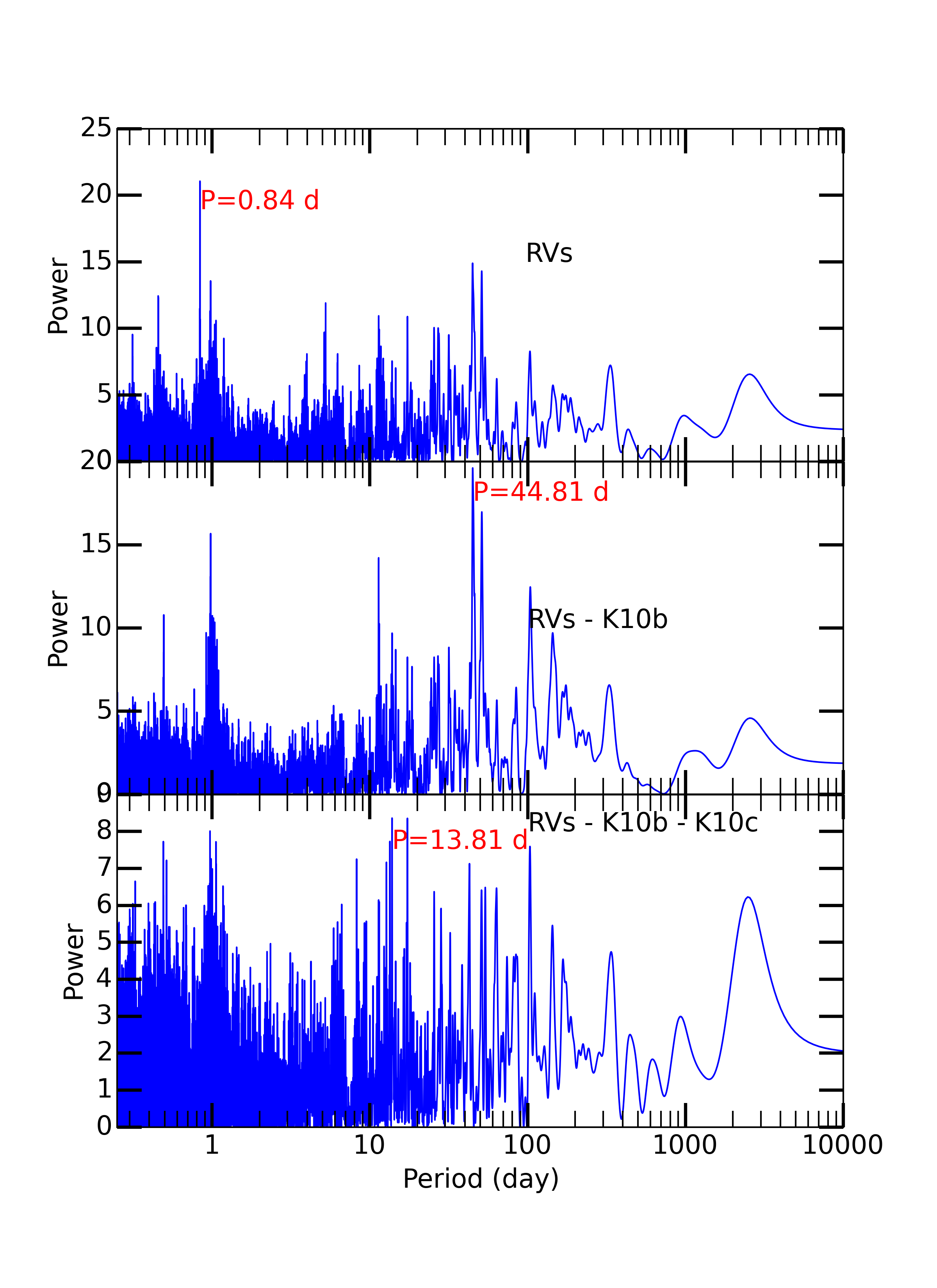}
\caption{Top: L-S periodogram of the combined RVs from HIRES and HARPS-N.  Center: L-S periodogram of the RVs after subtracting the model RVs for planet b.  Bottom: L-S periodogram of the RVs after subtracting the model RVs for planets b and c.}
\label{fig:pgram}
\end{center}
\end{figure}

We computed the Lomb-Scargle (L-S) periodogram of the combined HIRES and HARPS-N RVs using the fasper algorithm \citep[][see Figure \ref{fig:pgram}]{Press1989}.  The most prominent peak was at 0.84 days, the orbital period of planet b.  We subtracted the RV component from Kepler-10 b (as determined by our maximum-likelihood model) and computed the periodogram of the residuals, finding a pair of peaks at 44.8 and 51.5 days.  The orbital period of planet c is 45.3 days; the peak at 51.5 day is a one-year alias of the orbital period of planet c, intensified by noise in the manner described in \citet{Dawson2010}.  We subtracted the model RVs of planet c and computed the periodogram of the residuals, finding a forest of peaks from 13-100 days.

\subsection{Two-Planet Fit with Eccentricity for Planet c}
Kepler-10 b, which has an ultra-short orbital period of 0.84 days, very likely has a circular orbit because its circularization timescale is much shorter than the stellar age (B11).  However,  Kepler-10 c has a sufficiently long orbital period (45.3 days) to maintain a moderately eccentric orbit over the system age.  Because we cannot rule out a moderately eccentric orbit for Kepler-10 c, we explore possible two-planet fits in which the orbit of planet c (but not planet b) is allowed to be eccentric.

The two-planet fit in which planet c is allowed eccentricity has two free parameters in addition to the free parameters of the circular fit: \ecosom\ and \esinom.  These parameters are a combination of the Keplerian orbital parameters $e_c$ (the eccentricity of planet c) and $\omega_c$ (the argument of periastron passage of planet c).  We adopted a uniform prior on \ecosom\ and \esinom\ with the constraint $(\ecosom)^2 + (\esinom)^2 \le 1$.  The time of periastron passage is determined by a combination of the argument of periastron passage ($\omega_c$), the eccentricity ($e_c$), the time of transit (\TT), and the orbital period ($P$).

For the parameters $j_1$, $j_2$, $\gamma$, offset, $K_b$, and $K_c$, we adopt uniform priors with the same boundaries as listed for the two-planet circular fit.

\begin{figure*}[htbp]
\begin{center}
\includegraphics[width=6in]{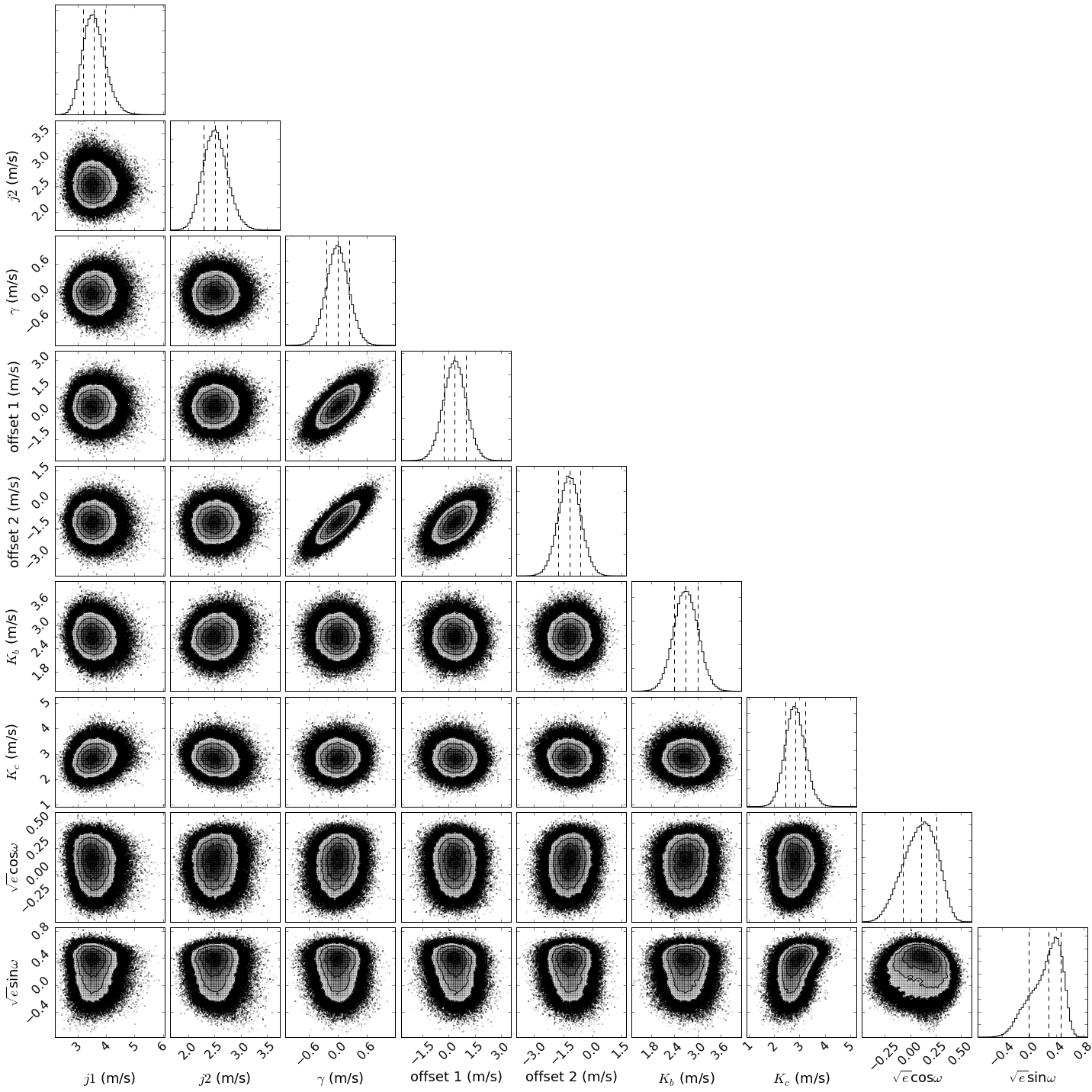}
\caption{Posterior distribution for the two-planet fit to Kepler-10 RVs, allowing eccentricity for planet c.  Variables are the jitter of the HIRES instrument ($j_1$), jitter of the HARPS-N instrument ($j_2$), velocity zero point ($\gamma$), velocity offset between HIRES and pre-upgrade HARPS-N RVs (offset 1), velocity offset between HIRES and post-upgrade HARPS-N RVs (offset 2), the semi-amplitudes of the RV curve for planets b ($K_b$) and c ($K_c$), and combinations of the eccentricity and argument of periastron of planet c, \ecosom\ and \esinom.  The dahsed lines indicate the 16$^{th}$, 50$^{th}$, and 84$^{th}$ percentiles.}
\label{fig:emcee2e}
\end{center}
\end{figure*}

\begin{figure}[htbp] 
   \centering
\includegraphics[width=3in]{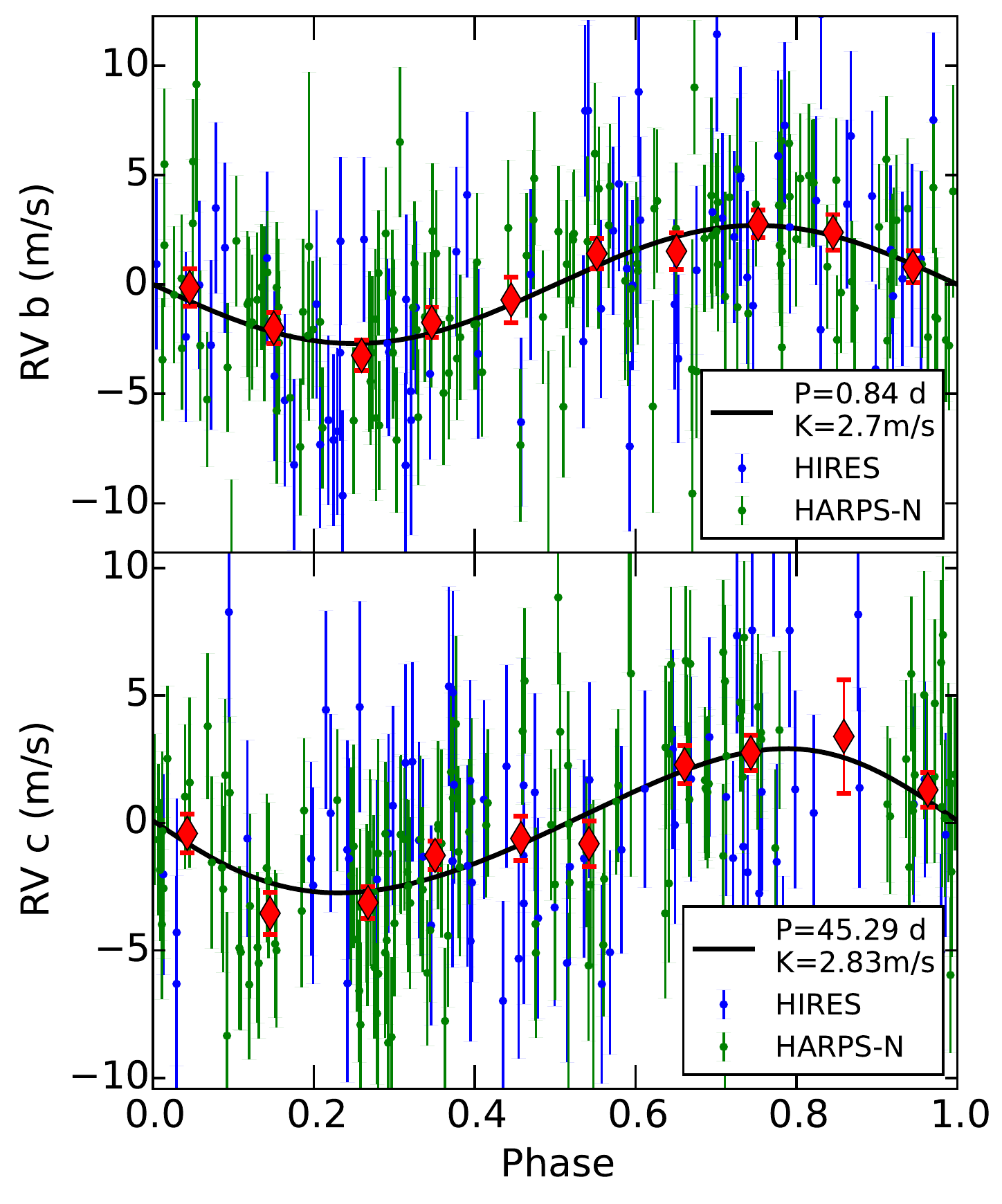}
   \caption{The RVs from HIRES (blue) and HARPS-N (green) phase-folded to the periods of Kepler-10 b (top) and c (bottom).  The red diamonds show the weighted mean RV of the HIRES and HARPS-N data combined in bins of 0.1 orbital phase.  The best two-planet fit in which the orbit of planet c ($P=45.3$ d) is allowed to be eccentric is shown in black.  The orbits of both planets are constrained by the \Kepler-determined transit times.  The best two-planet fit allowing eccentricity for planet c yields $m_b = \mbecc$, $m_c = \mcecc$, and $e_c = 0.17\pm0.13$.}
   \label{fig:2plecc}
\end{figure}

We perform an MCMC analysis over $j_1$, $j_2$, $\gamma$, offset 1, offset 2, $K_b$, $K_c$, \ecosom, and \esinom.  The posterior distribution of our sampler is shown in Figure \ref{fig:emcee2e}.  The two-planet fit using the parameters from the maximum likelihood of the posterior distribution is shown in Figure \ref{fig:2plecc}.  The best two-planet fit allowing eccentricity for planet c yields $m_b = \mbecc$, $m_c = \mcecc$, $\rho_b = \rhobecc$, $\rho_c = \rhocecc$, and $e_c = 0.17\pm0.13$.  Table \ref{tab:ecc} lists the median and $1\sigma$ uncertainties of the marginalized parameters and derived orbital and physical quantities.

\begin{table}[htbp]
\caption{Two Planet Eccentric MCMC Parameters}
\label{tab:ecc}
\tablenum{4}
\begin{tabular}{llllll}
\hline
\hline
\colhead{Paramter} & \colhead{Units} & \colhead{Median} & \colhead{$+1\sigma$} & \colhead{$-1\sigma$}& \colhead{Ref.}\cr
\hline
HIRES jitter  &\ms& 	3.57 &0.41& 0.36 & A \\
HARPS jitter  &\ms&2.51 &0.24 &0.22& A\\
$\gamma$  &\ms&-0.0&0.23&0.23& A\\
offset 1 &\ms &0.35&0.64 &0.64& A\\
offset 2  &\ms&-1.17&0.57&0.57& A\\
\Kb  &\ms&2.7&0.31 &0.31& A\\
\Kc  &\ms&2.83&0.4 	& 0.37& A\\
$\sqrt{e}$cos$\omega$  & 	&0.11 &	0.16 	&0.18& A\\
$\sqrt{e}$sin$\omega$  & & 	0.29 & 	0.18 	&0.29& A\\
$e_c$  & &0.13 	& 0.12 	& 0.09& A\\
$\omega_c$ & deg. & 66.81 	& 31.19 	& 68.51& A\\
$m_b$  &\mearth& 	3.76 	&0.43 	&0.42& A,B\\
$m_c$  &\mearth& 	14.59 &1.93&1.87& A,B\\
$r_b$ & \rearth & 1.47 & 0.03 & 0.02 & B\cr
$r_c$ & \rearth & 2.35 & 0.09 & 0.04 &B\cr 
$\rho_b$  &\gcc & 	6.53 &	0.76& 	0.74& A,B\\
$\rho_c$ &\gcc& 	6.21 	&0.82 &	0.8& A,B\\
\hline
\end{tabular}
\tablecomments{The priors on all the parameters were uniform.}
\tablerefs{A.This work.  B.\citet{Dumusque2014}}
\end{table}

\section{Transit Times of Kepler-10 c}
\begin{figure}[htbp]
\begin{center}
\includegraphics[width=3in]{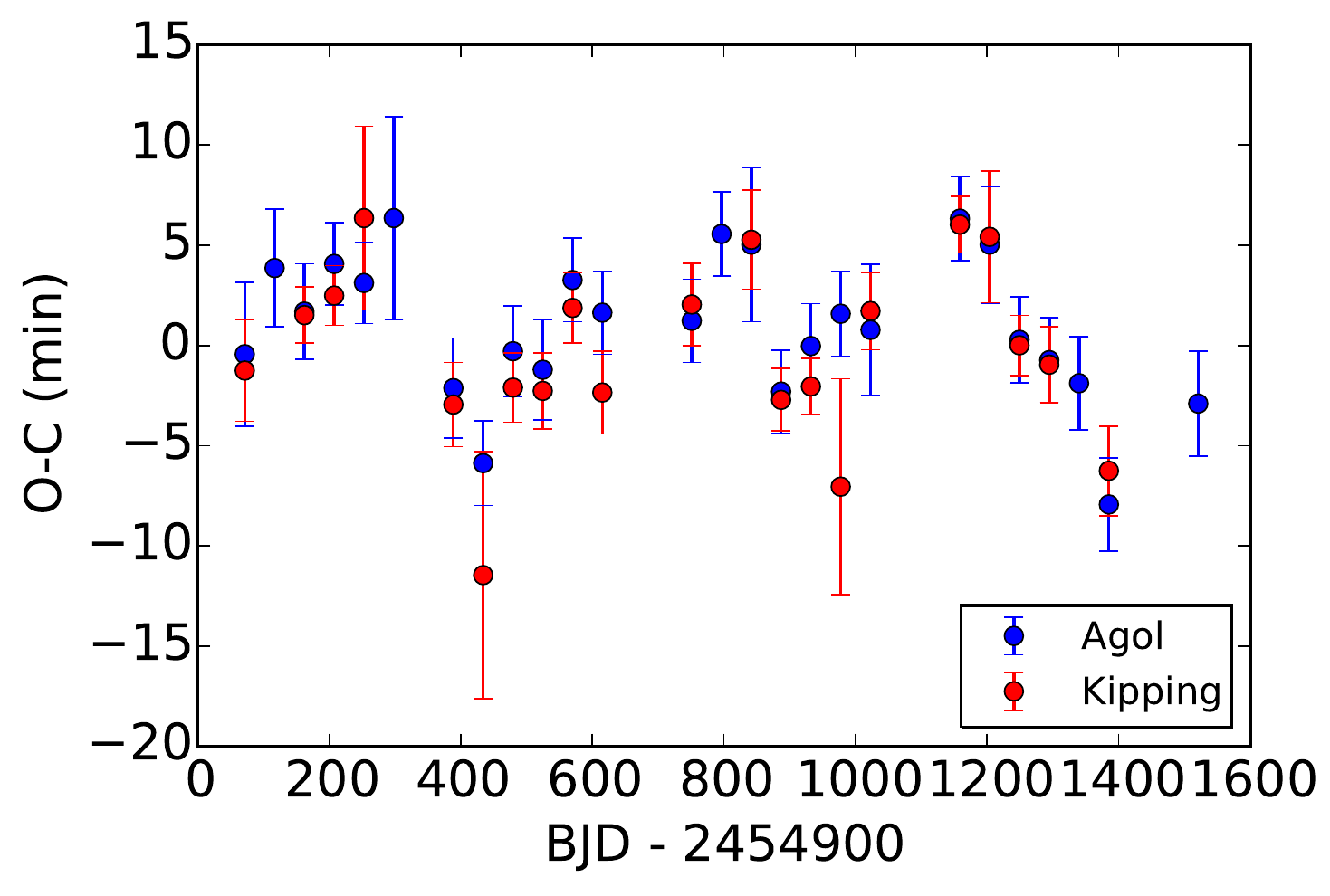}
\includegraphics[width=3in]{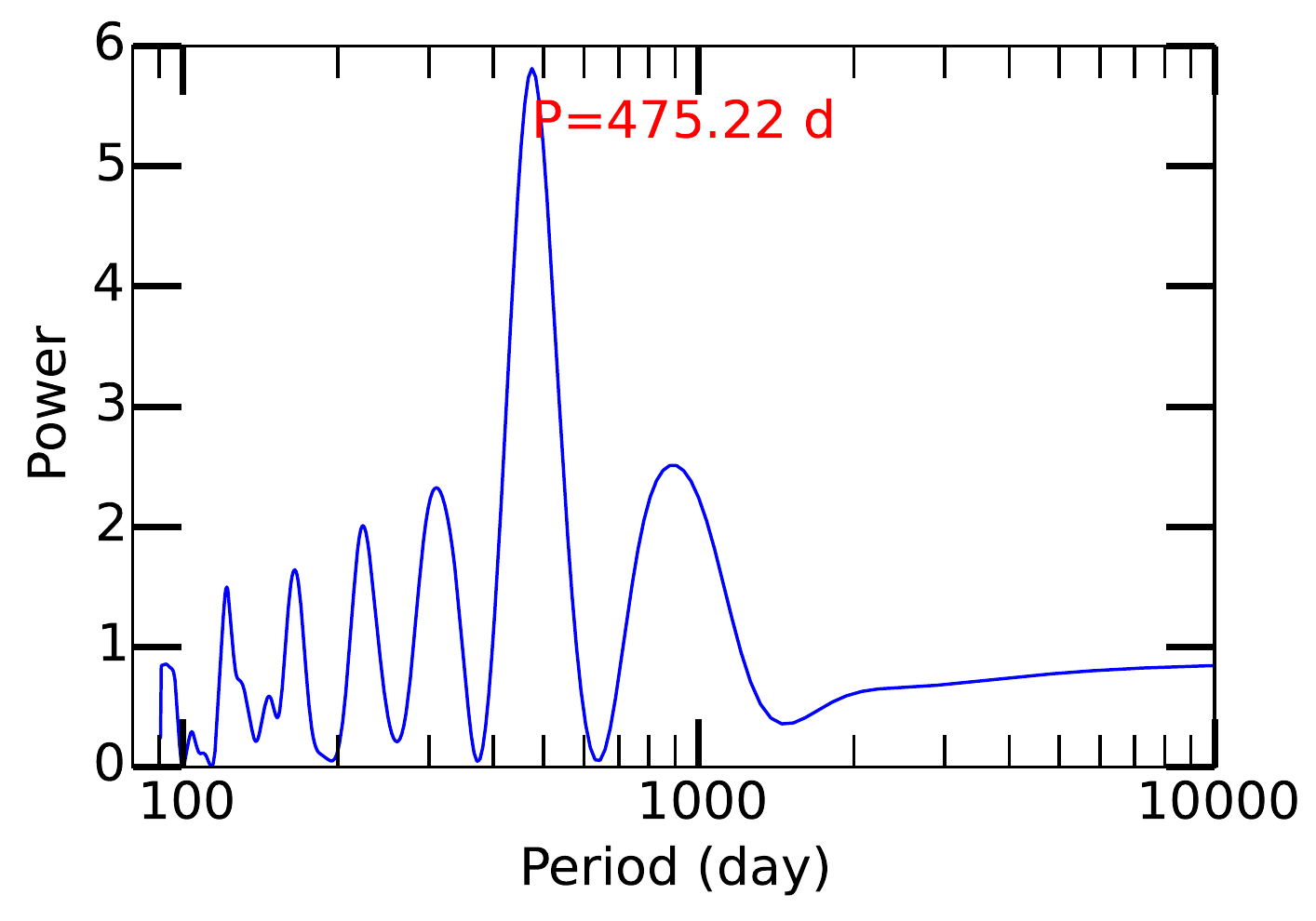}
\caption{Top: the TTVs of Kepler-10 c measured in independent analyses by Eric Agol and David Kipping.  The dependent axis (O-C) is the lateness of each observed transit with respect to a linear ephemeris.  Bottom: the periodogram of Eric Agol's TTVs.  The peak at 475 days corresponds to the observed sinusoidal period in the TTV time series and is the TTV super-period.}
\label{fig:TTVs}
\end{center}
\end{figure}

The transit times of Kepler-10 c vary with respect to a linear ephemeris.  \citet{Kipping2015} found transit timing variations (TTVs) in the long and short cadence data with $5\sigma$ confidence.  We independently measure the TTVs and find a solution consistent with the TTVs in \citet[][see Figure \ref{fig:TTVs}]{Kipping2015}.  The TTVs appear to have a sinusoidal period of about 475 days.

\subsection{Measuring the transit times}
From the photometry, we computed the TTVs twice.  David Kipping measured the transit times \TTKip, as documented in \citet{Kipping2015}, and Eric Agol measured the transit times \TTAgol\ with the method described here.  To include the impact of correlated noise on the transit timing
uncertainty measured from the short cadence data, we carried out the
following procedure:  1)  We did a joint fit to the transits of
both planets assuming white noise and polynomial detrending
near each transit. Overlapping transits of the two planets were
modeled simultaneously.  We let the transit times of 10 c vary, but
fixed the transit times of 10 b to a periodic ephemeris.  2) We
optimized this fit with a Levenberg-Marquardt model, and then subtracted it
from the short cadence data.  3) We computed the autocovariance
of the residuals to this initial fit for the short cadence light curve
as a function of the number of cadences, $a(n)$, where
$a(0)$ is the variance of the data and $a(1)$ is the covariance
between residuals separated by one cadence, etc.  We concatenated data
across gaps when computing the autocovariance as these gaps are
a small fraction of the entire dataset.  4) Using the computed
autocovariance
of the data, we computed the best-fit transit model to Kepler-10 c
with the model for Kepler-10 b subtracted.  We did not detrend
at this stage, but instead used a covariance matrix computed
from $a(n)$:  $\Sigma_{i,j} = a(\vert i-j\vert)$.  The likelihood
function was ${\cal L} \propto \exp{\left(-\frac{1}{2}({\bf r}^T {\bf
\Sigma}^{-1}
{\bf r})\right)}$, where ${\bf r}$ is the residual vector for each
transit after subtracting off the model component due to Kepler-10 b.
We computed the timing uncertainties, $\sigma_{t,i}$, from the covariance
of the model parameters at the best-fit value for the $i$th transit.
We then allowed transit time to vary by $\pm3\sigma_{t,i}$ for
each transit, and mapped out the effective chi-square,
$\chi^2 = -2\ln{\cal L}$, versus timing offset.  5) We found the upper and
lower time offsets at which the $\chi^2$ of the fit changed by
one, and chose the maximum of these offsets and $\sigma_{t,i}$ to
estimate the transit timing uncertainty.  The best-fit times of transit and
the uncertainty are reported in Table \ref{agol_times}.

\begin{center}
\begin{table}[htbp]
\caption{Transit times (\TTAgol) measured from short cadence
transits of Kepler-10 c using a correlated-noise analysis.}
\tablenum{5}
\tablewidth{0pt}
\begin{tabular}{lcc}
\hline
\hline
\colhead{Transit} & \colhead{Time} & \colhead{Uncertainty} \cr
             & \colhead{JD-2,455,000} & \colhead{(days)} \cr
             \hline
        0 &   62.2673 &  0.0017 \cr
        1 &  107.5632 &  0.0014 \cr
        2 &  152.8569 &  0.0014 \cr
        3 &  198.1534 &  0.0436 \cr
        4 &  288.7361 &  0.0014 \cr
        5 &  334.0278 &  0.0015 \cr
        6 &  379.3260 &  0.0016 \cr
        7 &  424.6197 &  0.0016 \cr
        8 &  469.9171 &  0.0014 \cr
        9 &  515.2102 &  0.0014 \cr
       10 &  651.0928 &  0.0014 \cr
       11 &  696.3902 &  0.0015 \cr
       12 &  741.6841 &  0.0015 \cr
       13 &  786.9733 &  0.0014 \cr
       14 &  832.2692 &  0.0015 \cr
       15 &  877.5646 &  0.0015 \cr
       16 &  922.8583 &  0.0015 \cr
       17 & 1058.7451 &  0.0015 \cr
       18 & 1104.0385 &  0.0016 \cr
       19 & 1149.3295 &  0.0015 \cr
       20 & 1194.6231 &  0.0015 \cr
       21 & 1239.9166 &  0.0016 \cr
       22 & 1285.2067 &  0.0016 \cr
       23 & 1421.0931 &  0.0016 \cr
       \hline
\end{tabular}
\label{agol_times}
\end{table}
\end{center}

\section{False Alarm Probability}
We used two diagnostics to explore the possibility that the apparent coherent signal of the Kepler-10 c TTVs were due to noise, rather than planetary dynamics.  First, we used a bootstrap test, which is a common method in the RV literature to assess whether an apparently coherent, sinusoidal signal could be produced by noise.  Second, we used a Monte Carlo test.  We applied each test to both \TTAgol\ and \TTKip.
\subsection{Scramble (Bootstrap) Tests}
We numbered the observed TTVs (O-C values) from 1-$N$ and produced 10,000 fake data sets of length $N$.  To construct each fake data set, we randomly drew a number $j$ between 1 and $N$ and used the $jth$ observed transit as the first transit in our fake data set.  We repeated this procedure (including the possibility of drawing $j$ again) until we had a fake data set of length $N$.  Thus, it would be possible to draw $j$ $N$ times, or to draw each number between 1 and $N$ exactly once, or any other combination from the $N^N$ possibilities.  

\begin{figure*}[htbp]
\begin{center}
\includegraphics[height=1.5in]{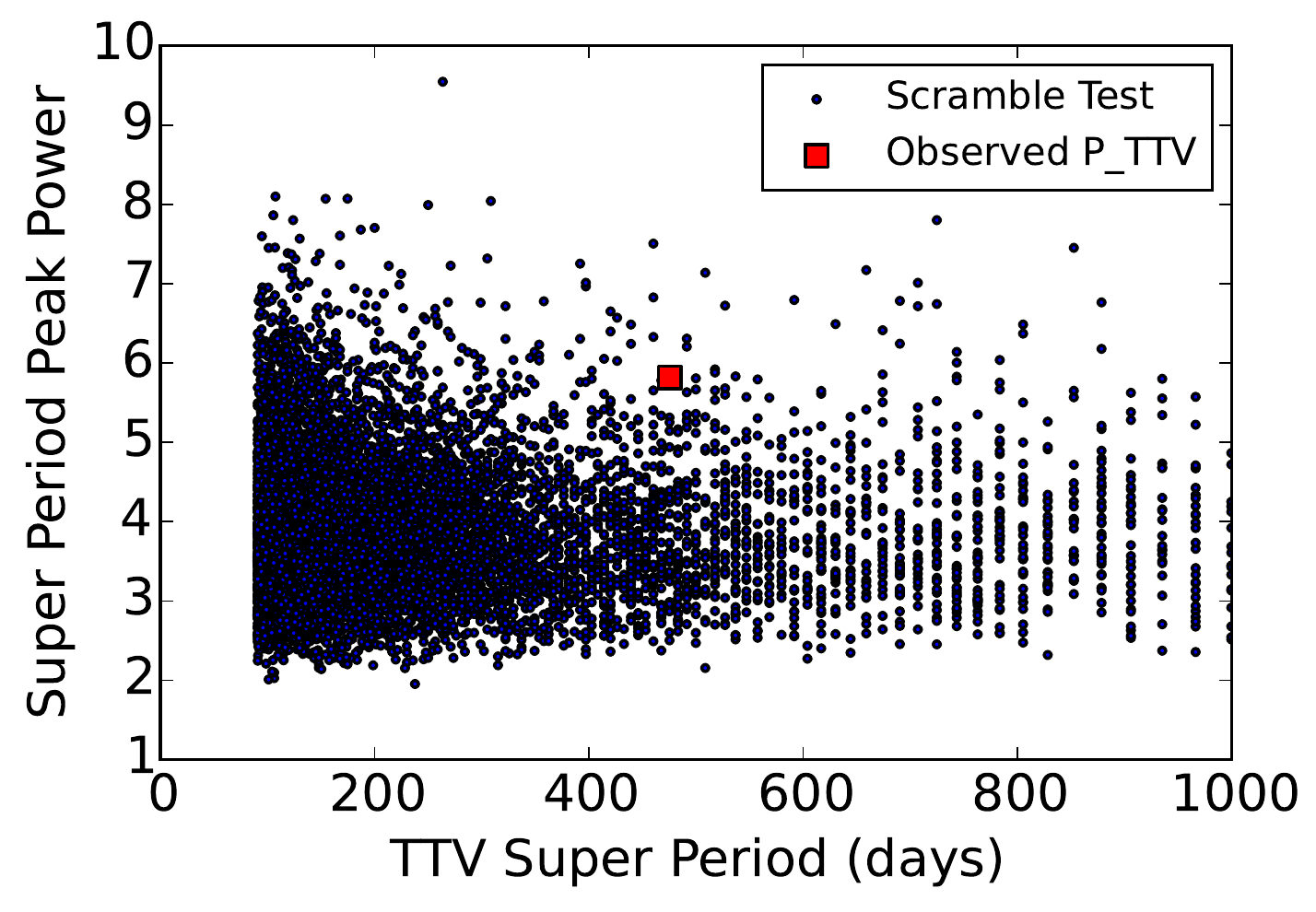}
\includegraphics[height=1.5in]{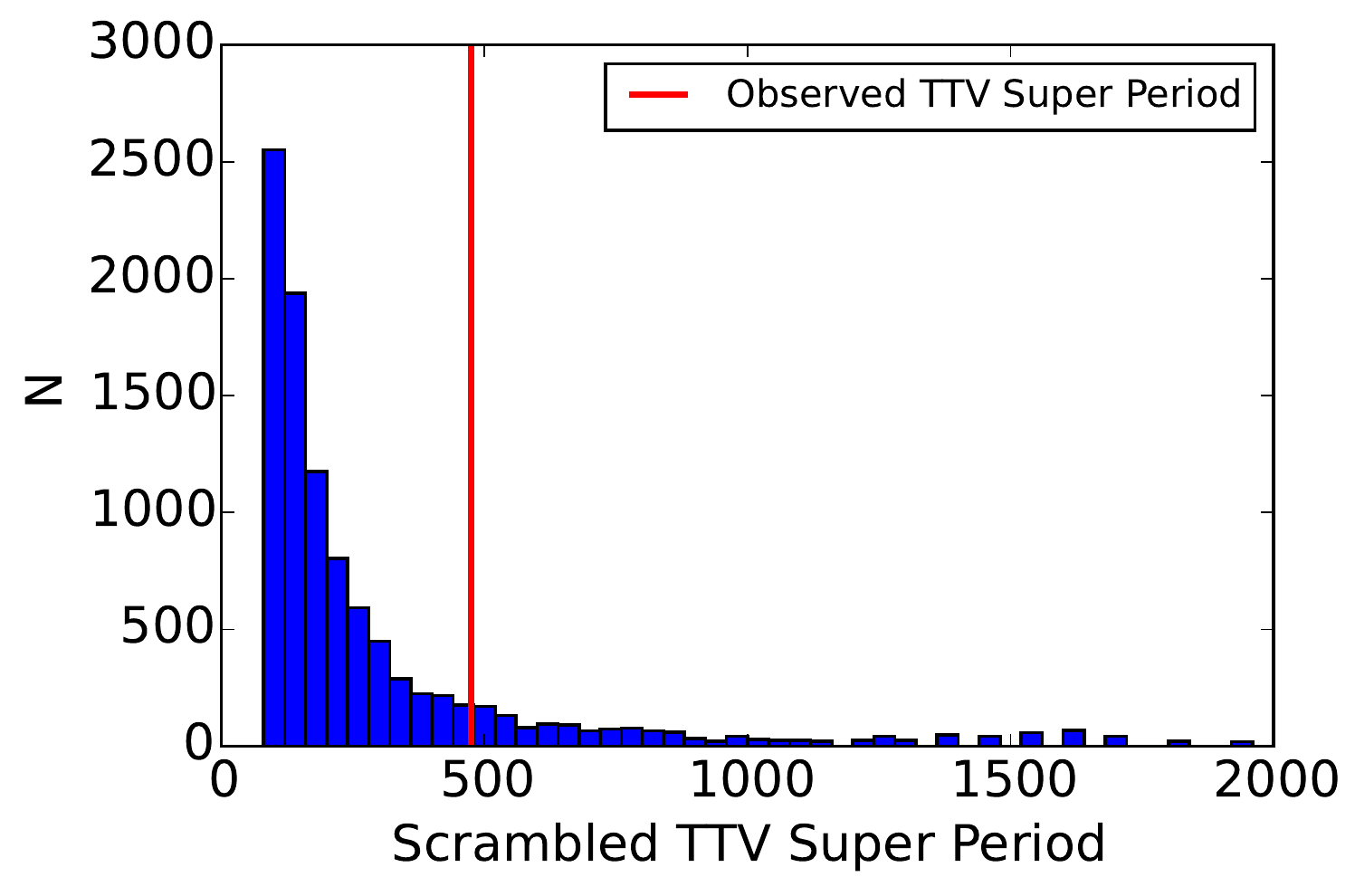}
\includegraphics[height=1.5in]{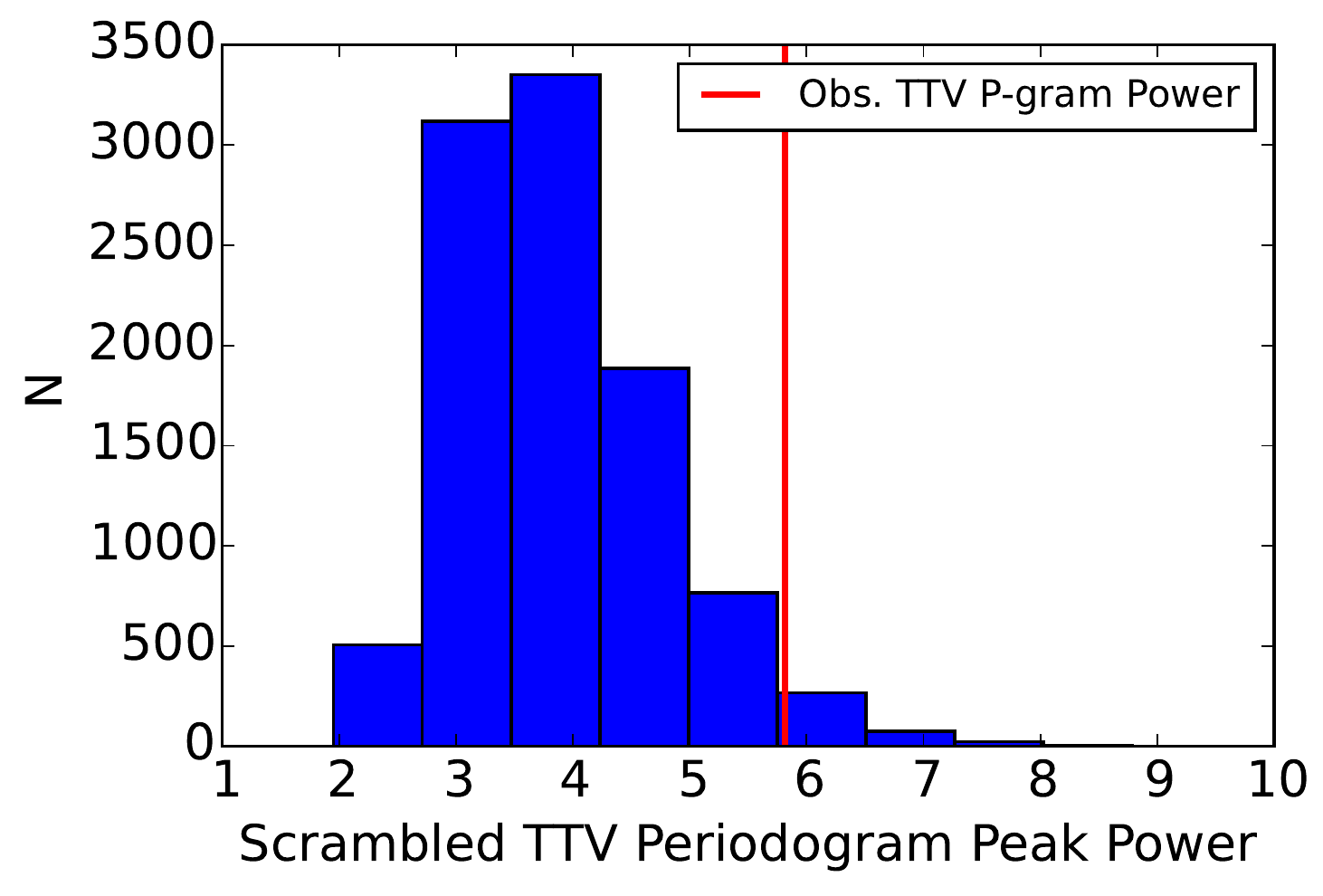}
\includegraphics[height=1.5in]{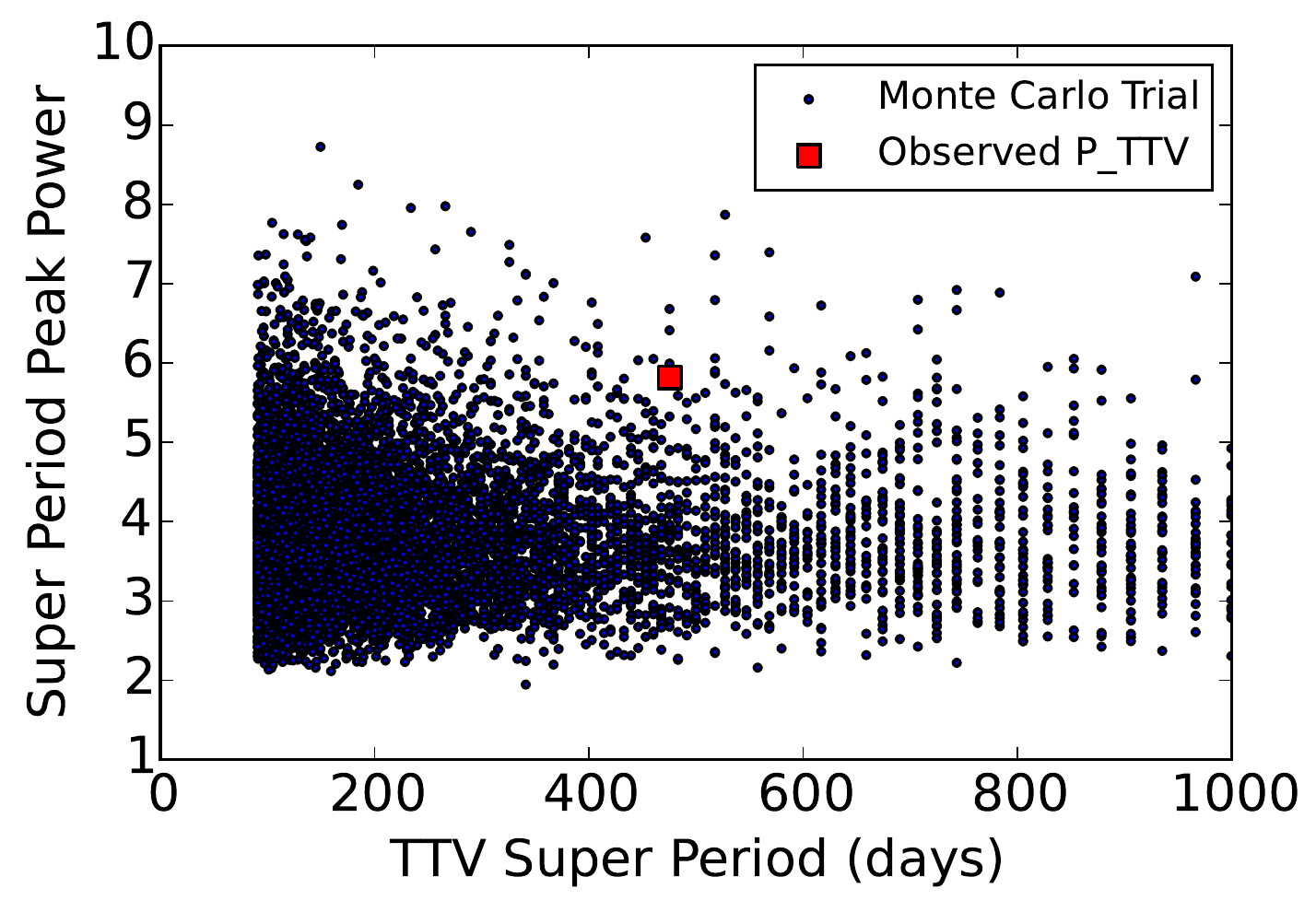}
\includegraphics[height=1.5in]{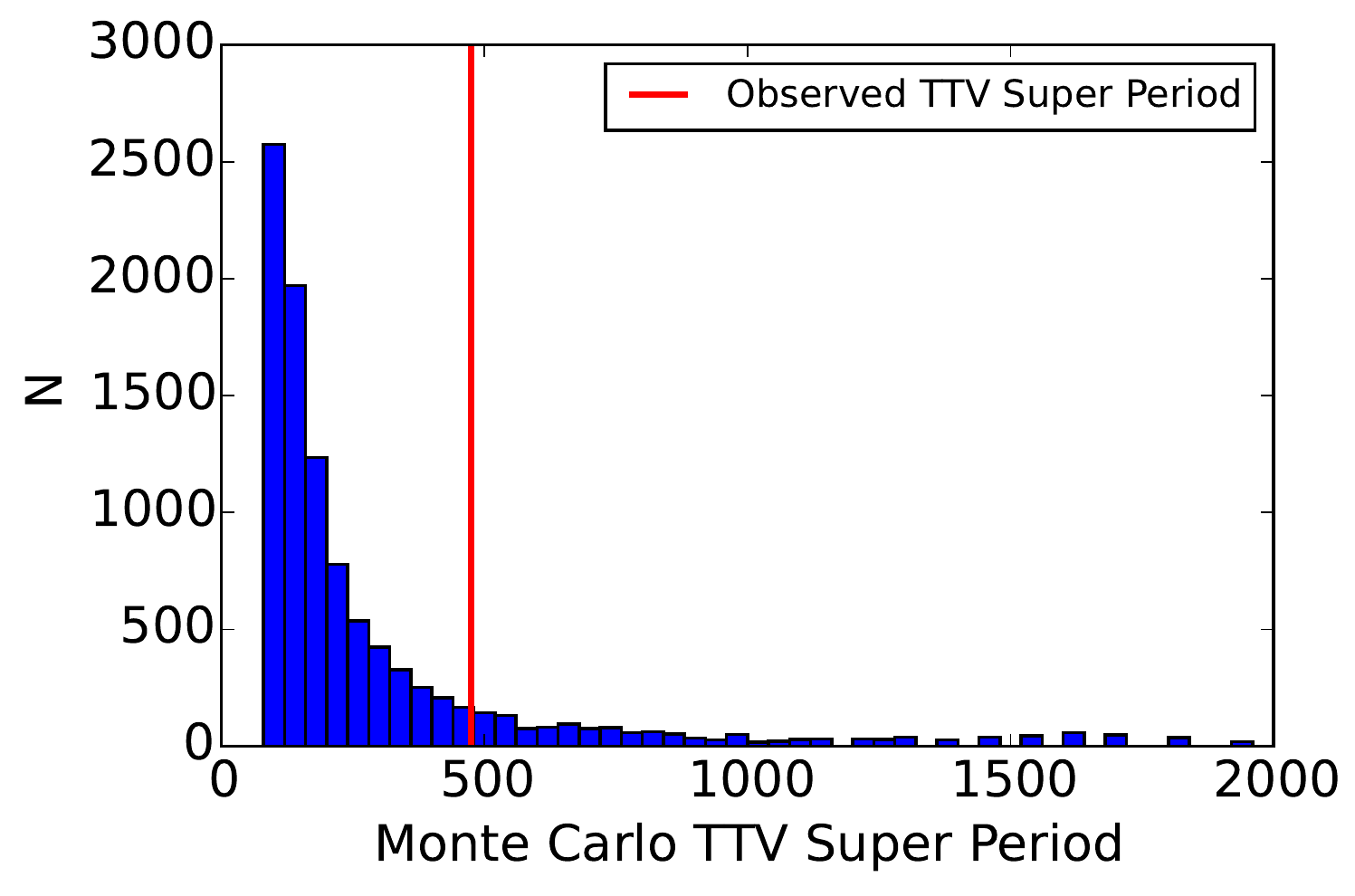}
\includegraphics[height=1.5in]{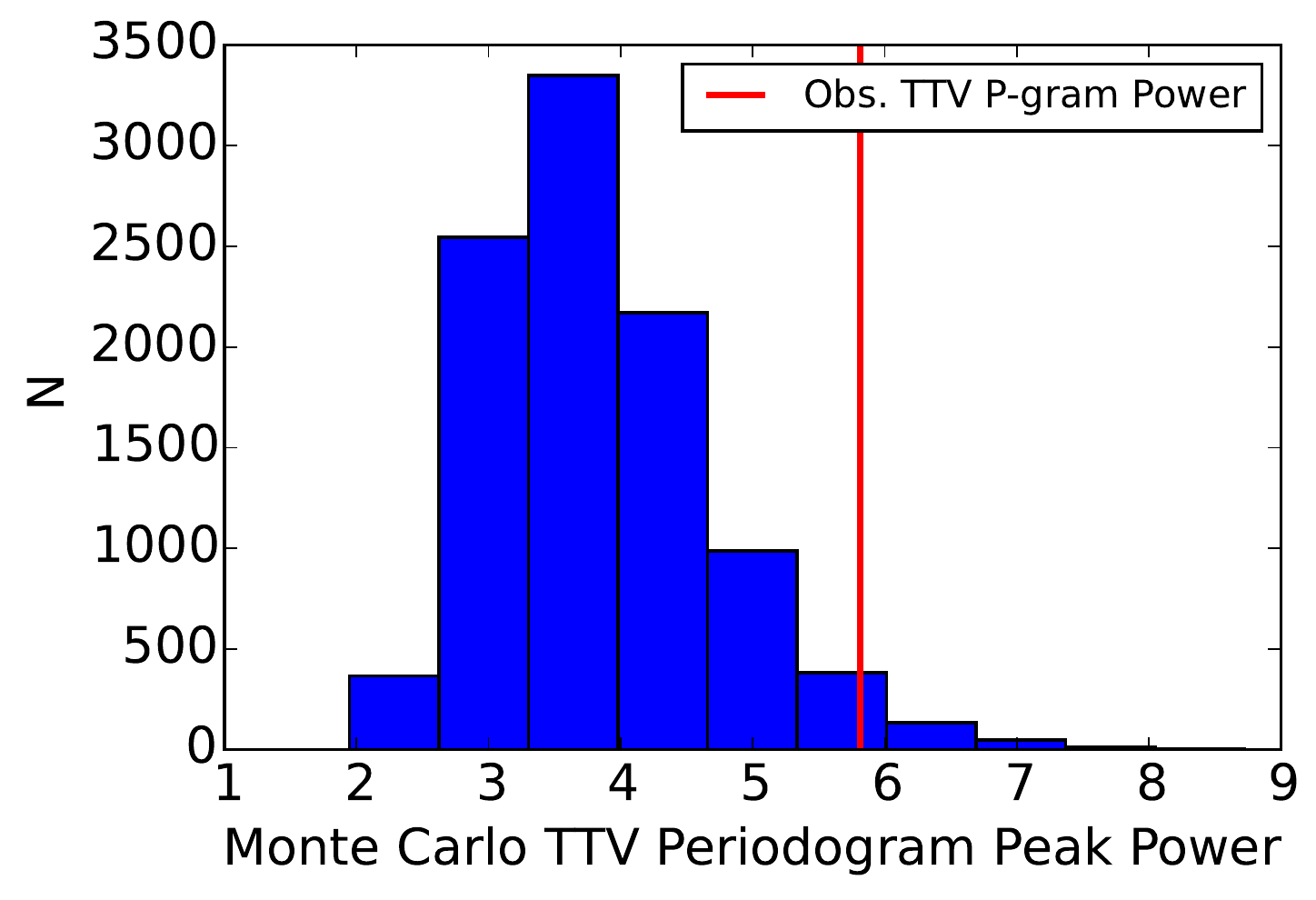}
\caption{False alarm probability tests for the Kepler-10 TTVs.  Top left: Peak periodogram power as a function of TTV super-period for 10,000 scramble tests (black), as compared to the observed peak periodogram power and super-period.  Top center: Histogram of the super-periods generated from 10,000 scramble tests, compared to the observed TTV super-period.  Top right: Histogram of the peak periodogram powers generated from 10,000 scramble tests, compared to the observed peak periodogram power of the TTVs.  Bottom row: same as top, but for 10,000 monte carlo tests.}
\label{fig:Agol_FAP}
\end{center}
\end{figure*}

For each fake data set, we computed the Lomb-Scargle periodogram \citep{Press1989} of the fake TTVs from the Nyquist period (90.5 days, i.e., twice the orbital period of planet c) to 10,000 days.  We found the period with the most power in the periodogram, and recorded this period (the ``TTV super period") and its associated power.  We then compared the periodogram of the observations to our suite of 10,000 data sets (see Figure \ref{fig:Agol_FAP}).  By counting the number of fake data sets that produce a peak with more power than the observed peak, we can estimate the false alarm probability, i.e., the probability that noise, rather than an astrophysical signal, is responsible for the observed peak.  For \TTAgol, we get FAP = 3.89\%.  For \TTKip, we get FAP =  3.71\%.

\subsection{Monte Carlo Tests}
We generated 10,000 Monte Carlo fake data sets of transit midpoint times.  To construct each fake data set, we drew a fake observation of each transit time from a Normal distribution with a mean of the observed transit time and a variance of the uncertainty in the transit time squared.  
We computed the false alarm probability in the same manner as for the scramble test, obtaining FAP = 2.6\% for \TTAgol, and FAP = 1.89\% for \TTKip\ (see Figure \ref{fig:Agol_FAP}).

The FAP values of 1.9\% to 3.9\% indicate a 3\% probability that the apparent TTV super period is spurious and that the apparent coherence of the TTVs is a chance occurrence due to noise in the TTVs.  This result differs from the Bayesian approach in \citet{Kipping2015}, which finds the TTVs with $5\sigma$ confidence.  Possible reasons for the difference between the FAP calculated in this work and the the $5\sigma$ confidence found in \citet{Kipping2015} are (1) we look for a sinusoidal signal, whereas \citet{Kipping2015} look for any type of variation from a flat line, and (2) we ignore the error bars (since the errors in \TTAgol\ are of nearly equal values), whereas \citet{Kipping2015} interpret the errors, allowing a few outliers with large errors to be down-weighted.

\section{Planetary Properties of Kepler-10: Three-Planet Solutions}
\subsection{Analytical Motivation}
If the observed coherent TTVs are due to the dynamical interactions of planets in the system, they indicate the presence of a third planet (KOI-72.X).  Known planets (b and c) cannot cause the TTVs because planet c cannot be its own perturber, and planet b is too far away in period space ($P_c/P_b = 54$) to perturb planet c at the amplitude observed.  Therefore, the best dynamical explanation of the TTVs is the existence of a third planet.  The periodogram of the residual RVs after the RVs due to planets b and c are subtracted does not present a strong peak (see Figure \ref{fig:pgram}).  Therefore we cannot unambiguously identify the orbital period KOI-72.X from the RVs alone.

We  use analytic theory to predict the most likely orbital periods of a third planet.  Equation 5 from \citet{Lithwick2012}, also called the TTV equation, relates the super-period of the TTVs to the orbital periods of the two planets:
\begin{equation}
\frac{1}{P_\mathrm{TTV}} = \big\lvert \frac{j}{P} - \frac{j-1}{P'}\big\rvert
\label{eqn:ttv}
\end{equation}
where \PTTV\ is the super-period measured from the TTV sinusoid, $P$ is the inner period, $P'$ is the outer period, and the planets are near the $j:j-1$ resonance.  Because the perturber can exist just inside or just outside the $j:j-1$ mean motion resonance, and because the perturber can be interior or exterior, there are 4 solutions for each $j:j-1$.  Using Equation \ref{eqn:ttv}, we can predict a series of likely orbital periods for the perturber, planet candidate KOI-72.X, which are enumerated in Table \ref{tab:pttv}.
\begin{table*}[hbtp]
\caption{Using Equation \ref{eqn:ttv}, we compute a series of likely orbital periods for planet candidate KOI-72.X.  The left three columns show the analytic values that solve the TTV equation for the observed super-period of 475 days.  The right columns show the parameters that minimize $\chi^2$ between the observed TTVs and RVs and the N-body model.  Each N-body model was seeded using the analytic prediction for KOI-72.X as an initial guess.  The last two rows show control trials, in which ($a$) the initial orbital period of KOI-72.X is far from the analytic solutions to the TTV equation, and ($b$) there is no KOI-72.X.}
\label{tab:pttv}
\tablenum{6}
\begin{center}
\begin{tabular}{cc|cccccccccccccc}
\hline
\hline
\multicolumn{2}{l|}{Analytic} & \multicolumn{11}{l}{Numerical -- Parameters at epoch $T_0 = 53.67844$ (BJD-2454900)}\cr
\hline
$j:j-1$ & $P_X$ (predicted, & $P_c$ & $P_X$ & $m_c$ & $m_X$ & ecc$_c$ & $\omega_c$ & ecc$_X$ & $\omega_X$ &  M$_c$ & M$_X$ & $\chi^2_{\mathrm{TT,c}}$ & $\chi^2_{\mathrm{RV}}$ & $\chi^2$ & $\Delta$BIC$^a$ \cr
 & days) & (days) & (days) & ($\mearth$) & ($\mearth$)& & (deg.) &  & (deg.) &  (deg.)  & (deg.)  &  &  & & \cr
\hline
2:1 & 100 & 45.295 & 101.36 & 13.94 & 6.84 & 0.09 & 79.7 & 0.19 & 96.5 & 225.5 & 222.9 & 19 & 198 & 217 & 0.0 \cr 
2:1 &  82.7 & 45.295 & 80.118 & 14.13 & 0.06 & 0.02 & 79.3 & 0.23 & 89.9 & 227.1 & 205.4 & 46 & 214 & 260 & 43.0 \cr 
 
2:1 & 23.8 & 45.295 & 23.761 & 15.37 & 0.86 & 0.28 & 89.1 & 0.24 & 50.8 & 217.4 & 8.6& 18 & 212 & 230 & 13.0 \cr 

2:1 & 21.6 &  45.296 & 21.619 & 14.74 & 0.54 & 0.11 & 89.0 & 0.26 & 107.9 & 217.7 & 360.0 & 28 & 212 & 240 & 23.0 \cr 
3:2 & 71.3 & 445.295 & 71.323 & 14.15 & 1.06 & 0.01 & 87.4 & 0.08 & 182.6 & 219.5 & 216.2 & 17 & 216 & 233 & 16.0 \cr 
3:2 & 64.9 & 45.294 & 63.56 & 14.29 & 0.93 & 0.06 & 90.0 & 0.02 & 41.0 & 216.9 & 116.0 & 27 & 213 & 240 & 23.0 \cr 
3:2 & 31.2 & 45.296 & 31.183 & 14.68 & 1.38 & 0.01 & 89.6 & 0.02 & 48.9 & 217.3 & 9.7 & 26 & 214 & 240 & 23.0 \cr 
3:2 & 29.7 & 45.3 & 28.777 & 15.83 & 3.3 & 0.19 & 92.5 & 0.22 & 89.6 & 215.3 & 270.6 & 42 & 204 & 246 &29.0 \cr 
\hline
ctrl. 1$^b$ & 90.5 & 45.296 & 90.433 & 13.81 & 3.21 & 0.0 & 79.6 & 0.07 & 230.1 & 227.3 & 90.2 & 47 & 222 & 269 & 52.0 \cr
ctrl. 2$^c$ & x & 45.295 & x & 13.98 & x & 0.00 & 60.9 & x & x & 245.9 & x  & 52 & 215 & 267 & 22.5 \cr 
\hline
\end{tabular}
\tablecomments{$^a$  The difference in the Bayesian Information Criterion (BIC) with respect to the best solution explored (at P=100 days).  According to \citet{Kass1995}, the favorability for the model with the lower BIC value is ``very strong" when $\Delta BIC > 10$, ``strong" when $6 < \Delta BIC < 10$, ``positive" when $2 < \Delta BIC < 6$, and ``not worth more than a bare mention" when $0 < \Delta BIC < 2$.}
\tablecomments{$^b$  KOI-72.X was placed exactly on the 2:1 exterior resonance, halfway between the 82 day and 101 day analytic solutions.}
\tablecomments{$^c$  A two-planet dynamic solution with only Kepler-10 b and c (i.e. no KOI-72.X).}
\end{center}
\end{table*}%

\subsection{Dynamical Solutions}
\begin{figure*}[htbp]
\begin{center}
\includegraphics[width=1.5in]{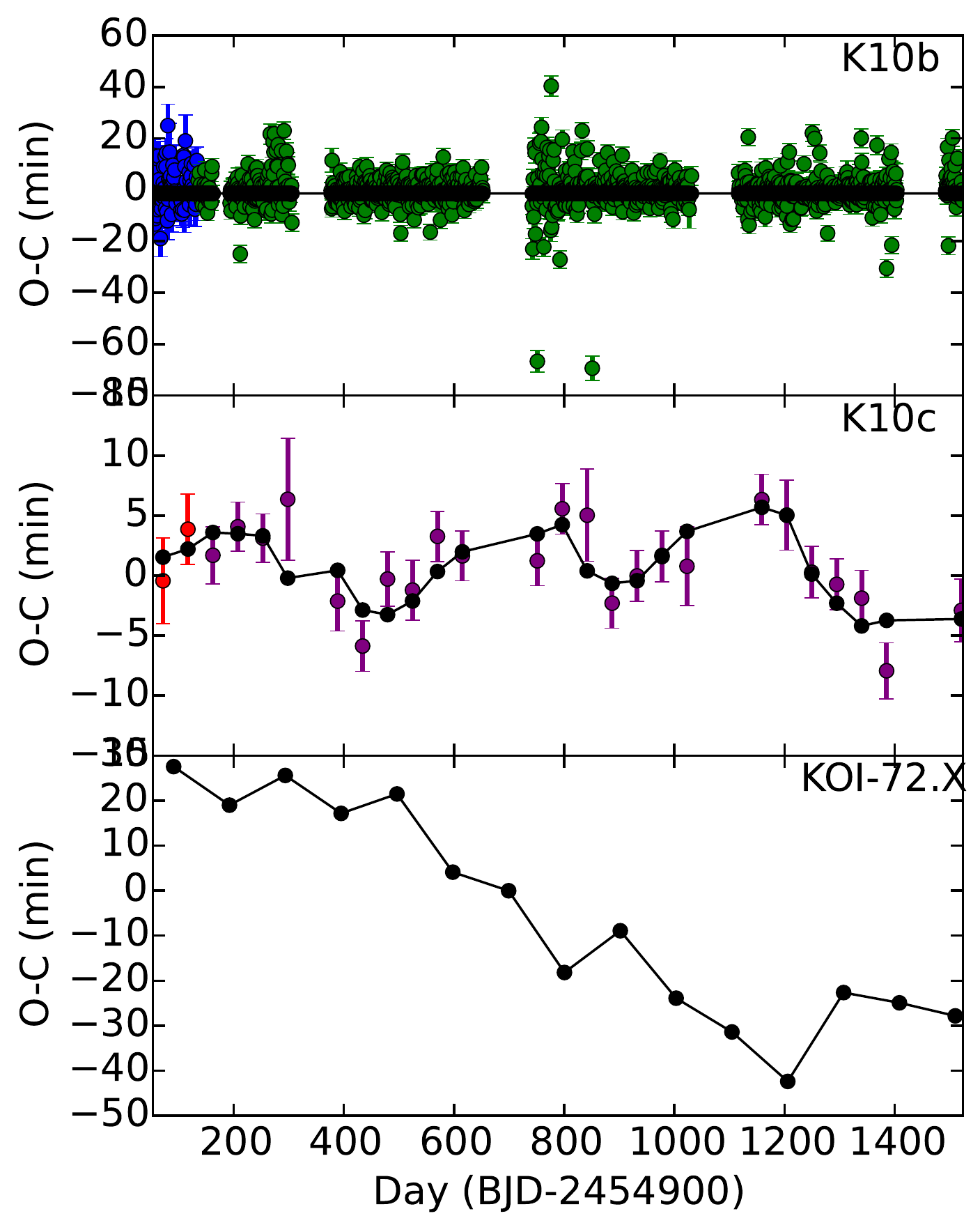}
\includegraphics[width=1.5in]{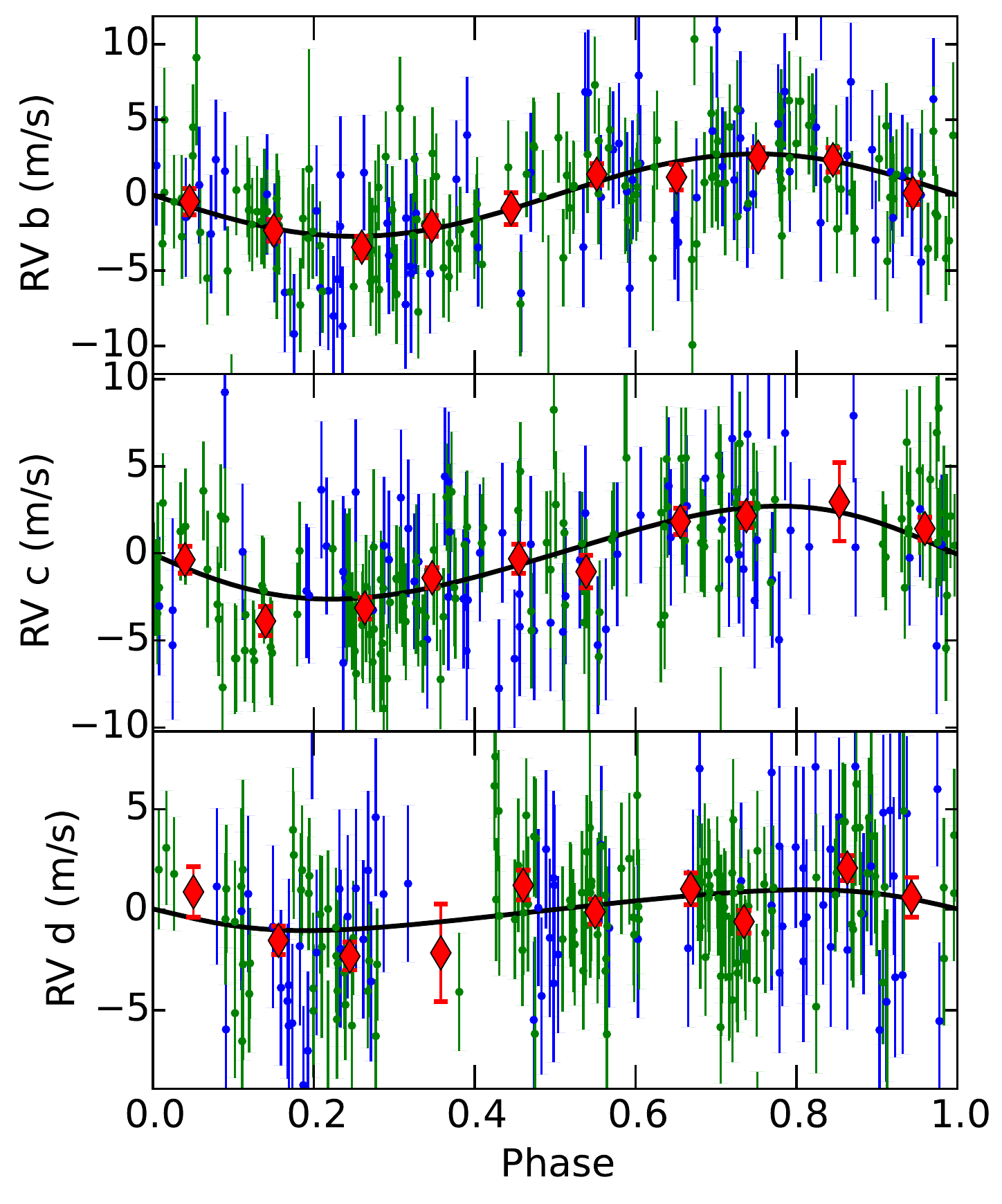}
\includegraphics[width=1.5in]{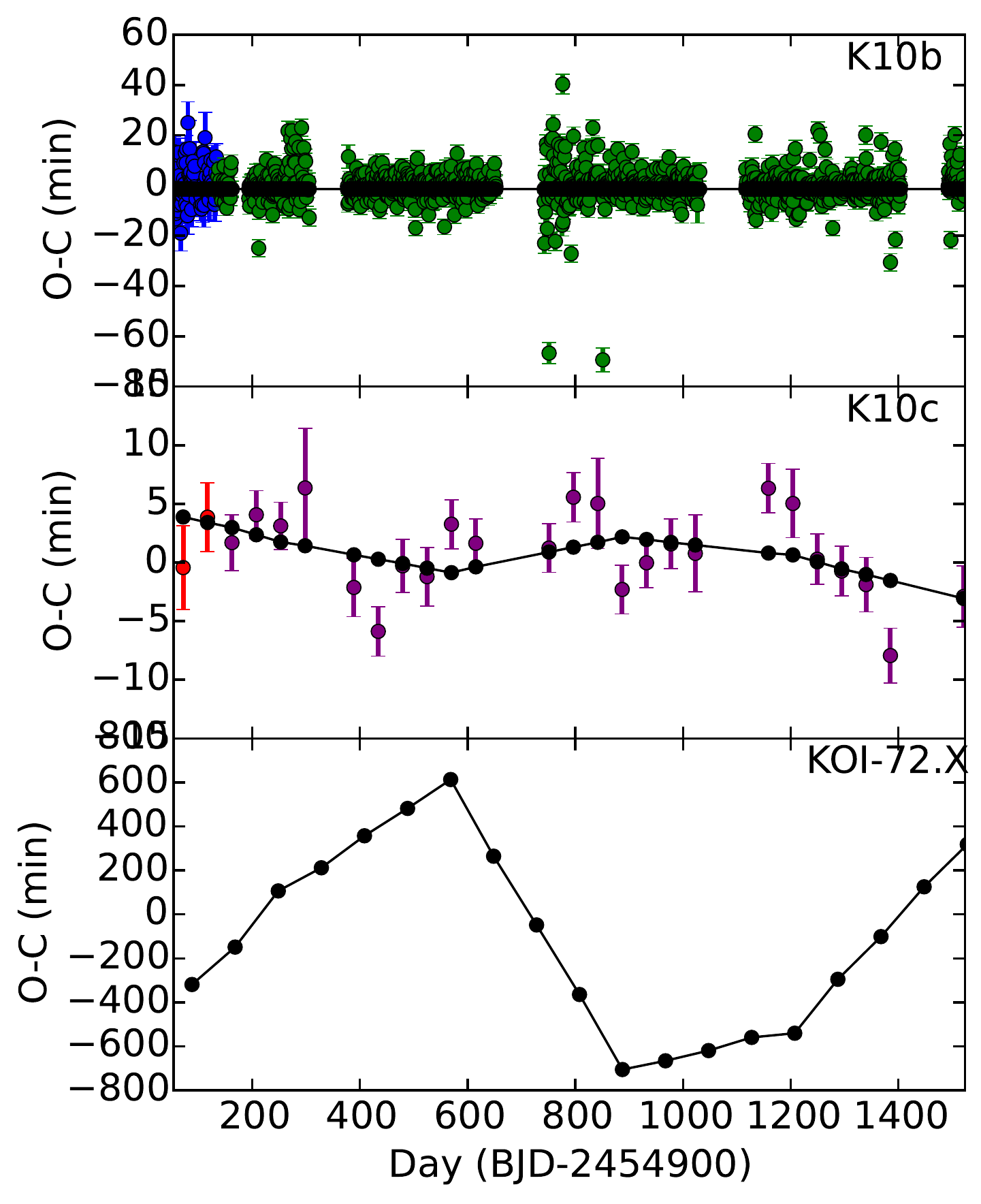}
\includegraphics[width=1.5in]{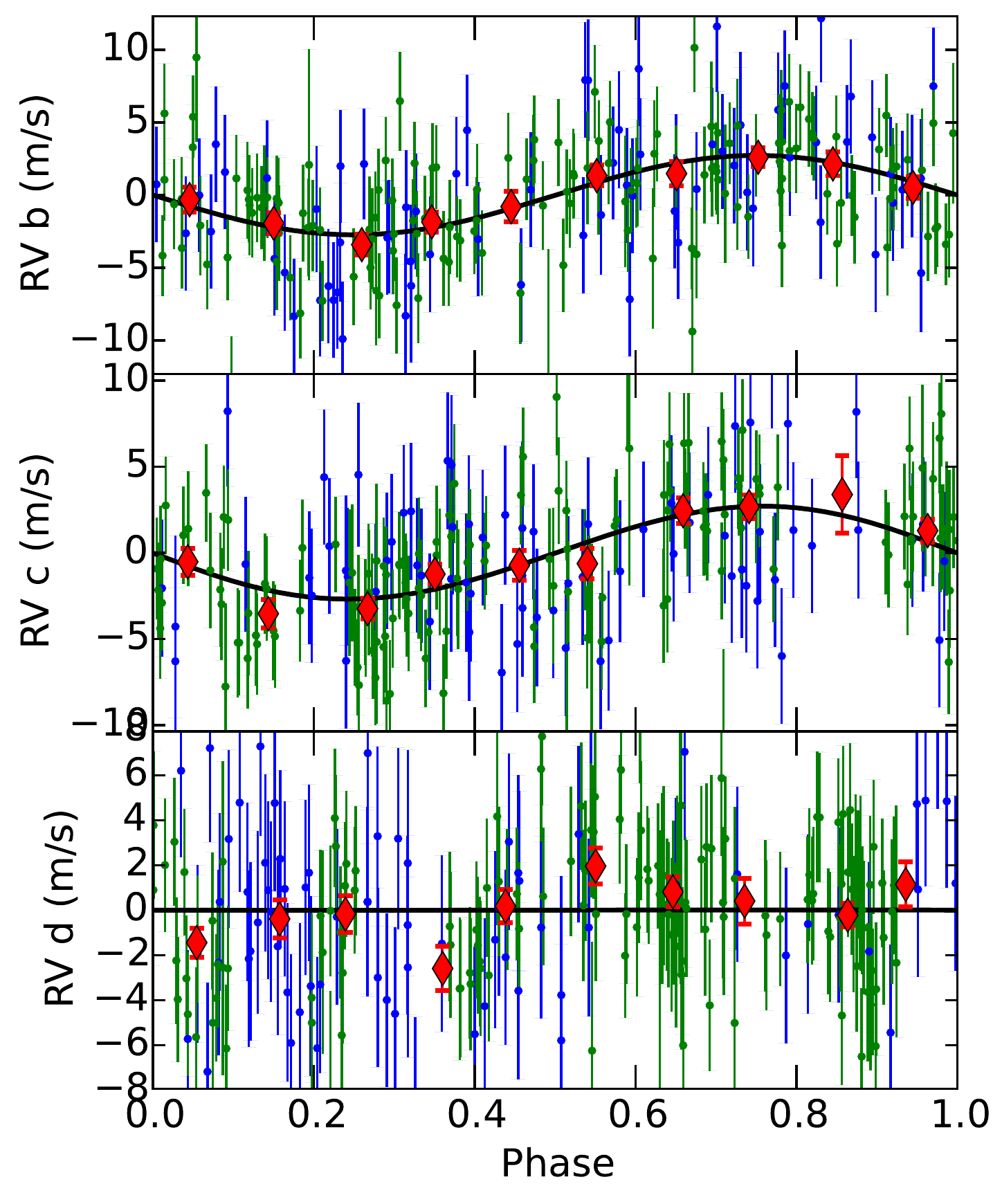}
\includegraphics[width=1.5in]{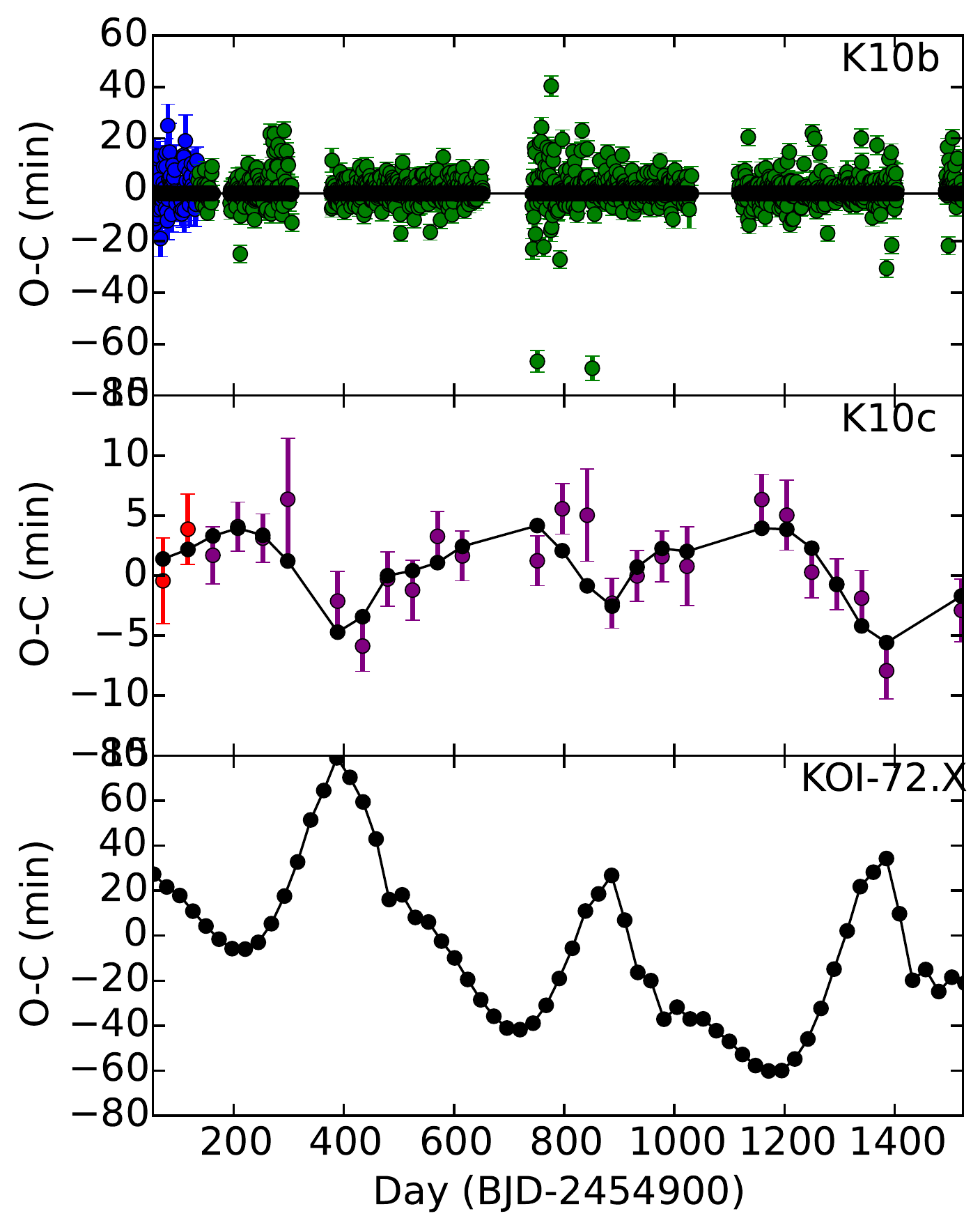}
\includegraphics[width=1.5in]{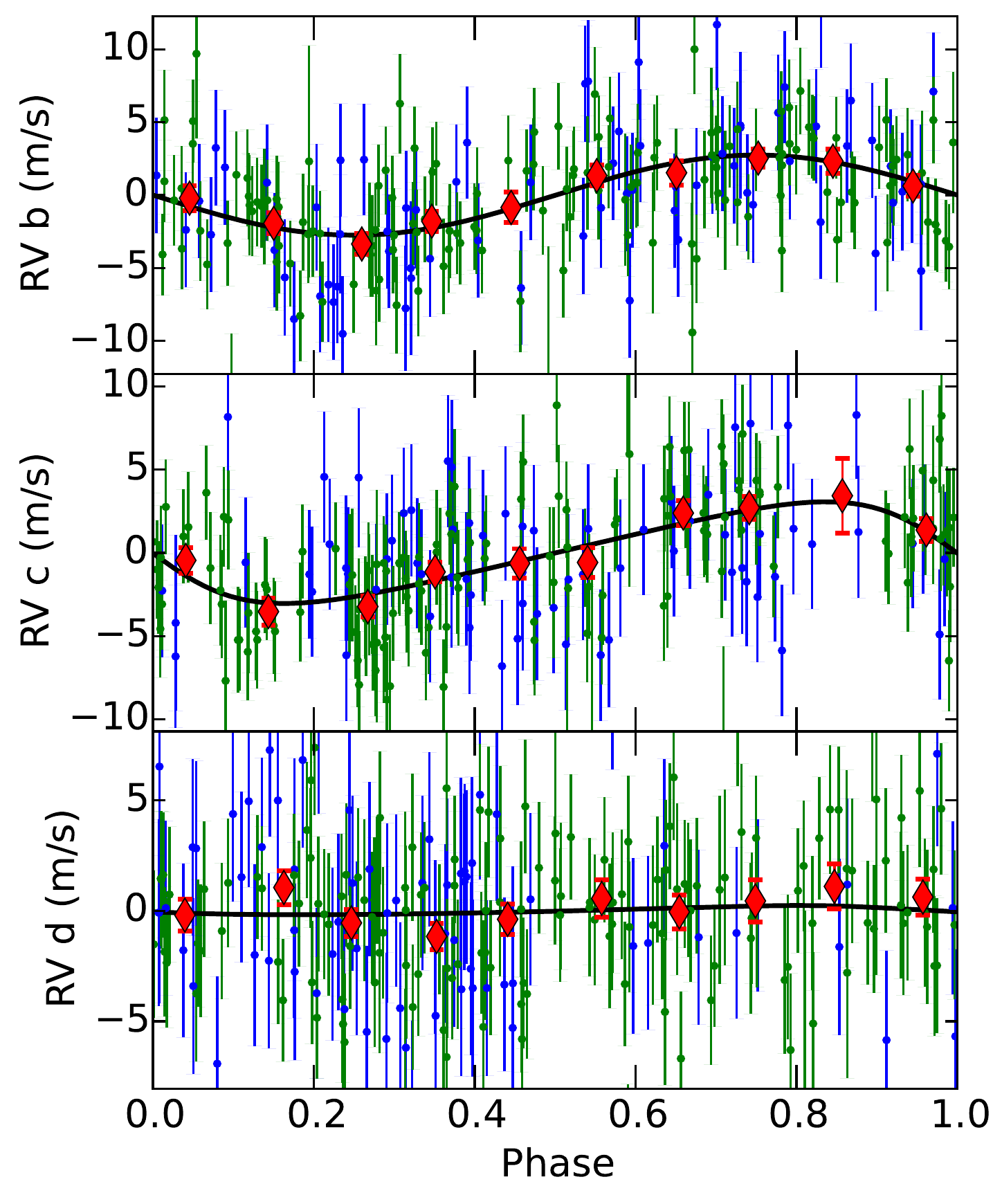}
\includegraphics[width=1.5in]{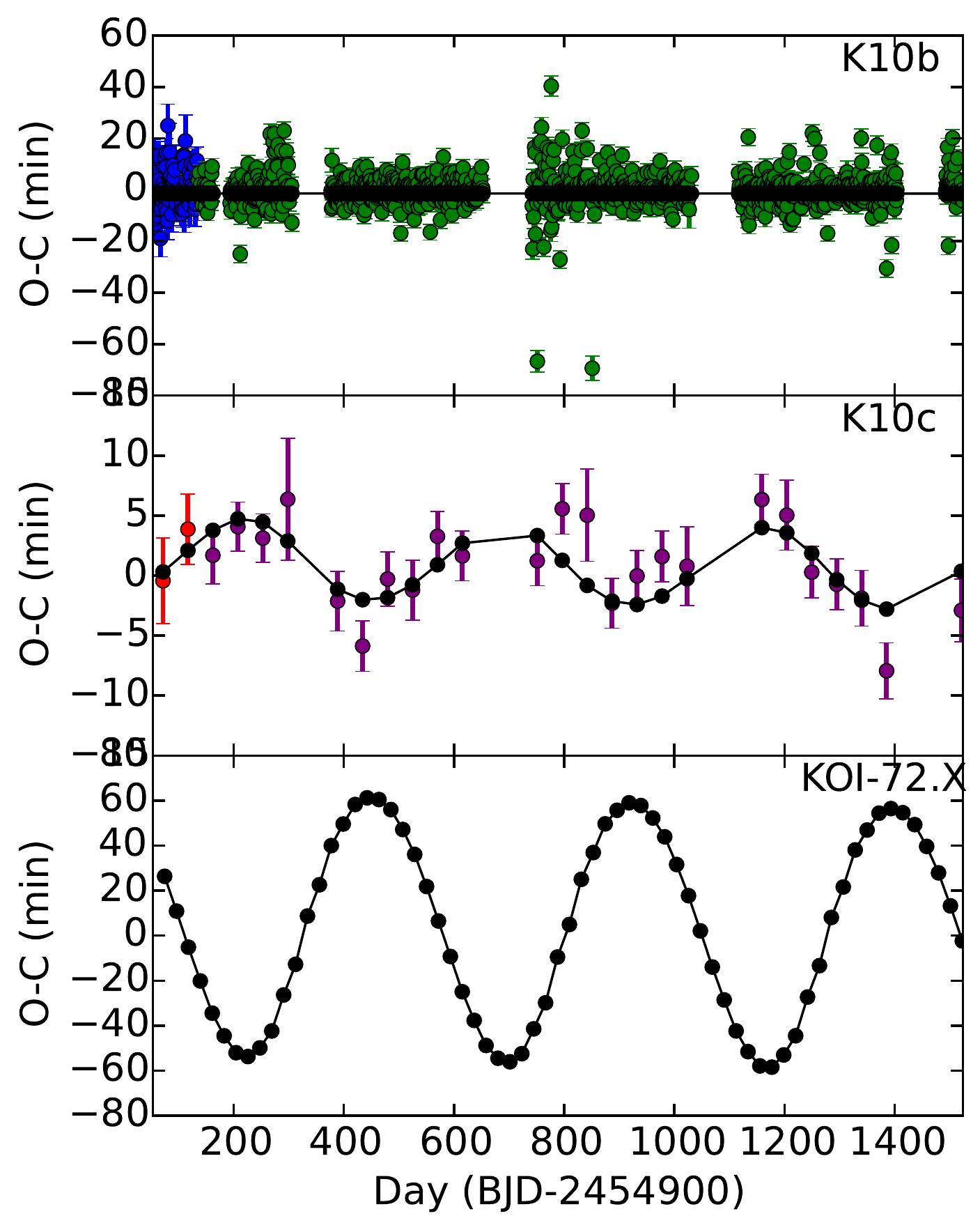}
\includegraphics[width=1.5in]{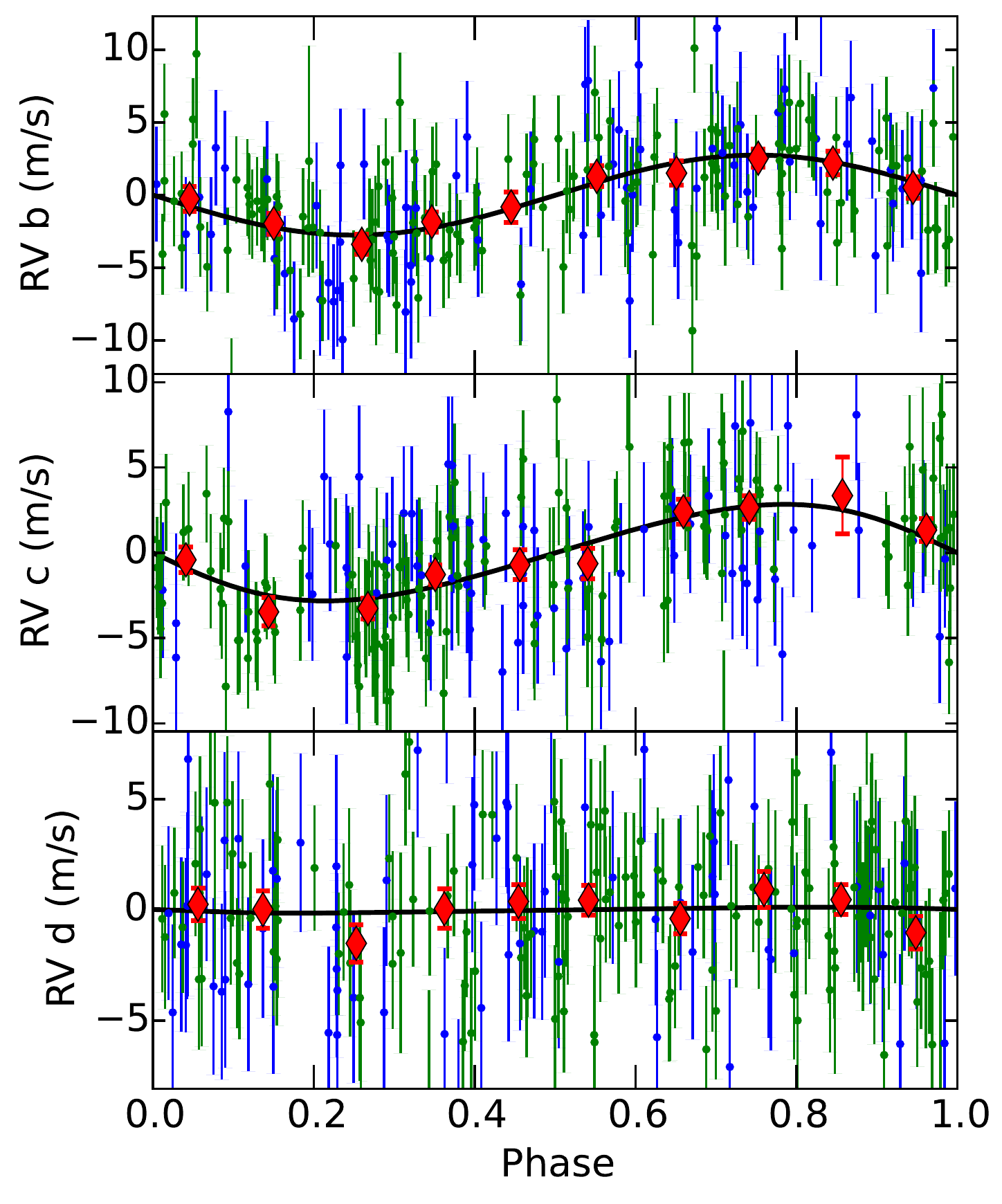}
\caption{Solutions for KOI-72.X near the 2:1 mean motion resonance.  Top left: O-C diagram for the solution with $P_X=$101 days, showing the transit times of planets b, c, and X in the three vertical sub-panels.  The colored points are the data; the black dotted line is the model.  Top, second from left: RVs of Kepler-10 decomposed into the orbits of planets b, c, and d (top to bottom), where $P_X=$101 days.  The blue points are from HIRES; the green are from HARPS-N, the red diamonds are the weighted mean RV in bins of 0.1 phase, and the black line is the model.  Top, second from right, and top right: the same as the top left two panels, but for P=82 days.  Bottom left and second from the left: The same as top left two panels, but for P=23 days.  Bottom second from right, and right: The same as top left two panels, but for P=21 days.}
\label{fig:ttvsoln_12}
\end{center}
\end{figure*}

\begin{figure*}[htbp]
\begin{center}
\includegraphics[width=1.5in]{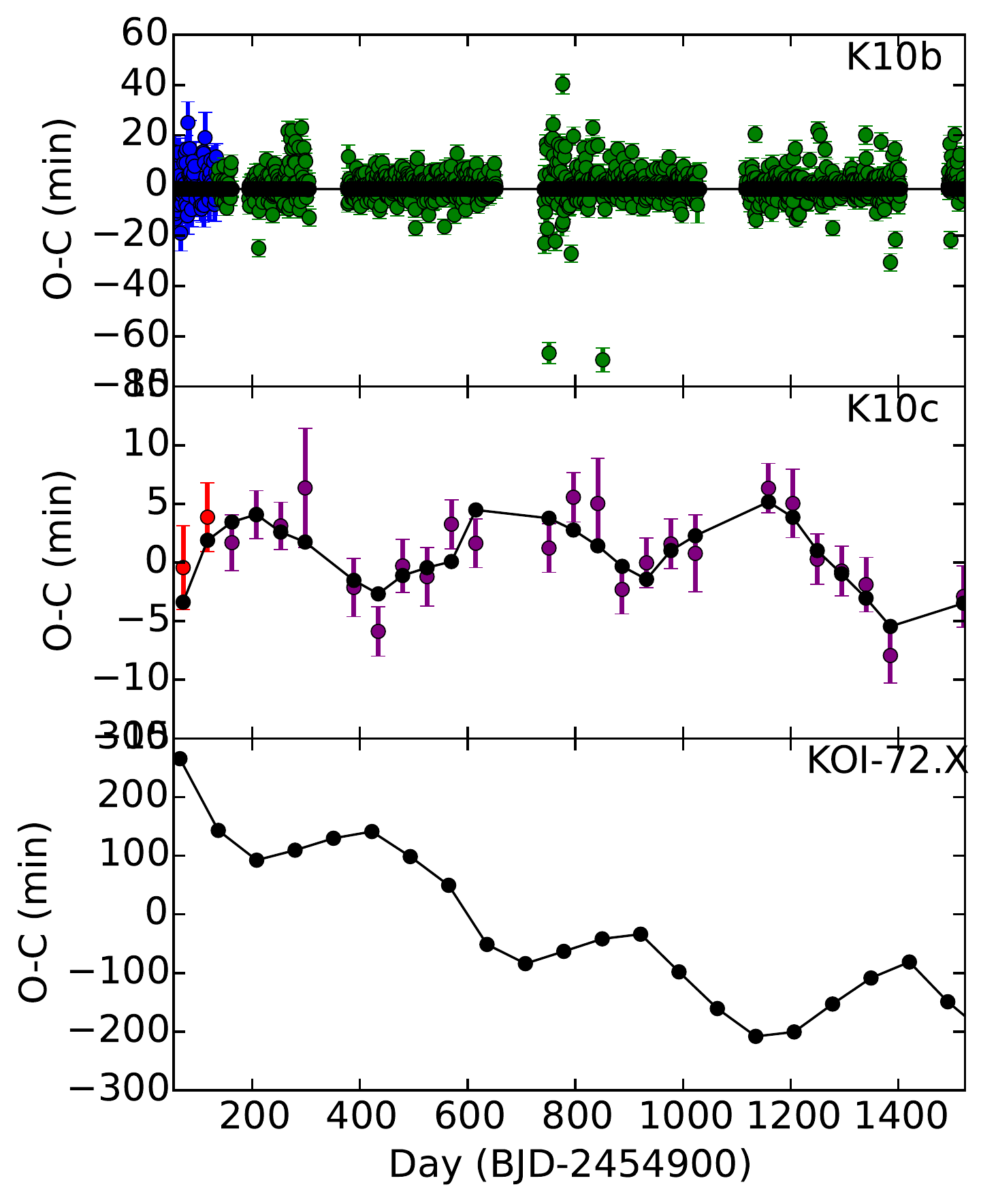}
\includegraphics[width=1.5in]{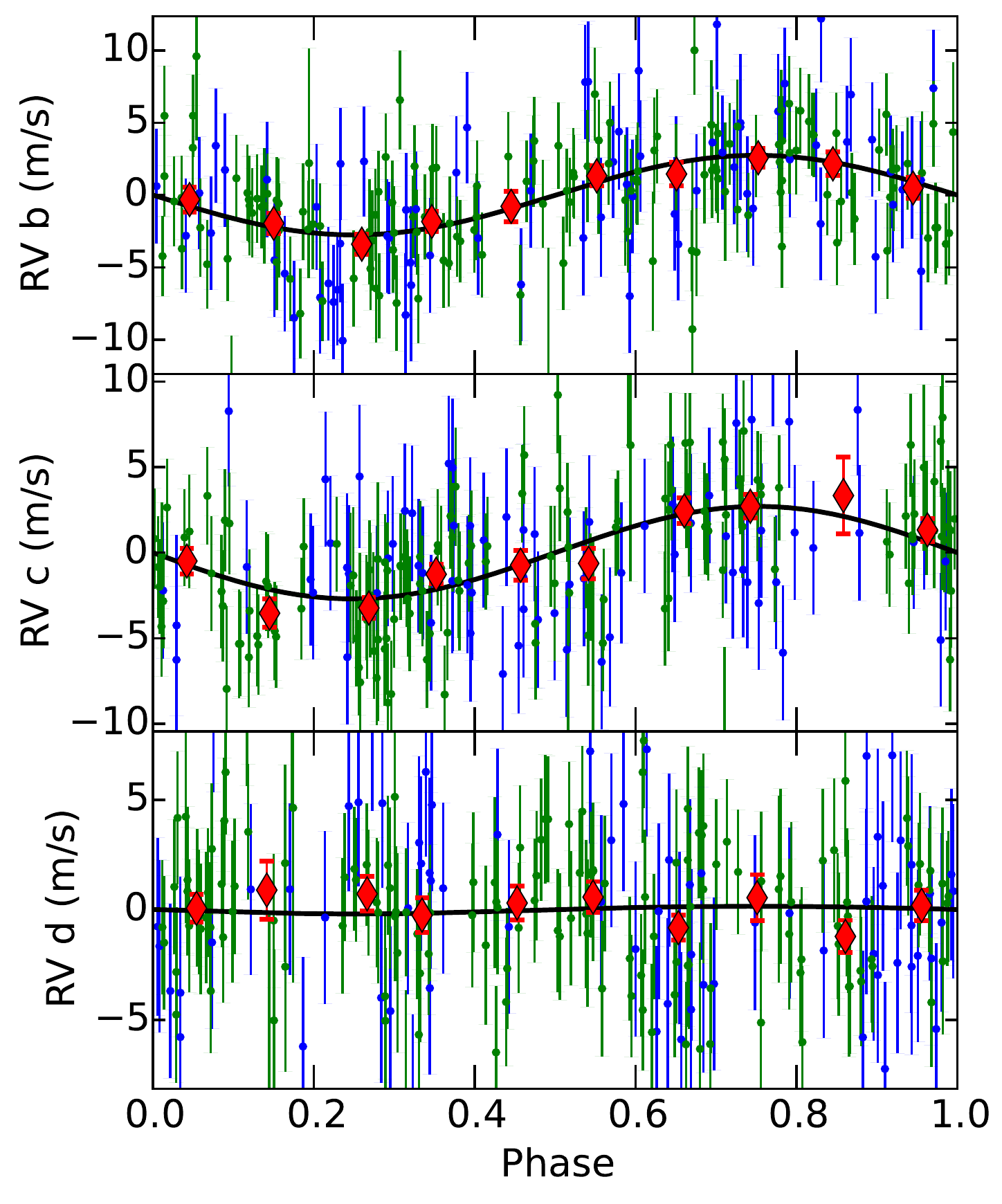}
\includegraphics[width=1.5in]{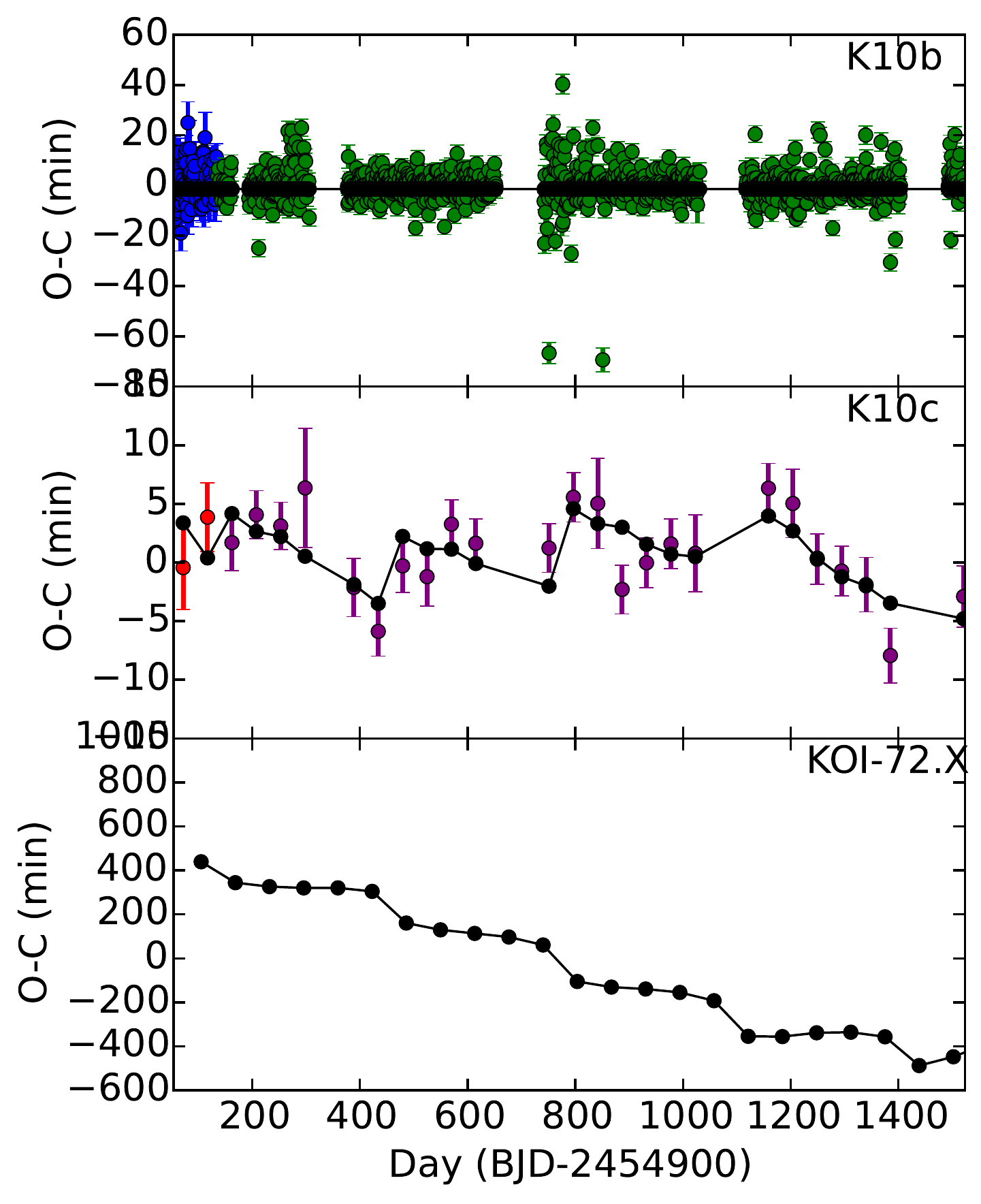}
\includegraphics[width=1.5in]{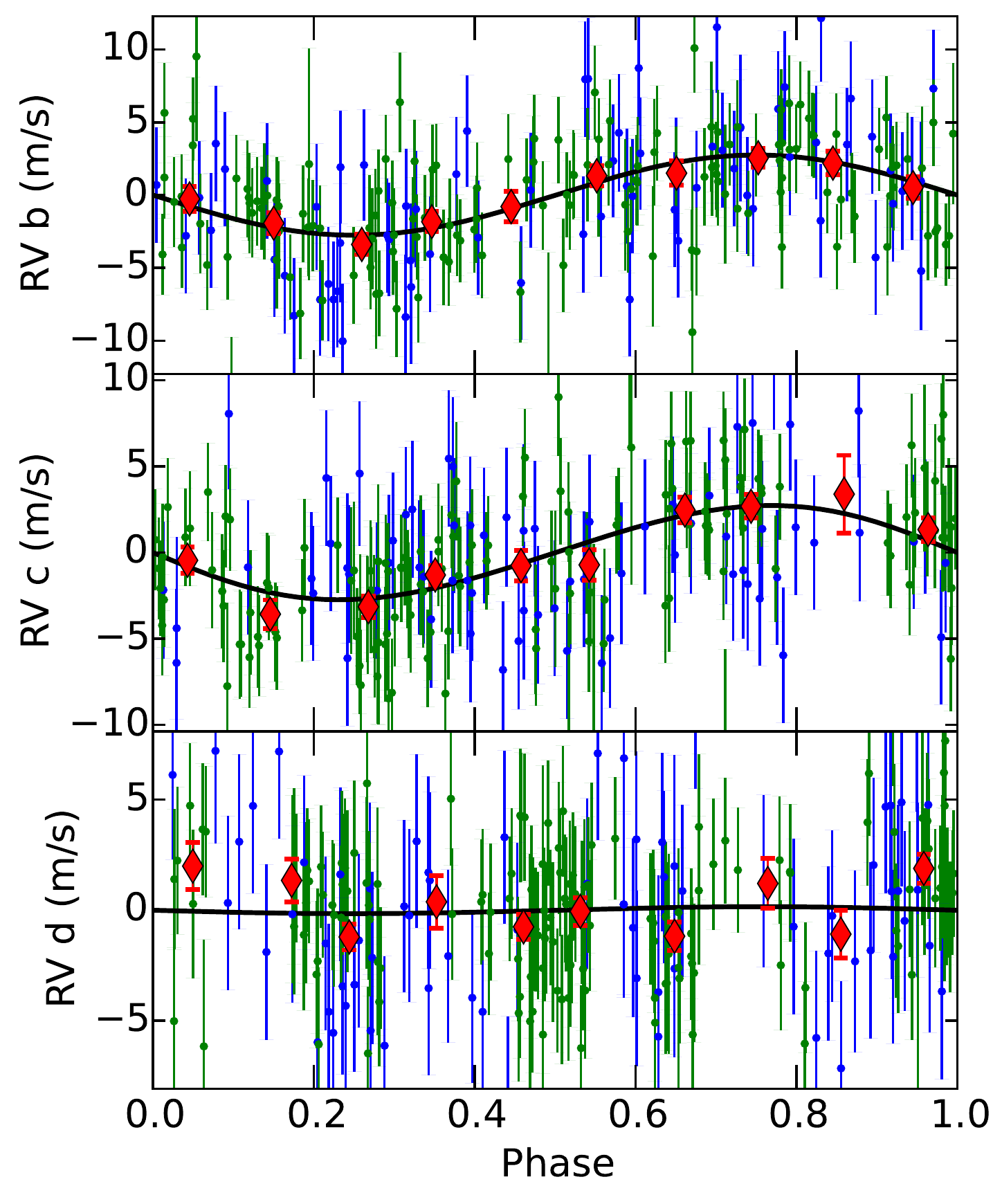}
\includegraphics[width=1.5in]{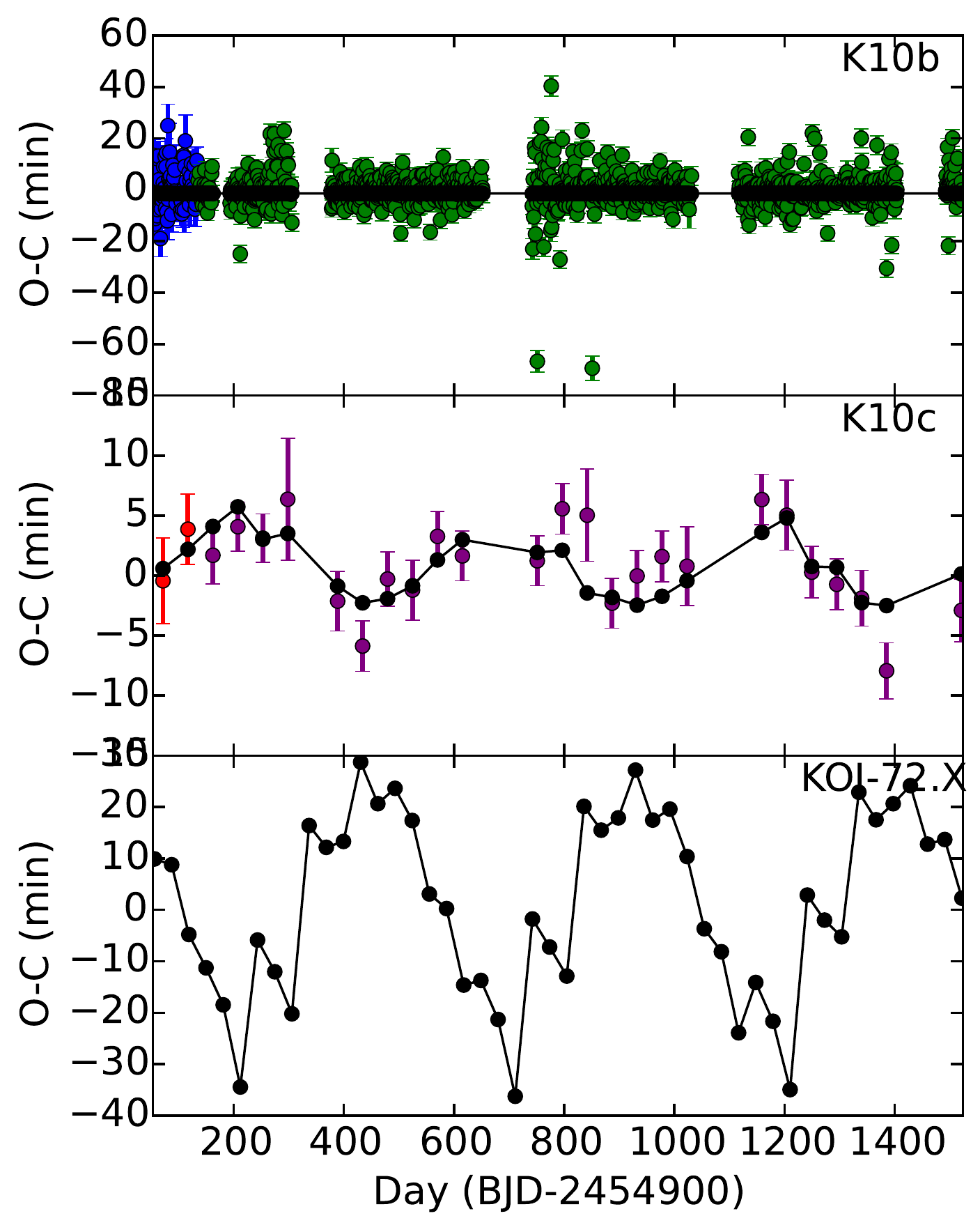}
\includegraphics[width=1.5in]{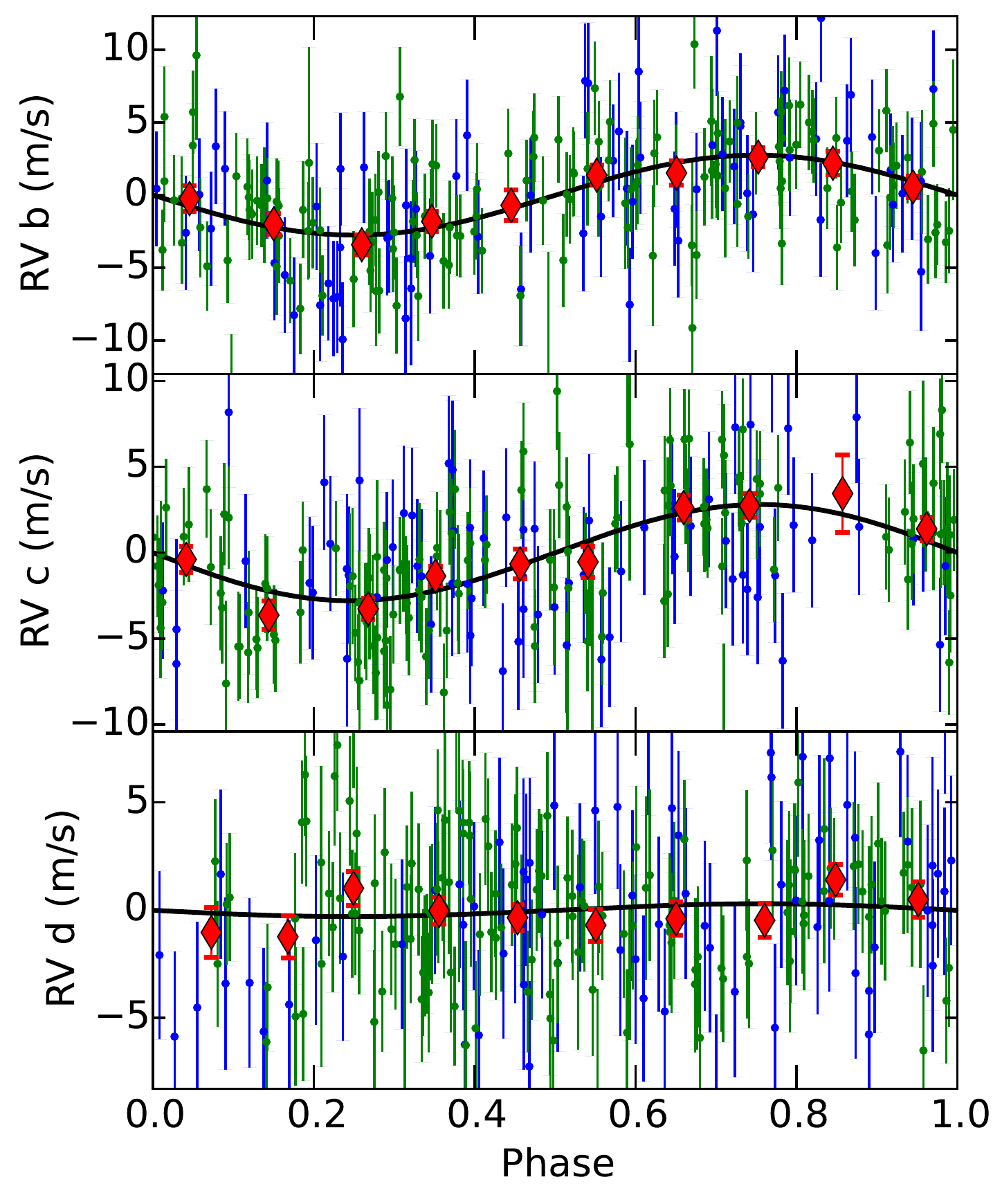}
\includegraphics[width=1.5in]{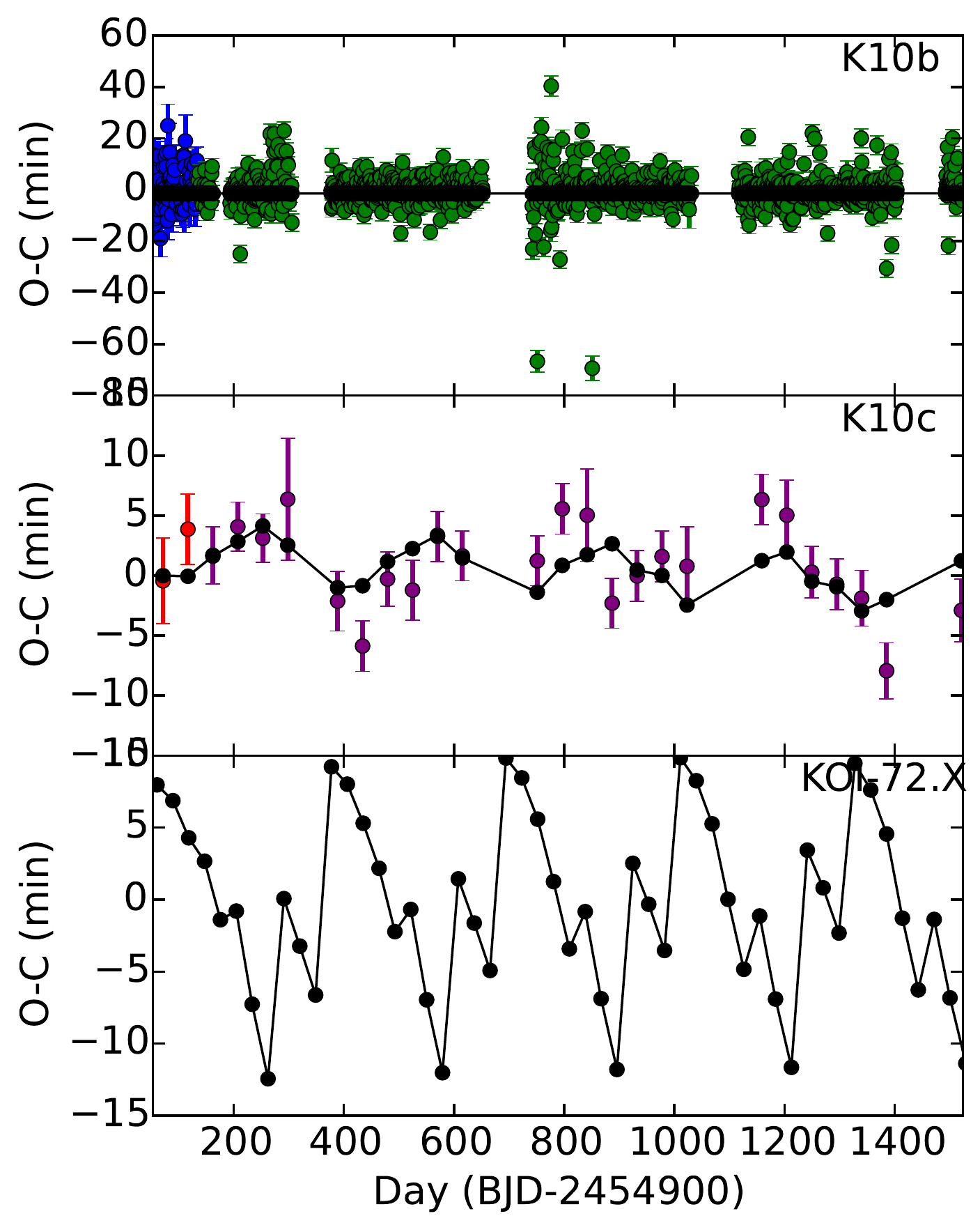}
\includegraphics[width=1.5in]{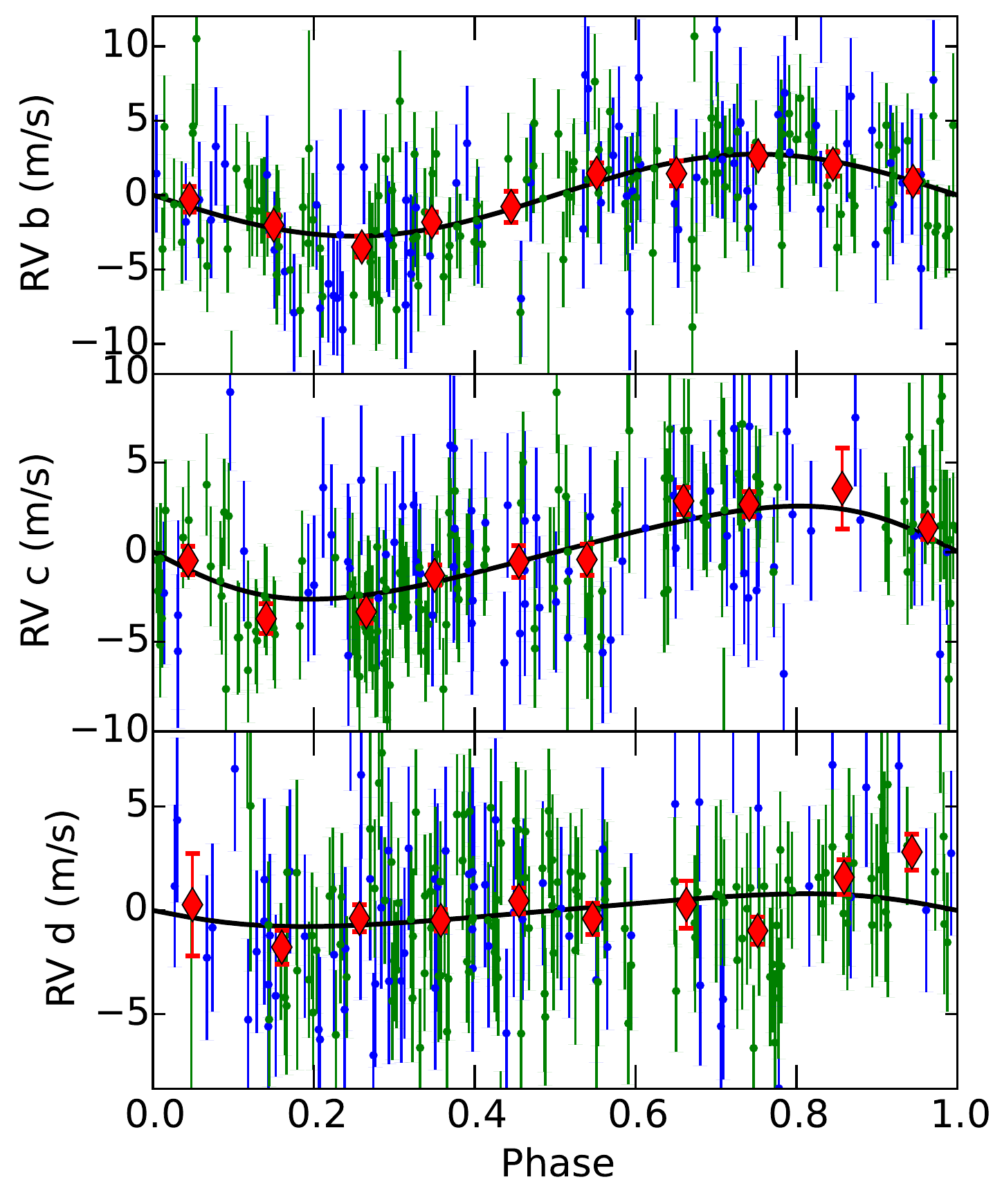}
\caption{Solutions for KOI-72.X near the 3:2 mean motion resonance.  The same as Figure \ref{fig:ttvsoln_12} but with with top left two panels: $P_X=$71 days; top right two panels: $P_X=$64 days; bottom left two panels: $P_X=$31 days; bottom right two panels: $P_X=$28 days.}
\label{fig:ttvsoln_23}
\end{center}
\end{figure*}

For each candidate orbital period for KOI-72.X listed in Table \ref{tab:pttv}, we perform a series of numerical N-body integrations using TTVFast \citep{Deck2014}.  We use the analytically predicted orbital parameters as inputs to the integrator at epoch $T0 (BJD - 2454900) = 53.67844$.  TTVFast predicts the times of transit of each planet in the N-body simulation, and the RVs of the star at the times of RV observation.  To determine the orbital parameters that best reproduce the observed transit times and RVs, we minimize the following statistic:

\begin{equation}
\chi^2 = \sum_k \frac{(\mathrm{TT}_{c,k} - \mathrm{TT}_{c,\mathrm{mod},k})^2}{\sigma_{\mathrm{TT},c,k}^2} + \sum_i \frac{(\mathrm{RV}_i - \mathrm{RV}_{\mathrm{mod},i})^2}{\sigma_{\mathrm{RV},i}^2}
\end{equation}

where the residuals between the observed and modeled transit times and RVs are simultaneously minimized.  We fit the TTVs of only planet c, since planet b does not have significant, coherent TTVs.  We use a combination of a Nelder-Mead and Levenberg-Marquardt (LM) minimization algorithm to find the orbital parameters and masses for planets c and d that produce a local minimum near the input orbital period.  We allow the period ($P$), eccentricity ($e$), argument of periastron passage ($\omega$), mean anomaly ($M$), and mass ($m$) of planets c and d to vary.  We fix the inclination $i=90^\circ$ and longitude of ascending node $\Omega = 0^\circ$ for all planets, for simplicity.  The orbital parameters and mass of planet b are fixed, since planet b is not interacting with planet c or d. The outcome of the LM minimization for each input orbital period of KOI-72.X in listed in the right half of each row of Table \ref{tab:pttv}.  Figure \ref{fig:ttvsoln_12} shows the numerical solutions in which KOI-72.X is near the 2:1 mean motion resonance.  For each solution, the predicted TTVs are overlaid on the observed TTVs in the left panel, and the RVs are phase-folded to the orbits of each of the planets in the right panel.  Figure \ref{fig:ttvsoln_23} is the same, but for solutions in which KOI-72.X is near the 3:2 mean motion resonance.

Considering that there are 246 data points (26 transit times plus 220 RVs), the reduced $\chi^2$ statistic is comparatively good for most of the orbital periods for KOI-72.X suggested by the TTV equation (top half of Table \ref{tab:pttv}); however, we achieve $\chi_\nu^2 \approx 1$ by construction based on how we treat the jitter in Equation \ref{eqn:rv}.  To convey how much some solutions are favored over others, we consider the change in the Bayesian Information Criterion ($\Delta$BIC) as a way to rank the possible solutions in order of preference.  The BIC is defined as:
\begin{equation}
BIC = \chi^2 + k~\mathrm{ln}(n)
\end{equation}
where $k$ is the number of free parameters in the model and $n$ is the number of observations \citep{Schwarz1978}.  The best solution for $P_X$ is at 101 days ($\chi^2=217$).  The $\Delta$BIC between the best and second-best model is 13, meaning that our best solution ($P_X$ = 101 days) is ``very strongly favored" over all the other solutions \citep{Kass1995}.  However, based on their low values of reduced $\chi^2\approx1$, the other solutions describe the data well enough that we cannot rule them out.  Furthermore, the solutions listed above do not fully explore the high-dimensional parameter space.  At each orbital period, multiple initial locations of KOI-72.X (as defined by $\omega_X$, $M_X$, and $e_X$) are possible, and produce comparatively good $\chi^2$ statistics to the values listed in Table \ref{tab:pttv}.  While the orbital solutions listed represent solutions that produce the lowest values of $\chi^2$ found in our Nelder-Mead algorithm, the parameters $\omega_X$, $M_X$, and $e_X$ are poorly constrained by the TTVs and RVs, even when a particular orbital period for KOI-72.X is chosen.  To fully explore the dynamical parameter space for the putative KOI-72.X would be very computationally intensive and is outside the scope of this paper.

Orbital periods for KOI-72.X that are not near solutions to the TTV equation produce significantly higher $\chi^2$ values than orbital periods that solve the TTV equation.  In the bottom half of Table \ref{tab:pttv}, the ``ctrl. 1" solution, we seed the orbital period for KOI-72.X with a value halfway between two solutions to the TTV equation ($P_X = 90.5$ days).  Even after using the Nelder-Mead minimizer to find the best orbital parameters for planets c and d with this initial guess, the lowest value of $\chi^2$ attained is 269, which is significantly worse (according to the BIC) than any of the analytically predicted solutions, with $\Delta$BIC = 52 (i.e. very strongly disfavored) with respect to the best three-planet solution.  Thus, we numerically demonstrate that the TTV equation accurately predicts orbital periods for KOI-72.X that best reproduce the observed TTVs.

In the last row of Table \ref{tab:pttv} (``ctrl. 2"), we perform an N-body integration in which we simulate only planet b (which still has fixed orbital properties and a fixed mass) and planet c (for which the orbital parameters and mass are allowed to vary).  The $\chi^2$ statistic for this control test is 267 ($\Delta$BIC = 22.5), demonstrating that the best three-planet solution is strongly favored over the two-planet solution.  In particular, the solutions for KOI-72.X at P=101, 24, and 71 days achieve significantly lower BIC values than the two-planet model, suggesting that these orbital periods are strongly favored over a two-planet model.

The transit duration variations (TDVs) of Kepler-10 c are negligible.  The various coplanar solutions for KOI-72.X that we examine all produce negligible TDVs as well, and so TDVs do not help break the degeneracy of the orbit for KOI-72.X in this particular system.  The absence of TDVs implies that KOI-72.X is not highly inclined with respect to planets b and c.

The various dynamical solutions for KOI-72.X yield $m_X$ between 1-7$\mearth$.  Without being able to choose the correct dynamical solution, we cannot pinpoint the mass of non-transiting planet candidate.  However, the masses that best fit the TTVs and RVs are consistent with the population of low-mass, small planets discovered by \Kepler\ and $\eta$-Earth surveys.

The family of solutions for KOI-72.X all yield similar masses for planet c (13.5-15~\mearth).  This is likely because (1) the mass of KOI-72.X is small, meaning that it contributes little to the RVs, (2) planet c cannot create its own TTVs, and so the presence of TTVs should not be correlated with planet mass.  For Kepler-10 c, modeling the TTVs with an N-body integrator does not result in an erroneously low planet mass.

\section{Interior Structure and Composition of Kepler-10 b and c} 

In this section, we estimate compositions of the planets Kepler-10 b and Kepler-10 c assuming the mass estimates from our formal fits using all available data.  However, given the issues discussed in Section \ref{sec:discrepancy}, actual uncertainties are larger than we assume for this analysis.  

Kepler-10 c falls near the high density extreme among the locus of sub-Neptune-sized planet mass-radius measurements characterized to date. As such, Kepler-10 c presents a valuable test case for planet formation theories and interior structure models. Comparing measured masses and radii to theoretical mass radius relations, which incorporate the behavior of materials at high pressure, reveals insights into the planets' possible bulk compositions. We employ planet interior structure models from \citet{Rogers&Seager2010ApJ, Rogers&Seager2010bApJ, RogersEt2011ApJ} to provide a mapping from planet mass and composition to radius, and apply the Bayesian approach described in \citet{Rogers&Seager2010ApJ} \citep[and applied in ][]{SchmittEt2014ApJ} to invert planet composition constraints from the radial velocity and transit observations. 

The first point to note about Kepler-10 c's bulk composition is that its transit radius is likely not defined by a ``rocky" or ``solid" surface. 
In the vast majority (98.4\%) of dynamical solutions to the combined {\it Kepler} photometry, HIRES spectroscopy, and HARPS-N spectroscopy dataset, Kepler-10 c has a sufficiently low mean density (less dense than a pure magnesium silicate sphere) that the planet must have a substantial complement of volatiles  (astrophysical ices and/or H/He) that contribute to the observed transit depth. Only 1.6\% of the the Kepler-10 c mass-radius posterior probability corresponds to scenarios in which the planet is sufficiently dense to have a stony-iron composition (comprised of an admixture iron and silicates alone). 
Since Kepler-10 c's measured density necessitates that the planet contains a volatile complement that contributes to the transit depth, the planetary composition is inconsistent with a ``solid" or ``rocky" makeup.  

The nature of Kepler-10 c's volatiles is ambiguous; the planet could have a H/He envelope contributing less than a fraction of a percent by mass to the planet, or a water envelope contributing a few tens of percent of the planet mass. The measured mass and radius of Kepler-10 c are consistent with a ``water-world" composition; less than 0.95\% of the Kepler-10 c mass-radius posterior PDF spills into a low density regime wherein an H/He envelope is required (corresponding to planet configurations that are less dense than even a 100\% pure water sphere with $T_{\rm eq}=550$~K). 
Assuming a 2-component planet interior structure model, where in the planet consists of an Earth-like stony-iron core (modeled as a 30:70 mix or iron and magnesium silicates) surrounded by a pure water envelope, the water mass fraction is constrained in this scenario to be $M_{\rm env,H_2O}/M_p=0.280^{+0.119}_{-0.102}$ (where the median, and range between the 16th and 84th percentiles are quoted) (Figure~\ref{fig:MR_K10c_Mpgmfh2o}).   
At the other extreme, the measured mass and radius of Kepler-10 c are also consistent with a dry (water-less) mixture of stony-iron material and H/He. Coupling the planet mass-radius posterior distribution to a 2-component planet interior structure model wherein the planet consists of a Earth-like stony-iron core surrounded by a 30 times enhanced metalicity solar composition envelope, the H/He envelope mass fraction is constrained to be $M_{\rm env, H/He}/M_p=0.0023^{+0.0017}_{-0.0012}$ (Figure~\ref{fig:MR_K10c_Mpgmf}). Whether the planet's volatile envelope is dominated by H/He or water, the radial extent of the envelope, $\Delta R_{\rm env}$, accounts for more than one tenth of the planet radius; in the end member scenarios described above, $\Delta R_{\rm env, H_2O}/R_p=0.285^{+0.077}_{-0.073}$ and $\Delta R_{\rm env, HHe}/R_p=0.162^{+0.038}_{-0.037}$ (lower panels of Figure~\ref{fig:MR_K10c_Mpgmfh2o} and \ref{fig:MR_K10c_Mpgmf}).

In contrast to Kepler-10 c, the measured mass and radius of the smaller inner planet, Kepler-10 b, is consistent with a rocky composition. Only 13.6\% of the posterior probability of Kepler-10 b spills into a low-density regime of mass-radius parameter space where volatiles are demanded. Under the assumption of a 2-layer rocky planet model consisting of an iron core and a silicate mantle (having molar magnesium fraction Mg$\#=90\%$), the iron core mass fraction of Kepler-10 b is constrained, $M_{\rm Fe, core}/M_p=0.169^{+0.115}_{-0.117}$ (Figure~\ref{fig:cmfK10b}). The measured mass and radius of Kepler-10 b is consistent with an Earth-like bulk composition (32.5\% iron core by mass), while an iron-enhanced Mercury composition (70\% iron core by mass) is strongly disfavored. 

All the composition constraints described above are based upon the combined analysis of the Keck HIRES and HARPS-N radial velocity measurements. The qualitative constraints on Kepler-10 b and c's bulk compositions are largely unchanged when only the Keck HIRES radial velocities are used to derive the planet mass.  
Omission of the HARPS-N dataset from the analysis leads to a downward shift in the density of Kepler-10 c, an upward shift in the density of Kepler-10 b, and an overall broadening of all the posterior PDFs. The downward shift in the density of Kepler-10 c further strengthens the conclusion that Kepler-10 c has a volatile envelope. Conditioned on the HIRES data alone, there is only a 0.4\% posterior probability that Kepler-10 c is more dense than a pure silicate sphere (compared to 4.1\% conditioned on both HARPS-N and HIRES). Further, the posterior probability that Kepler-10 c is sufficiently dense to have water envelope (instead of H/He) also decreases to 62\% (compared to 99\% conditioned on both HARPS-N and HIRES). On the other hand, the shift in Kepler-10 b posterior distribution toward higher masses and densities leads to increased posterior probability at higher Kepler-10 b iron core mass fractions  $M_{\rm Fe, core}/M_p=0.44^{+0.15}_{-0.19}$ (compared to $0.17\pm0.12$ based on both HARPS-N and HIRES).

\begin{figure}
\epsscale{1.0}
\plotone{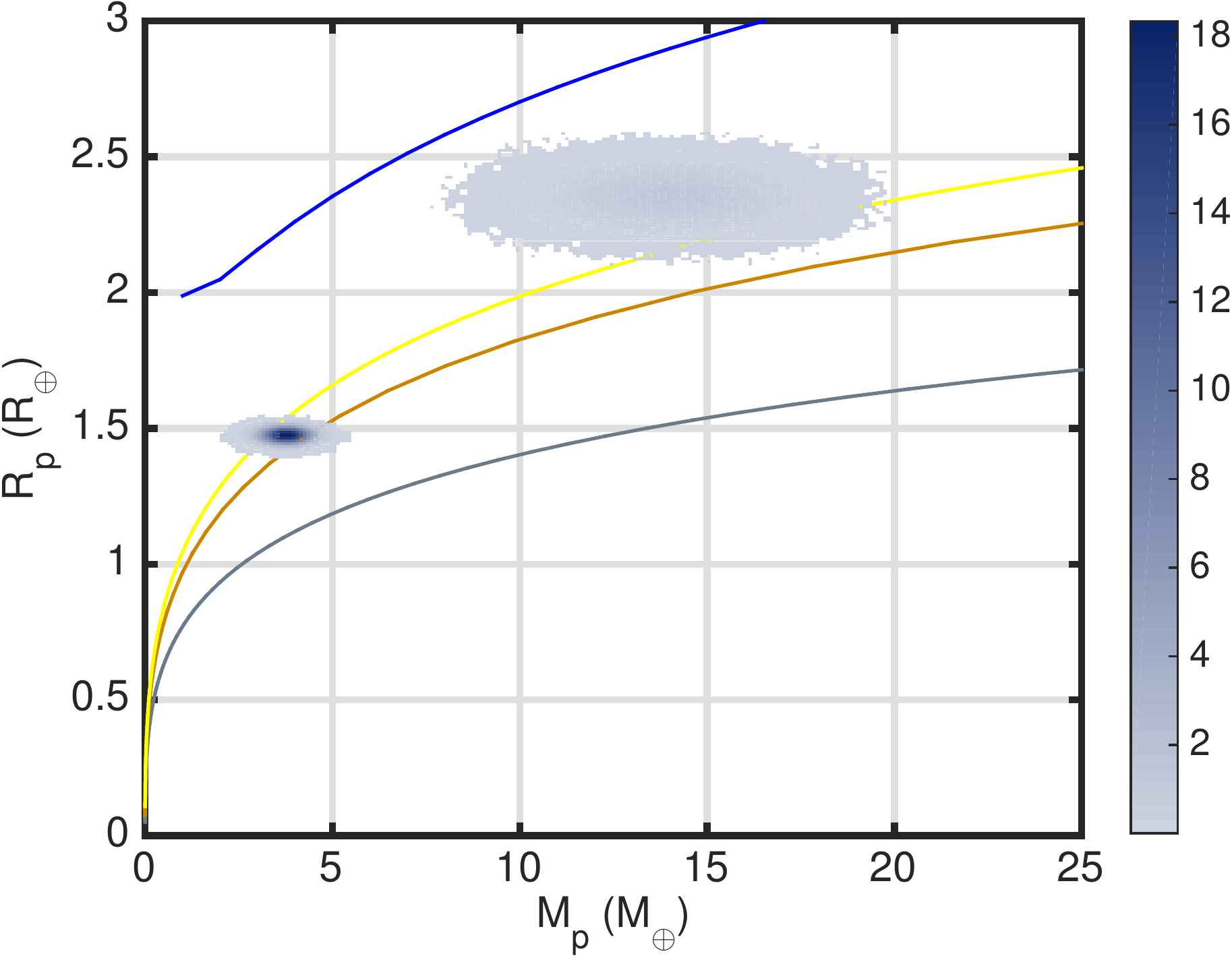}
\caption{Mass-radius posterior distribution for Kepler-10~b and Kepler-10~c, compared to theoretical mass-radius relations. The color bar indicates the mass-radius posterior probability density conditioned on the combined HIRES and HARPS dataset. The solid curves represent theoretical mass-radius relations for notable compositions: pure Fe (grey), Earth-like stony-iron (brown), pure silicate (yellow), pure water (blue). The water planet mass-radius relation includes thermal effects and the presence of a super-critical steam envelope.}
\label{fig:MR_K10bc}
\end{figure}

\begin{figure}
\epsscale{1.0}
\plotone{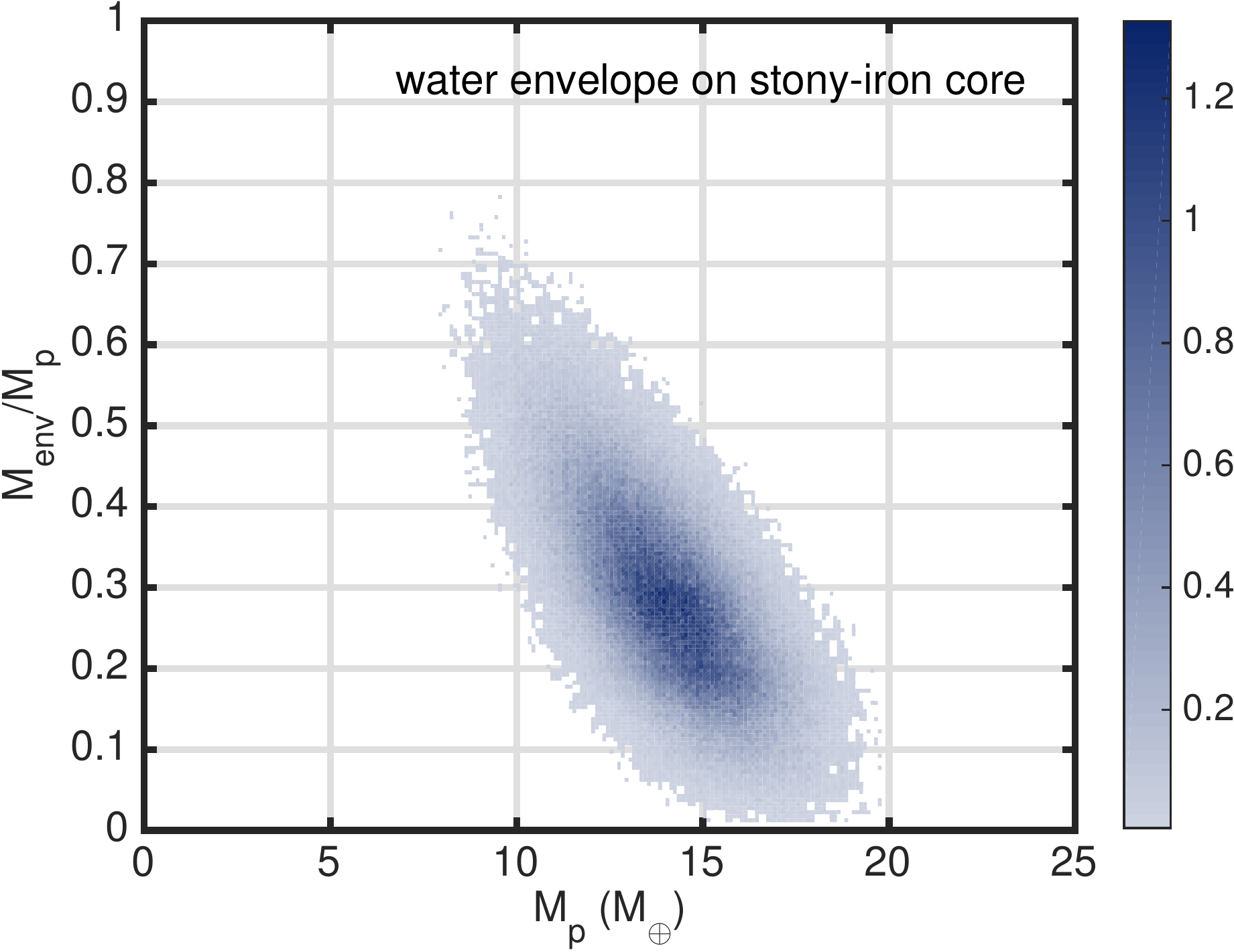}
\plotone{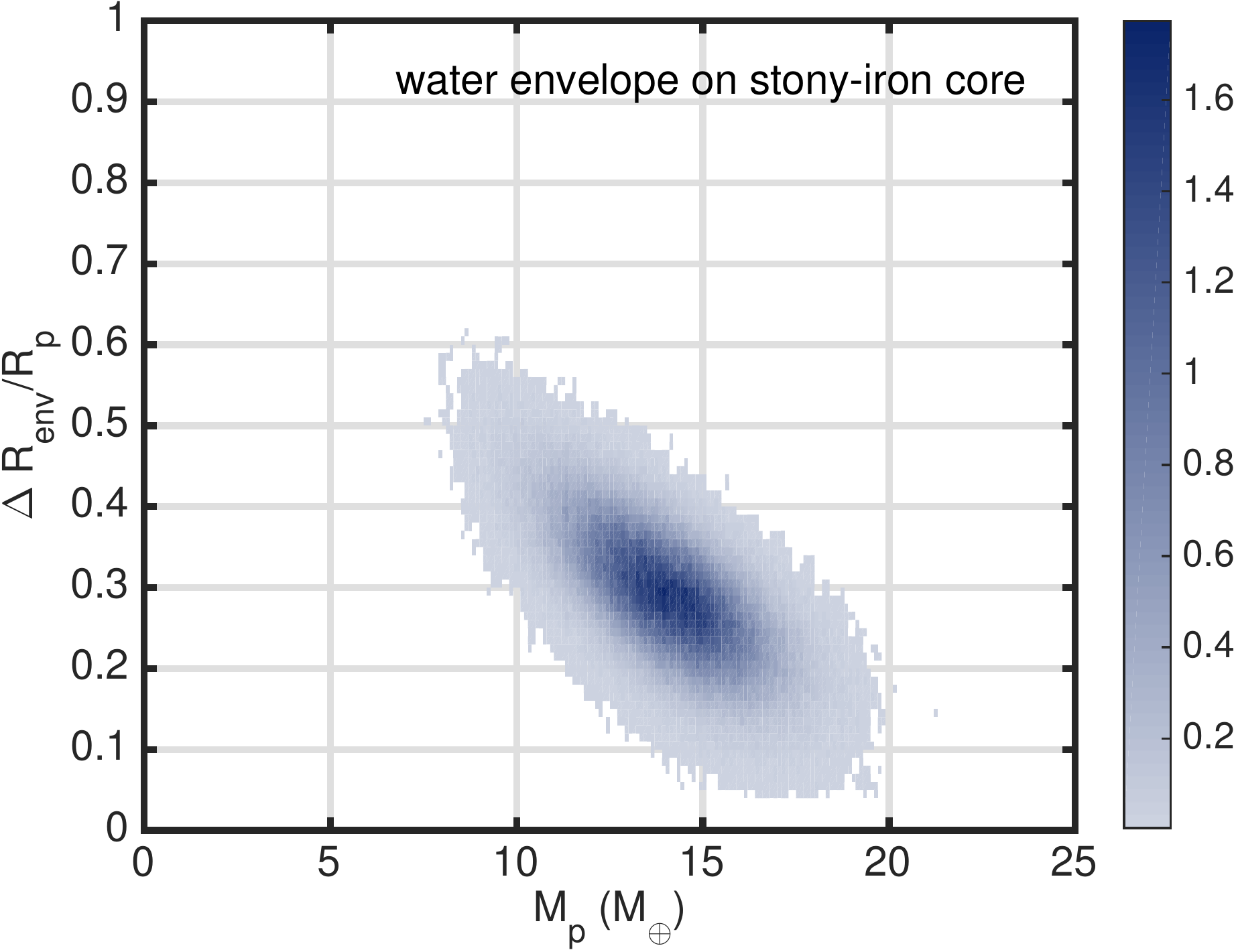}
\caption{Posterior probability density distribution for Kepler-10 c, as a function of planet mass, $M_p$ and the water envelope mass fraction $M_{\rm env, H_2O}/M_p$ (upper panel), and radius fraction $R_{\rm env, H_2O}/R_p$ (lower panel). Darker shades of blue represent higher probability. The posterior pdf obtained from the combined analysis of the Keck HIRES and HARPS-N radial velocity measurements is compared to planet interior structure models in which the planet is assumed to consist of an Earth-like stony-iron core (modeled as a 30:70 mix or iron and magnesium silicates) surrounded by a pure water envelope.}
\label{fig:MR_K10c_Mpgmfh2o}
\end{figure}

\begin{figure}
\epsscale{1.0}
\plotone{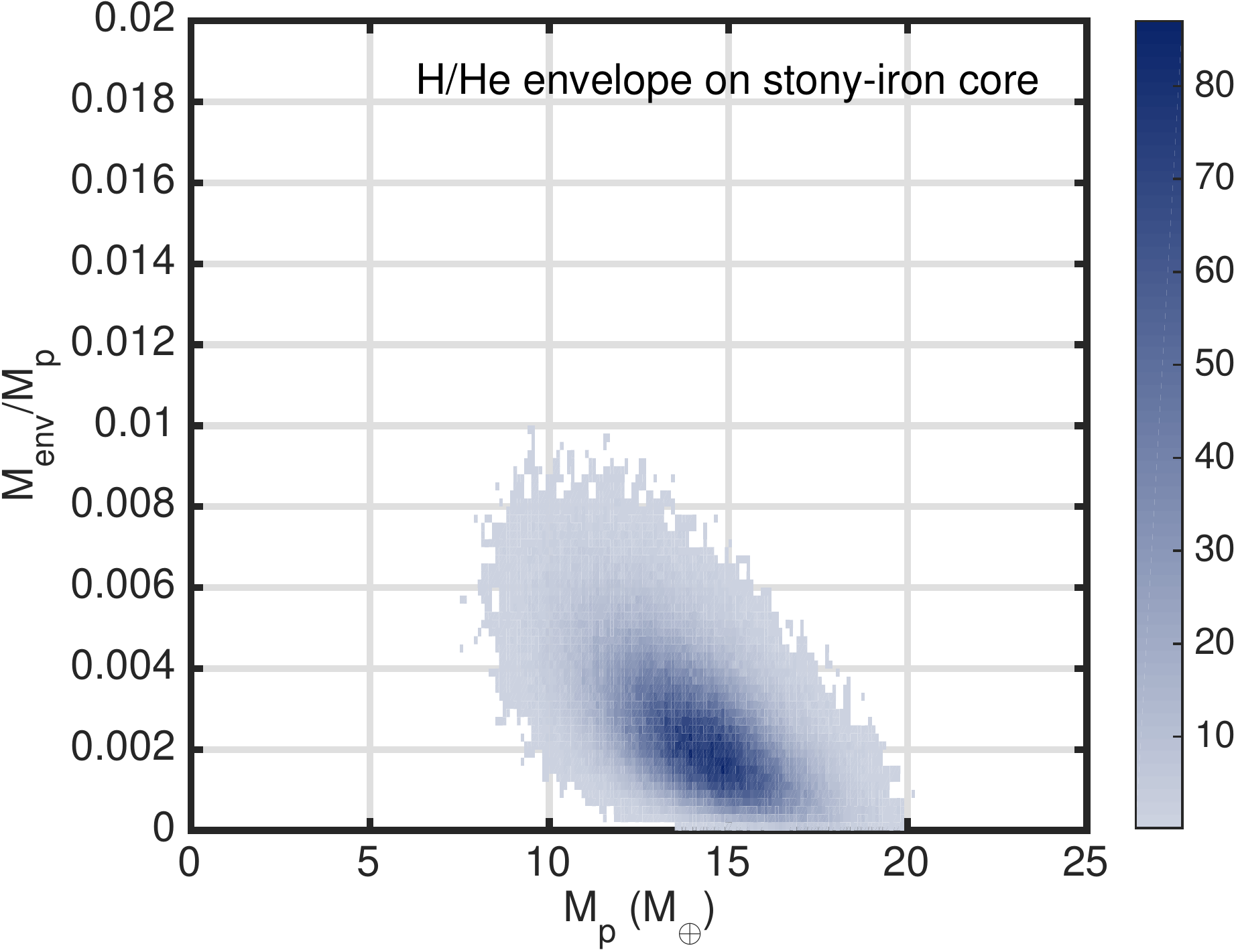}
\plotone{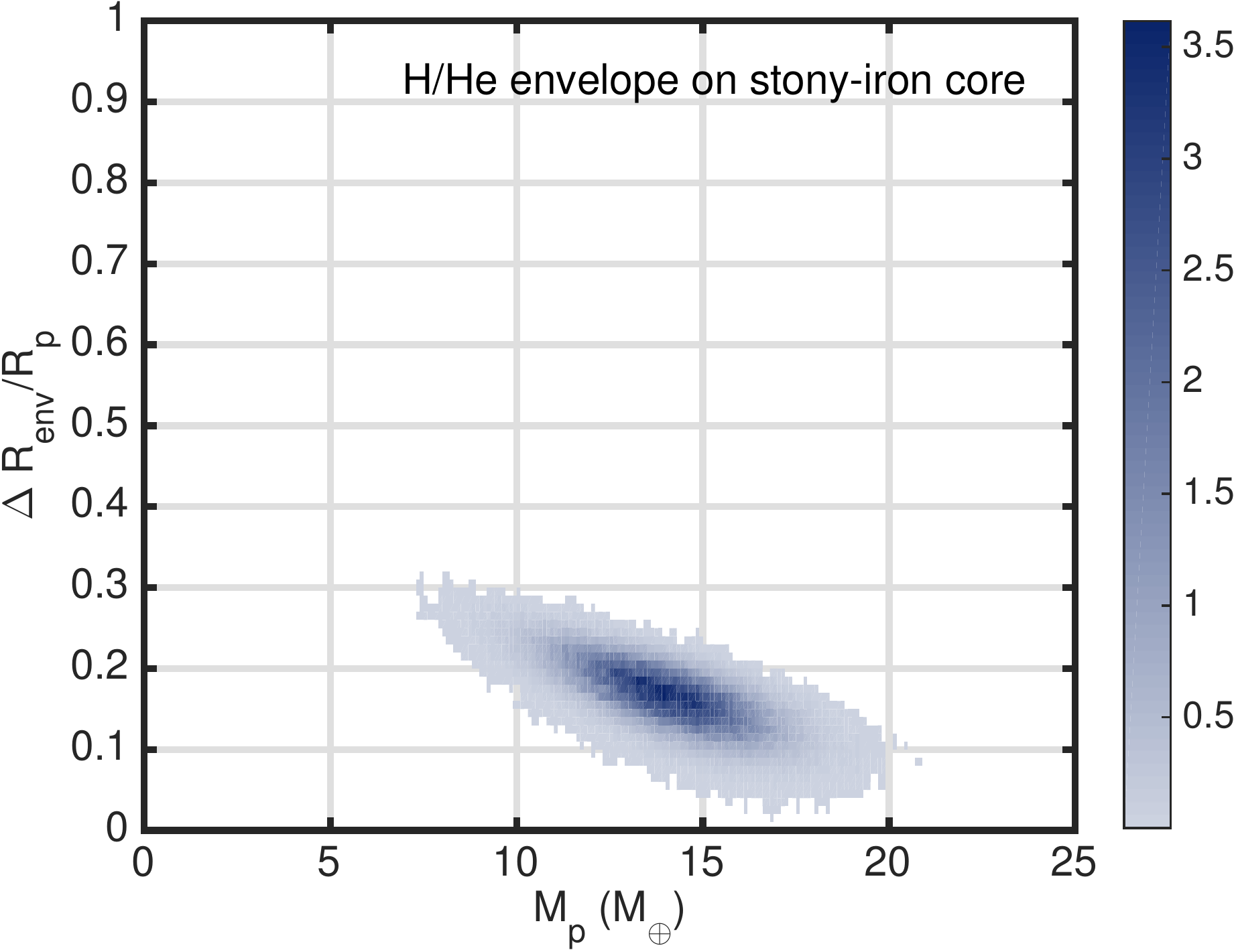}
\caption{Posterior probability density distribution for Kepler-10 c, as a function of planet mass, $M_p$ and the H/He envelope mass fraction $M_{\rm env, HHe}/M_p$ (upper panel), and radius fraction $R_{\rm env, HHe}/R_p$ (lower panel). This figure is analogous to Figure~\ref{fig:MR_K10c_Mpgmfh2o} except a H/He-dominated 30 times enhanced solar metalicity composition is assumed for the planet envelope instead of a pure water composition.}
\label{fig:MR_K10c_Mpgmf}
\end{figure}

\begin{figure}
\epsscale{1.0}
\plotone{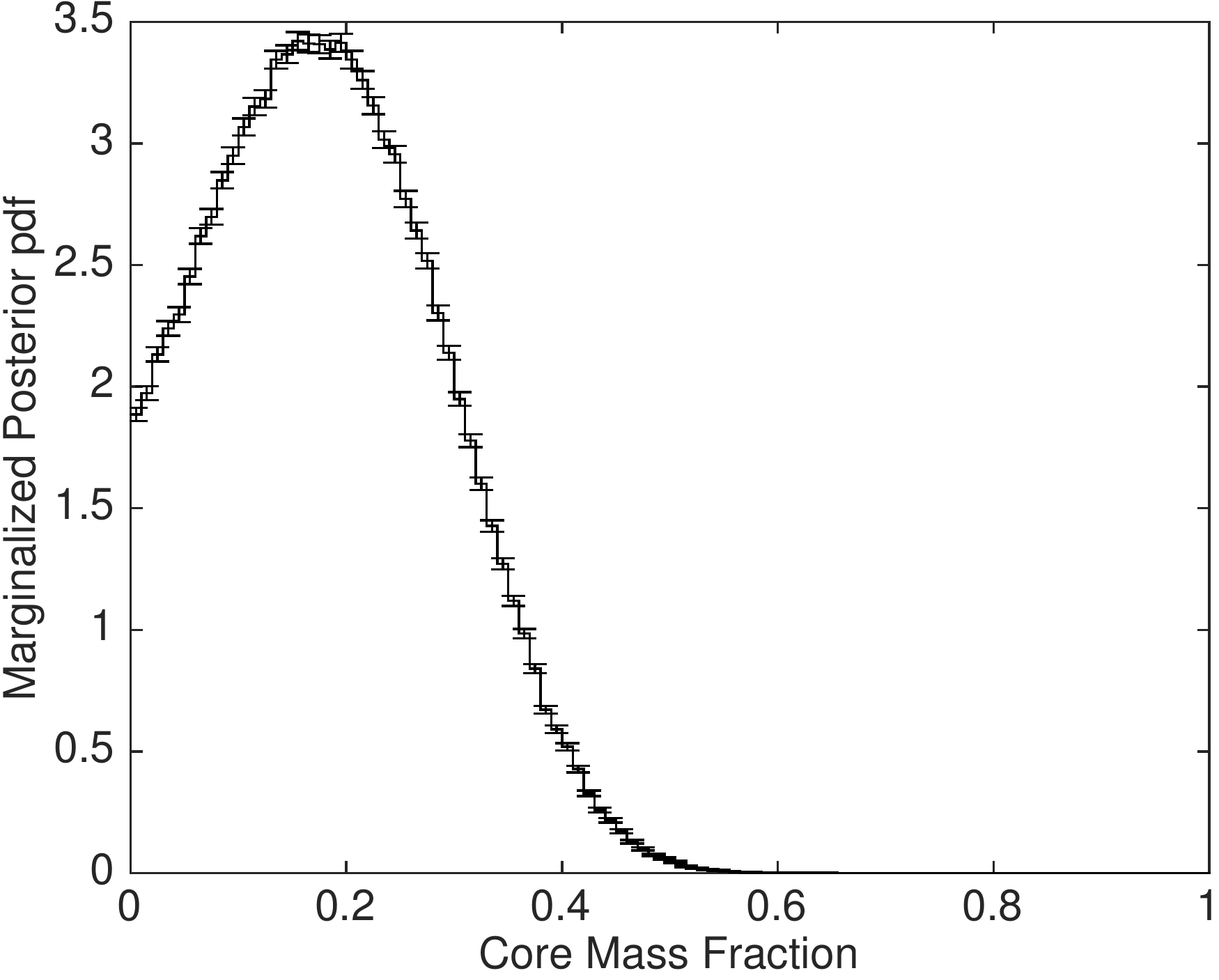}
\caption{Posterior probability distribution on the iron core mass fraction of Kepler-10 b. A 2-layer fully differentiated interior structure model wherein the planet consists of an iron core surrounded by a silicate mantle (Mg$\#=90\%$) is assumed.} 
\label{fig:cmfK10b}
\end{figure}

\section{Discussion}
\subsection{Insights on successful observing strategies for Kepler follow-up}
The source of the discrepancy between the HIRES and HARPS-N RVs is unknown.  However, the apparent time-variability of $K_b$ and $K_c$ on both instruments suggests that a time-correlated noise (or astrophysical signal) confounds both data sets on timescales of a summer observing season.  This time-correlated property could be from stellar rotation and the stellar magnetic cycle or from an additional planet in the system.  The most effective way to combat this time-correlated noise is to increase the quantity and baseline of observations of this star.  This strategy of employing a long observing baseline compared to time-correlated noise influences should be employed on other stars with low-mass planets to avoid confusion in the future.
\subsection{The masses and densities of the Kepler-10 planets}
This paper is the third study of the masses and densities of the Kepler-10 planets.  B11 and D14 previously measured the planetary and stellar properties of the Kepler-10 system.  The stellar and planetary parameters they found are shown in Table \ref{tab:comparison}.  The planetary properties obtained in this paper are also shown.  We include our solutions to the two-planet circular fit, the two-planet fit in which planet c is allowed eccentricity, and the best three-planet fit.

Table \ref{tab:comparison} reveals how the estimates of the masses of Kepler-10 b and c have changed based on which RVs were included in the analysis.  B11 used 52 RVs from Keck HIRES that were timed at the quadratures of planet b, and spanned 2 seasons (2009-2010).  They assumed that both planets had circular orbits.  They obtained masses of $m_b = 4.56\pm1.23 ~\mearth$, and an upper limit for planet c, $m_c < 20 ~\mearth$.  D14 used 148 RVs from TNG HARPS-N that were timed at the quadratures of planet c, and spanned 2 seasons (2012--2013).  They assumed that both planets had circular orbits.  They obtained $m_b = 3.3\pm0.49~\mearth$, a measurement $1\sigma$ lower than the B11 mass for Kepler-10 b, and a mass measurement for planet c of $M_c = 17.2\pm1.9~\mearth$.  Analyzing the 52 literature RVs plus 20 new RVs from HIRES, we obtain $m_b = 4.61\pm0.83\mearth$ and $m_c = 5.69^{+3.19}_{-2.90}\mearth$, which are $1.1\sigma$ and $3.1\sigma$ away from the values in D14, respectively.  We combine the 72 RVs from HIRES with 148 literature RVs from HARPS-N to obtain the best orbital coverage of both planets and the longest baseline.  In our two-planet circular fit (the most direct orbital comparison to the previous studies), we obtain $m_b = \mbcirc$ and $m_c = \mccirc$.  The mass we obtain for Kepler-10 b sits between the two previously published values.  Likewise, the mass we obtain for Kepler-10 c represents a compromise between the $2\sigma$ detection in the HIRES data and the massive planet obtained in the HARPS-N data.

When we allow the orbit of Kepler-10 c to be eccentric, or when we add a third planet to the system, the mass of Kepler-10 c does not change significantly.  The mass for Kepler-10 c obtained in the eccentric fit ($m_c = \mcecc$) is consistent with our circular fit.  Likewise, the masses for planet c obtained for various orbital solutions of candidate KOI-72.X all resulted in $m_c = 13$-$14.5 \mearth$, i.e., all within $1\sigma$ of the mass we obtained in the two-planet circular fit.  Therefore, allowing the orbit of planet c to be eccentric, or including a third planet at a variety of orbital periods, does not significantly affect the mass computed for Kepler-10 c.

\begin{table*}[htbp]
\caption{Kepler-10 Parameters in different studies}
\label{tab:comparison}
\tablenum{7}
\begin{center}
\begin{tabular}{rlllllll}
\hline
\hline
Parameter & units & B11$^a$ & D14$^b$ & \multicolumn{4}{l}{This Work} \\
& & & & HIRES only & 2 pl. circ. & 2 pl. ecc. & Best 3 pl.\\
\hline
\multicolumn{7}{l}{\it{Stellar}} \\
\hline 
\teff & K& $5627\pm44$& $5708\pm28$& \multicolumn{4}{l}{same as D14}\\
\logg & cgs & $4.341\pm0.012$ & $4.344\pm0.004$ &.&. &. \\
\mstar & \msun & $0.895\pm0.060$ & $0.910\pm0.0214$ &. & .&.\\
\rstar & \rsun & $1.056\pm0.021$ &$1.065\pm0.009$ &.& .&. \\
\lstar & \lsun & $1.004\pm0.059$ & $1.004\pm0.059$&.& .&. \\
\feh & dex & $-0.15\pm0.04$ & $-0.15\pm0.04$&.&. &. \\
\vsini & \ms & $0.5\pm0.5$ & $0.6\pm0.5-2.04\pm0.34$&.&. & .\\
age & Gyr & $11.9\pm4.5$ & $10.6\pm1.4$&.&. & .\\
\multicolumn{7}{l}{\it{Kepler-10 b}} \\
\hline
Period & days & 0.837495$\pm$4E-6& 0.8374907$\pm$2E-7 &\multicolumn{4}{l}{fixed, same as D14}\\
TT & BJD - 2,454,900 & $64.57375\pm0.0007$& 134.08687$\pm$0.00018 &\multicolumn{4}{l}{fixed, same as D14} \\
K & \ms & $3.3\pm0.9$& 2.38$\pm$0.35& 3.31$\pm$0.59 & 2.67$\pm$0.3&2.70$\pm$0.31 & 2.67 (fixed) \\
a & AU & 0.01684$\pm$0.00033& 0.01685$\pm$0.00013&\multicolumn{4}{l}{fixed, same as D14} \\
\mpl & \mearth & $4.56\pm1.23$ & 3.33$\pm$0.49& 4.61$\pm$0.83 & 3.72$\pm$0.42 &\mbecc & 3.70 (fixed)\\
\rpl & \rearth & $1.416\pm0.034$ & 1.47$\pm$0.03& \multicolumn{4}{l}{fixed, same as D14} \\
\rhopl & \gcc & $8.8\pm2.5$& 5.8$\pm$0.8&8.0$\pm$1.43& 6.46$\pm$ 0.72 & 6.53$\pm$0.75 &6.46 (fixed) \\
\multicolumn{7}{l}{\it{Kepler-10 c}} \\
\hline
Period & days & 45.29485$\pm$0.0007& 45.294301$\pm$4.8E-5&\multicolumn{3}{l}{fixed, same as D14}& 45.295\\
TT & BJD - 2,454,900 & 71.6761$\pm$0.0022 & 162.26648$\pm$0.00081 & \multicolumn{3}{l}{fixed, same as D14}&71.67\\
K & \ms & x & 3.26$\pm$0.36 & 1.09$\pm$0.58 &2.67$\pm$0.34 & 2.83 $\pm$0.38&2.66\\
a & AU &$ 0.2407\pm0.0048$  & 0.2410$\pm$0.0019& \multicolumn{3}{l}{fixed, same as D14}&0.24\\
e & & 0 (fixed) & 0 (fixed)& 0 (fixed) &0 (fixed) &0.17$\pm$0.13 & 0.09\\
$\omega_c$ &$^\circ$ & . & . &. &. &71$\pm$20 & 79.7 \\
\mpl & \mearth & $<20$& 17.2$\pm$1.9& 5.69$^{+3.19}_{-2.90}$ & 13.98$\pm$1.79 & 14.59$\pm$1.90 & 13.94\\ 
\rpl & \rearth & 2.277$\pm$0.054 & 2.35$\pm$0.06& \multicolumn{4}{l}{fixed, same as D14} \\ 
\rhopl & \gcc & $<10$ & 7.1$\pm$1.0 & 2.42$^{+1.36}_{-1.24}$& 5.94$\pm$0.75 & 6.21$\pm$0.81 &6.20\\ 
\multicolumn{8}{l}{\it{KOI-72.X}} \\
\hline
Period & days & .&. &. &.&. &101.360$^c$\\
e &  &.&.& .&. &. &0.19$^c$\\
\mpl & \mearth &  .&. &.&. &. &6.84$^c$\\
\end{tabular}
\end{center}
\tablecomments{$^a$ B11--\citet{Batalha2011}.}
\tablecomments{$^b$ D14--\citet{Dumusque2014}.}
\tablecomments{$^c$ The variety of orbital solutions with similar goodness of fit for planet candidate KOI-72.X prevents an accurate characterization of the planet's true orbital parameters and mass.  The numbers shown here reflect the best three-planet solution, which is strongly preferred over the two-planet solution and over other three-planet solutions based on the BIC.}  

\end{table*}%

\subsection{Updated mass-radius and density-radius relations}
\begin{figure*}[htbp]
\begin{center}
\includegraphics[width=6in,trim=25mm 5mm 25mm 15mm,clip]{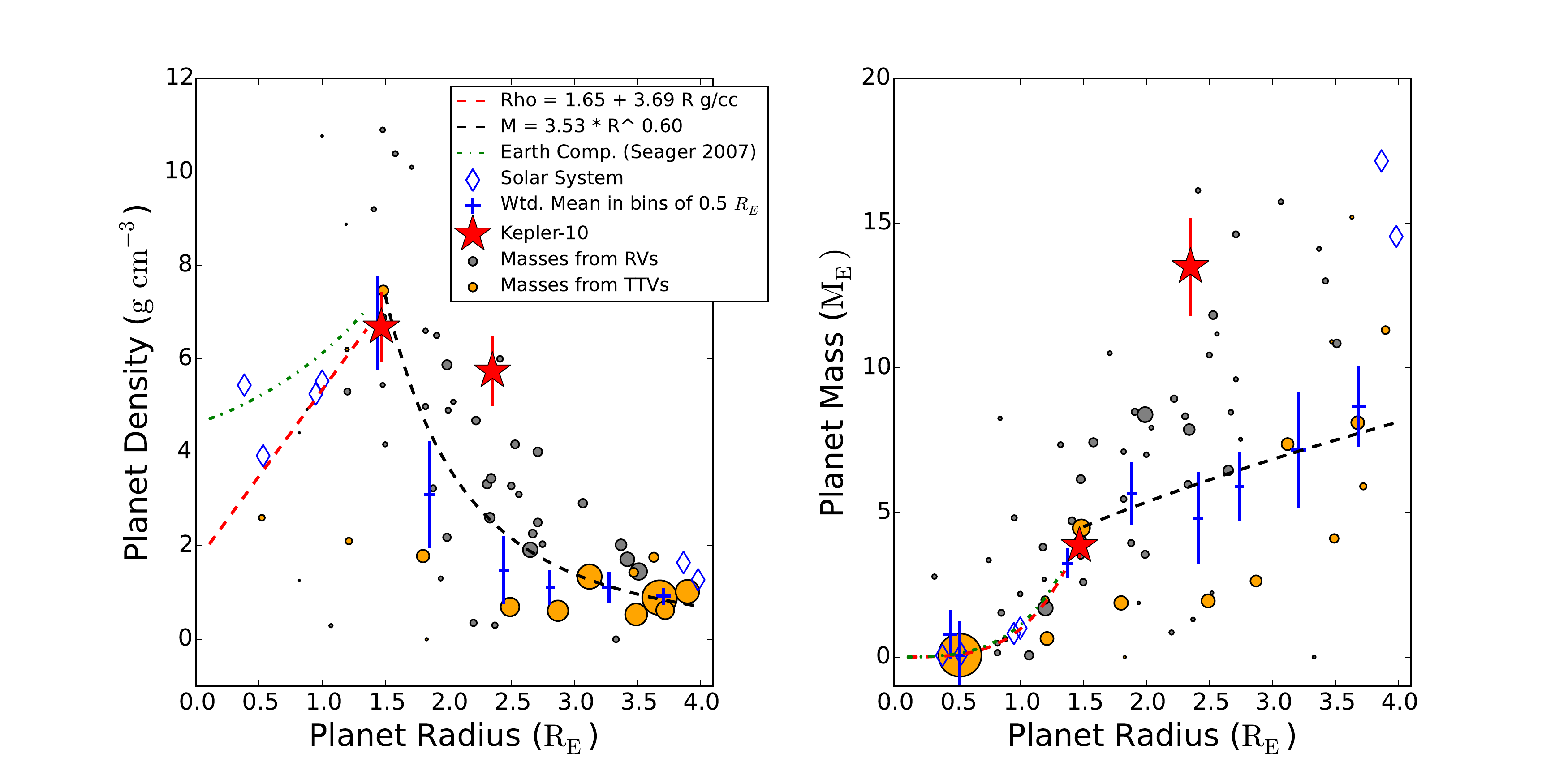}
\caption{Left: planet density as a function of planet radius for planets smaller than 4 \rearth, updated from \citet{Weiss2014}.  Planets with mass measurements from RVs are in gray; planets with mass measurements from TTVs are in gold.  The size of the dot corresponds to $1/\sigma_\rho^2$.  Kepler-10 b and c are shown as large red stars.  The solar system planets are shown as blue diamonds.  The blue crosses show the weighted mean density in bins of 0.5 \rearth\ to guide the eye, and their vertical error bars represent the RMS scatter of planet densities in that bin.  The green line is the predicted bulk density for a 32.5\% Fe, 67.5\% MgSiO$_3$ planet (like Earth) as a function of radius \citep{Seager2007}.  The red line is an empirical linear fit to planet density as a function of radius for $\rpl < 1.5 \rearth$.  The black line is an empirical power law fit to planet mass as a function of radius for $\rpl > 1.5 \rearth$.  Right: planet mass as a function of planet radius for $\rpl < 4 \rearth$.  The symbols and lines are the same as to the left, but the sizes of the dots corresponds to $1/\sigma_m^2$.}
\label{fig:mr_small}
\end{center}
\end{figure*}

We incorporate our newly derived mass measurements of the Kepler-10 planets, as well as several new planet discoveries, in an updated mass-radius relation.  The planets are the same as those in \citet{Weiss2014}, with the inclusion of planets listed in Table \ref{tab:mr}.  Figure \ref{fig:mr_small} shows how the weighted mean density and weighted mean mass of planets changes from 0-4 Earth radii.  

\begin{table*}[htbp]
\caption{Planetary Mass \& Radius Measurements from 2014-2015}
\label{tab:mr} 
\tablenum{8}
\begin{tabular}{lccccll}
\hline
\hline
\colhead{Name} & \colhead{Period} & \colhead{Mass} & \colhead{Radius} & \colhead{Insolation} & \colhead{First Ref.} & \colhead{Orbital Ref.}\\ 
\colhead{} & \colhead{(d)} & \colhead{($\mearth$)} & \colhead{($\rearth$)} & \colhead{($\searth$)} & \colhead{} & \colhead{}\\
\hline
Kepler-138 b & 10.3126 & 0.066 $\pm$ 0.048 & 0.522 $\pm$ 0.032 & 6.90 & \citet{Borucki2011} & Jontof-Hutter et al. 2015 \\
Kepler-138 c & 13.7813 & 1.97 $\pm$ 1.5 & 1.197 $\pm$ 0.070 & 4.75 & \citet{Borucki2011} & Jontof-Huttere al. 2015 \\
Kepler-138 d & 23.0881 & 0.640 $\pm$ 0.520 & 1.212 $\pm$ 0.075 & 2.35 & \citet{Borucki2011} & Jontof-Hutteret al. 2015 \\
HIP 116454 b & 9.120500 & 11.82 $\pm$ 1.33 & 2.530 $\pm$ 0.180 & 43.2 & \citet{Vanderburg2015} & \citet{Vanderburg2015} \\
Kepler-93 b & 4.726740 & 4.02 $\pm$ 0.68 & 1.478 $\pm$ 0.019 & 278 & \citet{Borucki2011} & \citet{Dressing2015} \\
KOI-273 b	& 10.600000 & 5.46	$\pm$ 2.50 & 1.820	$\pm$ 0.100 & 119 &\citet{Borucki2011} & \citet{Gettel2015}\\
Wasp-47 e & 0.789597 & $ < 22 $ & 1.829 $\pm$ 0.070 & 3998 & \citet{Becker2015} & \citet{Becker2015}\\
Wasp-47 d & 9.03081 & 15.2 $\pm$ 7 & 3.63 $\pm$ 0.14 & 155 & \citet{Becker2015}& \citet{Becker2015}\\
\hline
\end{tabular}
\end{table*}

Figure \ref{fig:mr_small} shows where Kepler-10 b and c sit in the density-radius and mass-radius diagrams for planets smaller than 4 \rearth.  As one of the first rocky exoplanet discoveries, Kepler-10 b has shaped our expectations for rocky exoplanets.  Like Kepler-10 b, most of the rocky exoplanets discovered so far are larger than the Earth.  This is because larger planets are easier to detect in transit, and because the masses of large, rocky planets are easier to measure than the masses of small, possibly rocky planets.  

Kepler-10 b, at 1.47 \rearth, sits very close to the peak of the density-radius diagram (at 1.5 \rearth).  Planet density increases with planet radius for $\rpl < 1.5 \rearth$, not only in the solar system terrestrial planets, but also among the few exoplanets with measured masses in this radius range (Kepler-78 b, Kepler-10 b, Kepler-36 b, Corot-7 b, Kepler-93 b).  We attribute the empirical linear increase in planet density with radius to the slight compressibility of rock.  The density of rock changes gradually with increasing planet mass or radius, and so the first-order (linear) Taylor expansion of the true density-radius relation of rock is sufficient to describe the observed terrestrial planets and exoplanets smaller than 1.5 \rearth.  Kepler-10 b (\rpl = 1.47 \rearth, \rhopl = 6.68 \gcc) sits exactly on the linear density-radius relation that describes the terrestrial planets. Kepler-10 b also achieves a density consistent with an Earth-like composition \citep[67.5\% MgSiO$_3$, 32.5\% Fe,][]{Seager2007}.

Kepler-10 c, at 2.35 \rearth, sits to the right of the peak of the density-radius diagram.  Its density (5.74 \gcc) is among the highest for planets between 2.0 and 2.5 \rearth.  However, its density is still lower than the density of Kepler-10 b.  Neither the empirical density-radius relation for rocky planets, nor the \citet{Seager2007} prediction for the densities of rocky planets, intersects with Kepler-10 c's position on the mass-radius diagram.  Extrapolating from both of these relations, we would expect the density of a 2.35 \rearth\ Earth-composition planet to be 11 \gcc, nearly twice the observed density for Kepler-10 c.  The mass and radius of Kepler-10 c are inconsistent with a stony-iron planet.  

Composition modeling for Kepler-10 c strongly disfavors a rocky interpretation.  Less than 2\% of the posterior distribution on the planet's mass and radius permits a planet denser than pure perovskite.  A stony-iron composition like the Earth and other known rocky exoplanets is excluded with high confidence.  Kepler-10 c may have a stony-iron interior overlaid with a 0.2\% mass (16\% radius) hydrogen-helium envelope.  Alternatively, Kepler-10 c may be a stony-iron interior covered with super-ionic water and a steam envelope.

Planet candidate KOI-72.X cannot be shown on the mass-radius diagram because it does not transit, and so its radius cannot be measured.

\section{Conclusions}
In this paper, we present the revised masses and densities of Kepler-10 b and c based on 220 RVs from two telescopes, 1.5 times as many RVs as were used in the most recent analysis of this planetary system.  The combined RVs yield a baseline 3 times as long as any previous RV publication for this system.  We find Kepler-10 b has $m_b=\mbcirc$ and $\rho_b=\rhobcirc$, and Kepler-10 c has $m_c=\mccirc$ and $\rho_c=\rhoccirc$.  However, we note that analysis of only HIRES data yields a higher mass for planet b and a lower mass for planet c than does analysis of the HARPS-N data alone, with the mass estimates for Kepler-10 c being formally inconsistent at the $3\sigma$ level.  While we cannot identify the source of the disagreement between the HIRES and HARPS-N RVs, we note that the apparent time-variability of $K_b$ and $K_c$ in both instruments suggests that time-correlated noise, in the form of either stellar activity or an additional planet, is responsible for the apparent discrepancy.  The time-correlated noise indicates that the uncertainties in the masses of the planets (especially planet c) likely exceed our formal estimates.  More RVs and/or better priors on the stellar rotation period are needed to adequately model the time-correlated noise without compromising our analysis of the planetary signal of Kepler-10 c.  We jointly analyze the TTVs and RVs of the Kepler-10 system to find planet candidate KOI-72.X.  Several possible orbital solutions exist for KOI-72.X, with orbital periods ranging from 21-100 days, and masses ranging from 1-7\mearth.  The existence of KOI-72.X has a negligible effect on the mass solutions for Kepler-10 b and Kepler-10 c.
\subsection{Compositions of Kepler-10 b and Kepler-10 c}
Kepler-10 b is very likely a planet with a rocky surface.  Its density is consistent with a stony-iron composition (90.8\% of the posterior distribution is denser than MgSiO$_3$).  Assuming a 2-layer model with an iron core and silicate mantle, the iron core mass fraction of Kepler-10 b is constrained at $0.17\pm0.12$.  Kepler-10 c is inconsistent with a purely rocky composition (only 1.6\% of its posterior distribution is denser than pure MgSiO$_3$).  Kepler-10 c may be a stony-iron interior overlaid with a 0.2\% mass (16\% radius) hydrogen-helium envelope.  Alternatively, Kepler-10 c may be a stony-iron interior covered with a 28\% mass (29\% radius) super-ionic water envelope.
\subsection{Non-transiting planet candidate KOI-72.X}
From the coherent TTVs of Kepler-10 c, we identify a third, non-transiting planet candidate KOI-72.X.  There is a 3\% probability that the coherent TTVs are due to noise rather than a third planet.  We explore possible solutions for the third planet, especially solutions that satisfy the TTV equation for the observed super-period of the TTVs, by running numerical N-body simulations over the duration of the \Kepler\ observations.  We demonstrate that orbital periods for the third planet that satisfy the TTV equation better reproduce the observed TTVs than other orbital periods do.  Some orbital solutions for the third planet are interior to Kepler-10 c, while others are exterior.  The most likely orbital periods for the non-transiting planet are 101 days, 24 days, and 71 days, which are near the 2:1, 1:2, or 3:2 mean motion resonance with planet c.  The mass for the non-transiting planet candidate ranges from $1-7\mearth$.

LMW gratefully acknowledges support from Kenneth and Gloria Levy.  This research was supported in part by the National Science Foundation under Grant No. NSF PHY11-25915 via the Dynamics of Exoplanets workshop at the Kavli Institute for Theoretical Physics in Santa Barbara, CA.  We thank Tsevi Mazeh for
informative discussions regarding time-correlated noise.  The authors wish to extend special thanks to those of Hawaiian ancestry on whose sacred mountain of Maunakea we are privileged to be guests. Without their generous hospitality, the Keck observations presented herein would not have been possible.

\bibliography{K10_v3_arxiv}{}
\bibliographystyle{apj}

\end{document}

%% file: physical_quantities.tex
\newcommand{\ms}{\ensuremath{\rm m\,s^{-1}}}

\newcommand{\gcmc}{\ensuremath{\rm g\,cm^{-3}}}
\newcommand{\gcc}{\gcmc}

\newcommand{\teff}{\ensuremath{T_{\rm eff}}}
\newcommand{\logg}{\ensuremath{\log{g}}}
\newcommand{\vsini}{\ensuremath{v \sin{i}}}
\newcommand{\feh}{[Fe/H]}

\newcommand{\rsun}{\ensuremath{R_\sun}}
\newcommand{\msun}{\ensuremath{M_\sun}}
\newcommand{\lsun}{\ensuremath{L_\sun}}

\newcommand{\rstar}{\ensuremath{R_\star}}
\newcommand{\mstar}{\ensuremath{M_\star}}

\newcommand{\lstar}{\ensuremath{L_\star}}

\newcommand{\rpl}{\ensuremath{R_{\rm p}}}
\newcommand{\mpl}{\ensuremath{M_{\rm p}}}

\newcommand{\rhopl}{\ensuremath{\rho_{\rm p}}}

\newcommand{\rearth}{\ensuremath{R_\earth}}
\newcommand{\mearth}{\ensuremath{M_\earth}}

\newcommand{\searth}{\ensuremath{S_\earth}}